\documentclass{dmathesis}
\usepackage{psfrag}
\usepackage{subfigure}
\usepackage{fancyhdr}
\usepackage{epsfig}
\usepackage{cite}
\usepackage{graphicx}
\usepackage{amsmath}
\usepackage{theorem}
\usepackage{amssymb}
\usepackage{latexsym}
\usepackage{epic}

\newcounter{ind}

{\theorembodyfont{\rmfamily}}
{\theorembodyfont{\rmfamily}}
{\theorembodyfont{\rmfamily}}
{\theorembodyfont{\rmfamily}}
{\theorembodyfont{\rmfamily}}
{\theorembodyfont{\rmfamily}}

\def\half{\textstyle\frac{1}{2}}
\def\quarter{\textstyle\frac{1}{4}}

\def\ie{{\it i.e.,}}
\newcommand{\be}{\begin{equation}}
\newcommand{\ee}{\end{equation}}
\newcommand{\bea}{\begin{eqnarray}}
\newcommand{\eea}{\end{eqnarray}}
\newcommand{\bml}{\begin{subequations}}
\newcommand{\eml}{\end{subequations}}

\begin{document}

\setcounter{tocdepth}{5}
\setcounter{secnumdepth}{5}


\pagenumbering{roman}

\setcounter{page}{1}

\newpage

\thispagestyle{empty}
\begin{center}
  \vspace*{1cm}
  {\Huge \bf Braneworld Cosmology and Holography}

  \vspace*{2cm}
  {\LARGE\bf Antonio Padilla}

  \vfill

  {\Large A Thesis presented for the degree of\\
         [1mm] Doctor of Philosophy}
  \vspace*{0.9cm}
  
   \begin{center}
   \includegraphics{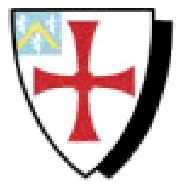}
   \end{center}

  {\large Centre for Particle Theory\\
          [-3mm] Department of Mathematical Sciences\\
          [-3mm] University of Durham\\
          [-3mm] England\\
          [1mm]  August 2002}

\end{center}

\newpage
\thispagestyle{empty}
\begin{center}
 \vspace*{2cm}
  \textit{\LARGE {For my Mum y para mi Papaito}}\\ 
\end{center}

\newpage
\thispagestyle{empty}
\addcontentsline{toc}{chapter}{\numberline{}Abstract}
\begin{center}
  \textbf{\Large Braneworld Cosmology and Holography}

  \vspace*{1cm}
  \textbf{\large Antonio Padilla}

  \vspace*{0.5cm}
  {\large Submitted for the degree of Doctor of Philosophy\\ August
  2002}

  \vspace*{1cm}
  \textbf{\large Abstract}
\end{center}
This thesis is devoted to studying two important aspects of braneworld
physics: their cosmology and their holography. We examine the Einstein
equations induced on a general $(n-2)$-brane of arbitrary tension,
embedded in some $n$-dimensional bulk. The brane energy-momentum
tensor enters these equations both linearly and quadratically. From
the point of view of a homogeneous and isotropic brane we see quadratic deviations from the FRW
equations of the standard cosmology. There is also a contribution from
a bulk Weyl tensor. We study this in detail when the bulk is
AdS-Schwarzschild or Reissner-Nordstr\"om AdS. This contribution can be understood
holographically. For the AdS-Schwarzschild case, we show that the
geometry on a brane near the AdS boundary is just that of a radiation
dominated FRW universe. The  radiation comes from a field theory
that is dual to the AdS bulk. We also develop a new approach which
allows us to consider branes that are not near the AdS boundary. This
time the dual field theory contributes quadratic energy
density/pressure terms to the FRW equations. Remarkably, these take
exactly the same form as for additional matter placed on the brane by
hand, with no bulk Weyl tensor. 

We also derive the general
equations of motion for a braneworld containing a domain wall. For the
critical brane, the induced geometry is identical to that of a
vacuum domain wall in $(n-1)$-dimensional Einstein gravity. We
develop the tools to construct a nested Randall-Sundrum scenario
whereby we have a ``critical'' domain wall living on an anti-de Sitter
brane. We also show how to construct instantons on the
brane, and calculate the probability of false vacuum decay.

\chapter*{Declaration}
\addcontentsline{toc}{chapter}{\numberline{}Declaration}
The work in this thesis is based on research carried out at the
Centre for Particle Theory, Department of Mathematical Sciences,
University of Durham, England.  No part of this thesis has been
submitted elsewhere for any other degree or qualification and is all
my own work unless referenced to the contrary in the text. 

Chapter~\ref{chap:RSbw} of this thesis is a review of necessary
background material. Chapter~\ref{chap:bwcosmo} is also a review,
although many of the results have been generalised to arbitrary
dimension for later convenience. Chapter~\ref{chap:bubbles} is
original work done in collaboration with my supervisor, Dr. Ruth
Gregory~\cite{Gregory:nested,
Gregory:instantons}. Chapter~\ref{chap:holography} contains some
initial review, but section~\ref{sec:CFTonnoncrit} onwards is all my
own work~\cite{Padilla:CFT}. Chapter~\ref{chap:exact} contains
original work done in collaboration with James Gregory~\cite{Padilla:exact}.

\vspace{2in}
\noindent \textbf{Copyright \copyright\; 2002 by Antonio Padilla}.\\
``The copyright of this thesis rests with the author.  No quotations
from it should be published without the author's prior written consent
and information derived from it should be acknowledged''.

\chapter*{Acknowledgements}
\addcontentsline{toc}{chapter}{\numberline{}Acknowledgements}
I think the first words of thanks have to go to my Mum and Dad. There
is no way I would have got this far without their love and support,
particularly since I came to Durham. I hope this thesis justifies
some of the faith you had in me, that I never even had myself. Thanks
also to Ram\'on and Susie. You've both looked out for your little
brother, even if he did used to break your things. And speaking of
breaking things, I have to mention Mr B., and of course my favourite
God-daughter, Big Face. I love you all, even if I don't always show
it.

Thanks  to everyone here in the department, particularly, Ruth, my
supervisor. You made me sweat early on but I know it did me
good, so thankyou. Thanks also to Clifford, Simon, Douglas,  Saffers,
Bert, Breacher, and Christos for
answering all my rubbish questions. Thanks to the lads in the office. Lord, thanks for mixing with us
commoners. Taff, thanks for the silly beard. Dave, thanks for helping
me annoy
Bostock and for trying to ruin this thesis. And James, thanks for not mixing me up with a budgie.

Thanks to G\'erard Houllier for the treble.

Thanks to all my mates at Grey for making it such a
laugh. Lewis, Robbie and Garbo: I'll never forget Madrid, even if
Garbo did miss the flea market. Thanks to Wokesie, Pease, John and Pinks, and all
those who love ``Bounce''. Thanks to Kerrie for being normal, when not
everyone around us was! Thanks to all my old mates
too: Hendo (for Aftershock), Neil (for Champagne), Bellers, Burrell,
Cheryl and  Sonia. Smarty, its been nice being able to use the grill pan again!

The last word goes to Perks. I'm so glad I came to Durham because here
I found you. You make me feel a much better man than I undoubtedly
am. Thankyou for being so lovely, and for making me so happy here. I
love you and I always will.

\tableofcontents
\clearpage


\pagenumbering{arabic}
\setcounter{page}{1}
\setcounter{equation}{0}
\chapter{Introduction} \label{chap:intro}
\section{From three to four dimensions}                                 
For centuries, physicists and philosophers have puzzled over the dimension of our
universe. Why is it we only experience three spatial
dimensions? Kepler~\cite{Kepler} reasoned that the threefold
nature of the Holy Trinity~\cite{God} was responsible. The advent of Special
Relativity~\cite{Einstein:SR} and Maxwell's theory of electromagnetism led to Minkowski's
suggestion~\cite{Minkowski} that we should understand physics geometrically in four-dimensional
\it spacetime \rm rather three-dimensional \it space\rm. As
observers, we only notice the ``mixing'' of space and time at very
high speeds, through phenomena such as length contraction and time
dilation. Ever since Minkowski's breakthrough, physicists have been
tempted to play with the dimensionality of our universe, either to
find new explanations to old problems, or to ``tidy up'' existing
theories. A particularly important example of this was {\it Kaluza-Klein
theory}~\cite{Kaluza, Klein:1, Klein:2}. For a nice introduction to higher
dimensions, see~\cite{Overduin:KKgravity}.
\section{Kaluza-Klein theory}
Kaluza's~\cite{Kaluza} aim was to unify
gravity and electrodynamics. Gravity is well described at a
classical level by the General Theory of
Relativity~\cite{Einstein:GR}. This states that matter causes
the universe to curve, with particles moving along geodesics in this
curved geometry. If matter is described by the four-dimensional
energy-momentum tensor, $T_{\mu\nu}$, and  $G$ is Newton's constant, then
\be \label{Eeqn}
R_{\mu\nu}-\half R g_{\mu\nu}=8\pi G T_{\mu\nu}
\ee
where $g_{\mu\nu}$, $R$ and $R_{\mu\nu}$ are the metric, Ricci scalar
and Ricci tensor of our universe. The Einstein equations
(\ref{Eeqn}) can be derived from the Einstein-Hilbert action
\be \label{EHaction}
S_G= S_m+\frac{1}{16\pi G}\int d^4 x ~\sqrt{g}R
\ee
where $g=\det g_{\mu\nu}$ and 
\be
T_{\mu\nu}=\left(\frac{2}{\sqrt{g}}\right)\frac{\delta S_m}{\delta g^{\mu\nu}}
\ee

Meanwhile, the Maxwell equations for a gauge potential, $A_{\mu}$,
coupled to a source of electromagnetic current, $j_{\mu}$, are given by
\be \label{Meqn}
\nabla_{\mu} F^{\mu\nu}=-\mu_0j^\nu,
\ee 
where
$F_{\mu\nu}=\partial_{\mu}A_{\nu}-\partial_{\nu}A_{\mu}$. Equation (\ref{Meqn}) can be derived from the following action
\be \label{EMaction}
S_{EM}=\tilde S_m-\frac{1}{4\mu_0}\int d^4 x ~\sqrt{g}F^2 
\ee
where 
\be
j^{\mu}=\left(\frac{1}{\sqrt{g}}\right)\frac{\delta \tilde S_m}{\delta A_{\mu}}
\ee
If we add together the  actions
(\ref{EHaction}) and (\ref{EMaction}) we get Einstein-Maxwell theory
for gravity coupled to an electromagnetic field. Kaluza's idea
was to consider pure gravity in five dimensions. Ignoring matter
terms, the five-dimensional action is simply
\be \label{KKaction}
S=\int d^4x dz \sqrt{\tilde g}\tilde R
\ee
where $\tilde g_{AB}$ is the five dimensional metric, and $\tilde R$
is the corresponding  Ricci scalar. Note that we have the original
four dimensions labelled with coordinates $x^{\mu}$ where
$\mu=0,1,2,3$. The fifth dimension is compactified on a circle and is
labelled by the coordinate $0 \leq z \leq L$.  

Now we can expand the metric as  a Fourier series of the form
\be
\tilde g_{AB}(x,z)=\sum_n \tilde g_{AB}^{(n)}(x)e^{inz/L}.
\ee
We find that we get an infinite number of fields in four
dimensions. Modes with $n \neq 0$ correspond to massive fields with
mass $|n|/L$. The zero mode corresponds to a massless field. As we take
$L$ to be smaller and smaller we see that the mass of the first
massive field becomes very large. This means that if we compactify on
a small enough circle we can truncate to massless modes in the
four-dimensional theory. We can only see the extra dimension by
exciting massive modes which are at energies beyond our reach.

Let us now focus on the zero mode, $\tilde g_{AB}(x)$.  We could
define $\tilde g_{\mu\nu}$, $\tilde g_{\mu z}$ and  $\tilde g_{zz}$ to
be the four-dimensional fields $g_{\mu\nu}$, $A_{\mu}$ and $\phi$. In
order that our results are more transparent we will actually define
the components of the metric in the following way:
\be
\tilde g_{\mu\nu}=e^{2 \alpha \phi}g_{\mu\nu}+e^{2 \beta
\phi}A_{\mu}A_{\nu}, \qquad
\tilde g_{\mu z}= e^{2 \beta
\phi}A_{\mu}, \qquad \tilde g_{zz}=e^{2 \beta
\phi}.
\ee
where $\alpha=1/2\sqrt{3}$ and $\beta=-1/\sqrt{3}$. Since we have
truncated to the massless fields, we can integrate out the $z$ part of the
action (\ref{KKaction}). We find that the four-dimensional effective
action is given by
\be
S_{\textrm{eff}}=L\int d^4x \sqrt{g} \left(R-\half(\partial
\phi)^2-\quarter e^{-\sqrt{3}\phi} F^2\right)
\ee
Although we had set out to obtain Einstein-Maxwell theory, we have
ended up with an additional coupling to the scalar field $\phi$. It
turns out we cannot consistently set this field to zero. This
was a worry to the original authors but today we are more comfortable
with the idea that scalar fields might exist, such as the Higgs. Here, $\phi$ is known as
the dilaton.

Kaluza-Klein type compactifications can be more complicated than simply
compactifying on a circle. The important thing is that the extra
dimension is small so that we do not excite
massive modes.  We can truncate to massless modes and read off the
effective theory in four dimensions.   

We need not restrict ourselves to just one extra dimension either. In
fact, higher dimensions have become very fashionable in the last
twenty years, mainly due to the success of \it string theory \rm as a possible
quantum theory of gravity. At the quantum level, bosonic string theory
is only consistent\footnote{Actually, bosonic string theory contains a
tachyon, but we will ignore that here.} in twenty-six (!) dimensions, although this figure
is reduced
to ten when we introduce supersymmetry. Furthermore, there
are five distinct string theories which can be viewed as different
elements of an embracing new theory, M-theory~\cite{Witten:Mtheory,Schwarz:Mtheory, Duff:Mtheory}. M-theory lives in eleven dimensions and has eleven-dimensional supergravity as
its low energy limit. 

Traditionally we achieve the reduction down to four dimensions using
Kaluza-Klein techniques. If we start with a $(4+n)$-dimensional
theory, we compactify on a small $n$-dimensional manifold. Different
manifolds generally give different effective theories in four
dimensions. The one thing all of these manifolds have in common is that
they are very small, and compact.

There is, however, an alternative to Kaluza-Klein
compactification. This is the idea that we live on something called a
{\it braneworld}, where the extra dimension can be infinite. 
\section{Introduction to braneworlds}
The idea is that our four-dimensional
world  is nothing more  than   an  infinitesimally thin 3-brane,
embedded       in          a        $(4+n)$-dimensional      spacetime~\cite{Rubakov:domain,Akama:braneworld, Pavsic:bw0, Pavsic:bw1, Pavsic:bw2, Pavsic:bw3,   Pavsic:bw4, Pavsic:bw5,  Pavsic:bw6, Pavsic:bw7, Gibbons:membrane, Visser:exotic}. All Standard Model fields are
bound   to the brane,  although gravity   may  propagate into the extra
dimensions. 

Of particular interest to us here are the Randall-Sundrum
braneworlds~\cite{Randall:hierarchy,Randall:compactification}. There
are in fact two models. The Randall-Sundrum I
model~\cite{Randall:hierarchy} is introduced in detail in
section~\ref{sec:RS1}.  Here we have two 3-branes of equal and
opposite tension separated by some five-dimensional anti-de Sitter bulk. In order to
preserve Poincar\'e invariance on the branes, we fine tune the brane
tensions against the bulk cosmological constant. 

The most important quality of the Randall-Sundrum I model is that it provides
an ingenious approach to the {\it hierarchy problem}. We will describe
what this is  in more detail at the beginning of
section~\ref{sec:RS1}. For now, we note that it is the
problem of the Planck scale being so much larger than the weak
scale. Braneworld models avoid this by stating that the
fundamental Planck scale is of similar size to the fundamental weak
scale. It is only when we examine the {\it effective} theory on the
brane that we see the hierarchy between scales emerge. Unfortunately,
the simplest braneworld models simply transfer the problem by
requiring that the extra dimensions be very large. The Randall-Sundrum
I model, however, is more subtle than this. By having anti-de Sitter space between the
branes we get an exponential warp factor in the metric.  This ensures
that the effective four-dimensional Planck scale is much larger than
the weak scale, even when there is no hierarchy
in the fundamental five-dimensional theory. Crucially, this is
achieved without the need for the extra dimension to be very large.

Despite this success of RS1, there are still some physical problems
with the model, such as how one should stabilise the extra
dimension. For this reason, we will focus on its  successor, the Randall-Sundrum II
model~\cite{Randall:compactification}, which we discuss in detail in
section~\ref{section:RS2}. This time there is only one brane and an
infinitely large anti-de Sitter bulk. The brane tension is positive
and is once again fine tuned against the bulk cosmological constant to
ensure Poincar\'e invariance on the brane. The warp factor in the bulk
metric does not play the role of solving the hierarchy problem like in
RS1. Here it ensures that gravity is localised on the brane.

Recall that standard Kaluza-Klein compactifications ensure that gravity looks
four-dimensional by stating that the extra dimensions should be
small. In Randall-Sundrum II, the extra dimension is
infinite! Gravity is allowed to propagate into the extra dimension so we
would expect it to look five-dimensional even to an observer on the
brane. However, the warp factor causes metric
perturbations to be damped as they move away from the brane. This has
the effect that gravity looks four-dimensional, at least
perturbatively, to a braneworld observer. Randall-Sundrum II offers an
interesting ``alternative to compactification''.

RS2 branes are often referred to as {\it critical}
because the brane tension is fine tuned to a critical
value. This ensures that the metric induced on the brane is
Minkowski. If we relax this fine tuning we obtain {\it non-critical}
branes, which are discussed in section~\ref{sec:noncrit}. Branes whose
tension exceed the critical value have a de Sitter induced
metric. Those with a tension smaller than the critical value have an
anti-de Sitter induced metric. The de Sitter brane in particular is
important because our universe may have a small positive cosmological constant~\cite{Perlmutter:astro, Riess:astro}. 
\section{Braneworld cosmology}
The initial success of RS2, from a gravitational point of view, sparked
off a lot of interest, especially amongst cosmologists. In
particular, Shiromizu {\it et
al}~\cite{Shiromizu:einstein} calculated the Einstein equations induced
on the brane. In
chapter~\ref{chap:bwcosmo}, we generalise their work  to arbitrary dimensions. By this we mean
considering the geometry induced on an $(n-2)$-brane in an $n$-dimensional
bulk. We start by writing the energy-momentum tensor for the brane in
the following way:
\be
S_{ab}=-\sigma h_{ab}+\mathcal{T}_{ab}
\ee
where $\sigma$ is the brane tension, $h_{ab}$ the brane metric and
$\mathcal{T}_{ab}$ the energy-momentum of additional matter on the
brane. In the linearised analysis of chapter~\ref{chap:RSbw}, we take
$\mathcal{T}_{ab}$ to be small and ignore quadratic
contributions. However, from a cosmological point of view, it is
important to consider situations where $\mathcal{T}_{ab}$ is not
small. In this instance, we
use the Gauss-Codazzi formalism to derive the Einstein tensor on the
brane. Leaving the details until chapter~\ref{chap:bwcosmo}, we will
give a rough version of the result. If $R_{ab}$ and
$R$ are the Ricci tensor and scalar on the $(n-2)$-brane,
then 
\be \label{braneEtensor}
R_{ab}- \half R h_{ab}=-\Lambda_{n-1}h_{ab}+8 \pi
G_{n-1}\mathcal{T}_{ab}+\mathcal{T}^{(2)}_{ab}-E_{ab}.
\ee
The first two terms on the right hand side are what we would have
expected from Einstein gravity in $(n-1)$ dimensions: a cosmological
constant term and a linear matter term. The brane
cosmological constant depends on $\sigma$ and the bulk cosmological
constant. As we stated at the end of the last section, it vanishes for
critical branes, but not for non-critical branes. The Newton's
constant on the brane, $G_{n-1}$, turns out to be proportional to the
bulk Newton's constant, $G_n$, and the brane tension. This dependence
on the brane tension is often ignored although it turns out to be
very important when we study braneworld holography on non-critical
branes in chapter~\ref{chap:holography}. 

The last two terms on the right hand side of equation
(\ref{braneEtensor}) are the most interesting. The $E_{ab}$ term is
often referred to as the electric part of the bulk Weyl tensor. It
vanishes for a pure anti-de Sitter bulk, but can be non-zero if (say)
we have a bulk black hole. This term is best understood from a
holographic point of view so we will postpone its discussion until the
next section. 

The $\mathcal{T}^{(2)}_{ab}$ term is actually quite complicated. The important thing is that it is quadratic in
$\mathcal{T}_{ab}$.  In section~\ref{sec:FRWbrane}, we consider a
Friedmann-Robertson-Walker brane. The $\mathcal{T}^{(2)}_{ab}$ terms show
up in the FRW equations as quadratic terms in energy density and
pressure. If these quantities are small, we can neglect the
quadratic contribution. However, this might not be the case in the early
universe so the  $\mathcal{T}^{(2)}_{ab}$ terms could be important.

Braneworld cosmology deviates slightly from pure Einstein gravity in
$(n-1)$ dimensions. In
chapter~\ref{chap:bubbles}, we  consider non-perturbative gravity on
the brane in a
different way. We investigate what happens when we have a strongly
gravitating object such as a domain wall on the
brane~\cite{Gregory:nested, Gregory:instantons}. We can think of
this as a
domain wall within a domain wall. It turns out that the equations of
motion for this kind of configuration are completely integrable.

The most interesting solutions are the following: the domain wall
living on a critical RS brane, the nested Randall-Sundrum scenario,
and the Coleman-De Luccia instantons. The first of these yields a remarkable result. It turns out that the geometry induced on the
$(n-2)$-brane agrees exactly with what we would have expected from
$(n-1)$-dimensional Einstein gravity. Let us make this a little
clearer: suppose we have a domain wall of tension, $T$,  sitting in
$(n-1)$-dimensional flat space. If we do Einstein gravity in
$(n-1)$-dimensions we find that our flat spacetime has a certain
geometry. This geometry is {\it exactly} the same as the geometry on
an $(n-2)$-brane containing a nested domain wall, also of tension,
$T$. We see that  we have exact Einstein gravity on the
brane, even at a non-perturbative level.   

Although the original motivation was to look at strong gravity on the
brane, we have developed tools that enable us to construct other interesting
configurations. The nested Randall-Sundrum scenario has a ``critical''
nested domain wall living on an anti-de Sitter brane. The geometry
induced on the brane is the traditional RS2 geometry, in $(n-1)$ dimensions.

Staying with the cosmological theme, in section~\ref{sect:inst} we show how to construct
gravitational instantons on the brane. These are the braneworld
analogue of the Coleman-De Luccia
instantons~\cite{Coleman:vacuumdecay}. In this paper, the authors
calculate the probability of (say) a flat bubble spacetime nucleating in a de Sitter
false vacuum. This kind of instanton describes a first order phase
transition in the early universe. We show how to patch together our
solutions so as to create these instantons on a brane. We do the same
probability calculations and find that they agree
with~\cite{Coleman:vacuumdecay}, at least in certain limits.   
\section{Braneworld holography}
Having examined brane cosmology and strong brane gravity, we change
direction in chapter~\ref{chap:holography}, and discuss {\it
braneworld holography}. We begin by reviewing the holographic
principle.  For now, all we need to say is that this involves
projecting all the degrees of freedom in some volume on to its
boundary surface. The AdS/CFT correspondence~\cite{Maldacena:adscft, Witten:adscft, Gubser:adscft} is the first
concrete example of this principle in action. We find that a gravity
theory on $AdS_5 \times S^5$ is dual to a conformal field theory on
the boundary. Braneworld holography is slightly different to
AdS/CFT. The bulk gravity theory is conjectured to be dual to a field
theory on the brane. This field theory is cut-off in the ultra-violet,
and unlike in the AdS/CFT correspondence, it is coupled to gravity on
the brane.

The difficulty with braneworld holography is that we do not know the
precise nature of the dual field theory. We can, however, make use of
the coupling to gravity. If we place a black hole in the bulk, the
Hawking radiation causes the brane to heat up. Any dual field theory
that lives on the brane should absorb energy which we can try to calculate.

This procedure was first carried out for critical
branes~\cite{Savonije:braneCFT}, and is reviewed in detail in
section~\ref{sec:critical}. To summarise, we place a black hole of
mass, $M$, in an $n$-dimensional bulk, and consider a critical FRW brane near the
boundary of AdS. $M$ is measured by an observer  using the
bulk time coordinate, $t$. This should translate into the
energy of the dual field theory~\cite{Witten:thermal}. However, the
field theory lives on the brane, so we should use the brane time
coordinate, $\tau$. To find its energy, we need to scale the black hole mass with some
red-shift factor, $\dot t$, where dot denotes differentiation with
respect to $\tau$. By using conservation of
energy, we can also calculate the pressure on the brane.

Given that we have a FRW brane, we can write down FRW equations for
its cosmological evolution. If $Z(\tau)$ is the scale factor, and
$H=\dot Z/Z$ is the Hubble parameter, then
\bml\bea 
H^2 &=&-\frac{1}{Z^2} +\frac{c}{Z^{n-1}} 
\\
\dot H &=&
\frac{1}{Z^2}-\left(\frac{n-1}{2}\right)\frac{c}{Z^{n-1}} 
\eea \eml
where $c$ is proportional to $M$. This black hole mass term comes from the non-trivial bulk Weyl tensor, $E_{ab}$. Using the ideas just described, we
can calculate the energy density, $\rho$, and the pressure, $p$, of
the dual field theory, in terms of $M$, or equivalently, $c$. We find
that we can rewrite the FRW equations entirely in terms of field
theory quantities:
\bml\bea 
H^2 &=& -\frac{1}{Z^2}+\frac{16 \pi G_{n-1}}{(n-2)(n-3)}\rho  \\ 
\dot{H} &=& \frac{1}{Z^2}- \frac{8 \pi G_{n-1}}{(n-3)}(\rho+p)
\eea \eml
These are the FRW equations of the standard cosmology in $(n-1)$
dimensions. We see that we do indeed have a holographic description. On
the one hand the brane cosmology is driven by the bulk black hole. On
the other hand  it is driven by the energy-momentum of a dual field
theory. It turns out that for an uncharged black hole in the bulk,
this field theory behaves like radiation.

In section~\ref{sec:CFTonnoncrit}, we attempt to extend these ideas to
de Sitter and anti-de Sitter branes~\cite{Padilla:CFT}. This is not as straightforward as
we might have thought. We have to be more careful than to say that the
bulk energy is given by the black hole mass. Our calculation of the
bulk energy is affected by 
cutting the spacetime off at the brane. We use Euclidean quantum gravity
techniques to calculate the bulk energy from first principles, and
then multiply by a red-shift factor to get the energy of the field
theory. It turns out that various factors combine to give us a similar
holographic description to before. The only difference is that the FRW
equations now contain a cosmological constant term corresponding to
the de
Sitter or anti-de Sitter brane, as appropriate.
  
The main problem with all the analysis of
chapter~\ref{chap:holography} is that its relies on a number of
approximations. In particular, we assume that the brane is near the
AdS boundary. This has two implications. The first is that it enables
us to get a reasonable approximation for the bulk energy. The second is
that it means the cut-off in the field theory is fairly
insignificant. The dual field theory is nearly conformal, which is
consistent with it behaving like radiation. However, a general brane
trajectory does not need to go near the AdS boundary. In chapter~\ref{chap:exact}, we take a completely
different approach to braneworld holography~\cite{Padilla:exact}. We
modify the Hamiltonian technique of Hawking and
Horowitz~\cite{Hawking:hamiltonian} to calculate the energy of the
dual field theory {\it exactly}, with no assumptions made about the
position of the brane. As a result, we can also get an exact
expression for the pressure. We end up with a highly non-trivial
equation of state that simplifies to radiation only as the brane gets
nearer to the AdS boundary. The really interesting result,
however, lies
in the effect on the FRW equations. When we express these equations
using the exact braneworld quantities, we find that they take the
following form: 
\bml
\bea
H^2 &=& a - \frac{1}{Z^2} + \frac{8 \pi G_n \sigma_n}{n-2} \rho + \left (
\frac{4 \pi G_n}{n-2} \right)^2 \rho^2  \\
\dot{H} &=& \frac{1}{Z^2}-4 \pi G_n \sigma_n (\rho+p) -(n-2)\left (
\frac{4 \pi G_n}{n-2} \right)^2\rho  (\rho+p) 
\eea
\eml
where where we have included the possibility of a brane cosmological
constant in the $a$ term, and $\sigma_n=4 \pi G_n/(n-2)$. Although
these equations do not correspond to the FRW equations for the
standard cosmology, they have exactly the same form as the
unconventional braneworld cosmology we discussed in the last
section, complete with quadratic energy-momentum terms. When these
equations are encountered in chapter~\ref{chap:bwcosmo}, they
correspond to a brane moving in a pure anti-de Sitter bulk, with
additional matter placed on the brane by hand. In
chapter~\ref{chap:exact}, they have a very different origin. There is
no additional matter on the brane although we now have a black hole in
the bulk. When we derive properties for the dual field theory from the
black hole, we find that the field theory behaves exactly as if it had
been placed on the brane by hand. This means that the dual descriptions
of chapter~\ref{chap:holography} are merely an approximation of this
larger relationship. 

We conclude this thesis in chapter~\ref{chap:discuss} with some
general thoughts and discussion. The main results are
stated and interpreted as we go along.

\chapter{Randall-Sundrum Braneworlds} \label{chap:RSbw}
\setcounter{equation}{0}
\section{Randall-Sundrum I (RS1)} \label{sec:RS1}
In a four-dimensional world there  are at least two fundamental energy
scales: the weak scale, $m_{EW} \sim 10^3 ~\textrm{GeV}$ and the Planck
scale, $m_{pl} \sim 10^{19}  ~\textrm{GeV}$. Physics is well described
by the Standard Model at least up to $100$ GeV or so.  At the Planck scale,
gravity becomes as strong as the SM interactions  and a quantum theory
of gravity  is required. Why is  there such a vast  difference between
the two scales?  This question is the essence of the
{\it hierarchy problem}. Consider the  Higgs  boson whose physical  mass,
$m_H  \sim m_{EW}$. Now  suppose our theory   is cut-off at some large
scale $\Lambda$, where  $m_H \ll \Lambda$.  When  we calculate the one
loop correction for the Higgs mass we find that $\delta m_H^2 \sim
\Lambda^2$. The bare mass must then be of order $-\Lambda^2$ to give a
renormalised mass  near  the  weak scale.   If  we believe   that  our
fundamental  theory contains scales as high  as the Planck scale, then
the cancellation just  described  is disturbingly precise,  given  the
huge numbers  involved.  What  is   more, this  bizarre precision   is
required again at all subsequent orders of perturbation theory.

Traditionally, it is  thought that this vast  desert  between the weak
and the Planck  scales must be   populated with new theories,  such as
supersymmetry. Above the scale of supersymmetry breaking, the problems
with radiative corrections to the  Higgs mass are solved, although  we
may still ask why the desert exists at all. There is, however, another
solution  to the hierarchy  problem  that  is radically  different  to
supersymmetry.  We assume that  there  is only one fundamental  energy
scale, the weak scale.  The  large (effective) Planck scale comes from
extra dimensions, beyond  the traditional four.  As observers, we are bound to a braneworld embedded in a $(4+n)$-dimensional spacetime.  The $(4+n)$-dimensional  Planck  scale, $M$, is  now  the
fundamental  scale of gravity, and  is taken to  be  of order the weak
scale.   The extra dimensions are  given by an $n$-dimensional compact
space  of      volume    $V_n$.      In    the      simplest     cases
~\cite{Arkani:hierarchy,Arkani:phen, Antoniadis:newdim}, our effective
four-dimensional Planck scale is given by
\begin{equation}
m_{pl}^2=M^{n+2}V_n.
\end{equation}
By taking $V_n$  to be sufficiently large  we can recover $m_{pl} \sim
10^{19} ~\textrm{GeV}$.  However, in  some sense the hierarchy problem
has not gone away. There is now a new hierarchy between the weak scale
and   the          compactification    scale,     ~$1/V_n^{1/n}    \ll
m_{EW}$.  Fortunately, the Randall-Sundrum  I (RS1)
model~\cite{Randall:hierarchy} is an extension of  these ideas that
does not  appear to transfer the problem in this
way\footnote{Actually, the hierarchy problem remains if we
consider fluctuations in the ``radion'' field. We will comment on this
later.}.

\subsection{The model}
\begin{figure}
\begin{center}
\includegraphics[width=8cm]{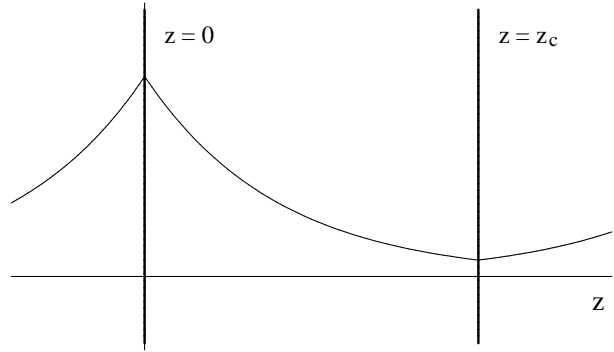}
\vskip 5mm
\caption{The behaviour of the warp factor in the RS1 model} \label{fig:RS1}
\end{center}
\end{figure}

In RS1, we have two 3-branes embedded in a
five dimensional anti-de Sitter bulk spacetime. We define $x^{\mu}$ to
be the familiar  four-dimensional coordinates
while  $0   \leq  z  \leq  z_c$   is the   coordinate   for   the extra
dimension. Since  our spacetime clearly fails  to fill out all of
the five  dimensions we need  to specify boundary conditions: identify
$(x^{\mu}, z)$ with $(x^{\mu},  -z)$ and take  $z$ to be periodic with
period $2z_c$.  The orbifold fixed points at $z=0, ~ z_c$ are the
positions of the  two branes,  which  we  will  take to  have  tension
$\sigma_0, ~\sigma_c$ respectively.   These fixed points may  also  be
thought of as the boundaries
of the five-dimensional spacetime so that the action describing this model is given by
\begin{equation} \label{eqn:RS1action}
S = M^3\int d^4x \int_{-z_c}^{z_c} dz \sqrt{g}\left(R - 2\Lambda
\right)~ -\sigma_0 \int_{z=0}  d^4x \sqrt{h_0} ~ -\sigma_c \int_{z=z_c}
d^4x \sqrt{h_c}~.
\end{equation}
where $g$ is the bulk  metric and  $h_0, ~h_c$ are the metrics on the
branes at $z=0, ~z_c$ respectively. $M$ is of course the five-dimensional
Planck scale.  We    now  require  the  3-branes to      exhibit
four-dimensional   Poincar\'e  invariance and choose the metric to
take the following form
\begin{equation} \label{eqn:RS1ansatz}
ds^2=a^2(z)\eta_{\mu\nu}dx^{\mu}dx^{\nu}+dz^2
\end{equation}
The bulk equations of motion with orbifold
boundary conditions  impose  a  fine tuning of the brane tensions against the bulk cosmological constant
\begin{equation} \label{eqn:finetune}
\sigma_0=-\sigma_c=12M^3k, ~\Lambda=-6k^2
\end{equation}
We are also free to set $a(0)=1$ so that we arrive at the following solution for the metric
\begin{equation} \label{eqn:RS1metric}
ds^2=e^{-2k|z|}\eta_{\mu\nu}dx^{\mu}dx^{\nu}+dz^2 \quad \textrm{for
} -z_c \leq z \leq z_c.
\end{equation}
The $\mathbb{Z}_2$ symmetry about $z=0$ is explicit whereas the other
boundary conditions should be understood. We also note that the
constant $z$ slicings exhibit Poincar\'e invariance as required. The
metric (\ref{eqn:RS1metric}) contains an exponential warp factor which
is seen graphically in figure~\ref{fig:RS1}. 
Notice the peak in the warp factor at the positive tension brane and
the trough at the negative tension brane. At this point we should
emphasize that RS1 is really only a toy model. It is, however,
possible to construct string theory/supergravity models that have similar
properties~\cite{Horava:heterotic, Horava:sugra, Lukas:domain, Duff:susyRS}.

\subsection{Tackling the hierarchy problem}

In order to  tackle the hierarchy problem,  we will need to derive the
(effective) four-dimensional  Planck scale, $m_{pl}$  in terms  of the
five-dimensional  scales $M, ~k, ~z_c$. We  do this by identifying the
four-dimensional low energy effective theory. This comes from massless
graviton fluctuations. In principle,  we should also  include massless
fluctuations in the brane separation~\cite{Charmousis:radion}, often referred to as the radion
field.   This does    not    affect  the   calculation  of    $m_{pl}$
directly~\cite{Pilo:radion} so we will
ignore the radion  in this section and  assume the brane separation is
stabilised at $z_c$. The gravitational zero modes now take the form
\begin{equation} \label{eqn:RS1graviton}
ds^2=e^{-2k|z|}\bar{g}_{\mu\nu}(x)dx^{\mu}dx^{\nu}+dz^2 \quad
\textrm{where } \bar{g}_{\mu\nu}=\eta_{\mu\nu}+h_{\mu\nu}(x)
\end{equation}
and we   interpret   $h_{\mu\nu}$ as  the   physical  graviton in  the
four-dimensional    effective   theory.  We   now   substitute equation
(\ref{eqn:RS1graviton})  into  the    action (\ref{eqn:RS1action})  to
derive the effective  action. Focusing   on  the
curvature term we find that
\begin{equation}
S_{eff} = M^3 \int d^4x ~\sqrt{\bar{g}}  \bar{R}\int_{-z_c}^{z_c}
dz~ e^{-2k|z|}+\ldots
\end{equation} 
where  $\bar{R}$  is        the   Ricci  scalar    built   out      of
$\bar{g}_{\mu\nu}(x)$. We now perform the $z$-integral to obtain
\begin{equation} \label{eqn:RS1planck}
m_{pl}^2=\frac{M^3}{k}\left[1-e^{-2kz_c}\right].
\end{equation}
This tells us that $m_{pl}$ depends weakly on $z_c$ in the limit
of large $kz_c$. We will see that this is not the case for the physical masses in the
SM.

Suppose we live on the negative tension brane at $z=z_c$. Consider a
fundamental  Higgs   field  bound   to  this brane.      If it  has  a
five-dimensional  mass parameter, $m_0$,  then  the matter part of the
action near the brane is given by
\begin{equation} \label{eqn:Higgsmetric1}
S_c                  =           \int_{z=z_c}                d^4x
\sqrt{g_c}\left[g_{c}^{\mu\nu}\nabla_{\mu}H^{\dagger}\nabla_{\nu}H-\lambda\left(|H|^2-m_0^2\right)^2
\right]
\end{equation}
where  $\nabla_{\mu}$    is  the   covariant   derivative  corresponding to
$g_c$. The metric at  $z=z_c$ is
$\bar{g}_{c\mu\nu}=e^{-2kz_c}\bar g_{\mu\nu}$ so that
\begin{equation} \label{eqn:Higgsmetric2}
S_c                  =          \int_{z=z_c}                d^4x
\sqrt{\bar g}e^{-4kz_c}\left[e^{2kz_c}\bar g^{\mu\nu}\nabla_{\mu}H^{\dagger}\nabla_{\nu}H-\lambda\left(|H|^2-m_0^2\right)^2
\right]
\end{equation}
We now renormalise the Higgs wavefunction, $H \to e^{kz_c}H$, to
derive the following part of the effective action
\begin{equation}
S_{eff}                  =            \int_{z=z_c}                d^4x
\sqrt{\bar g}\left[\bar g^{\mu\nu}\nabla_{\mu}H^{\dagger}\nabla_{\nu}H-\lambda\left(|H|^2-e^{-2kz_c}m_0^2\right)^2
\right]+\ldots
\end{equation}
An observer on the brane will therefore measure the physical mass of
the Higgs to be
\begin{equation} \label{eqn:Higgsmass}
m_{H}=e^{-kz_c}m_0.
\end{equation}
This result generalises to any mass parameter on the negative tension brane.

We shall now address the hierarchy problem directly. Assume that the
bare Higgs mass, $m_0$, and the fundamental Planck mass, $M$, are both
around $10^{19}$ GeV, thereby eliminating any hierarchy between
the two scales in the five-dimensional theory. The physical masses in
the effective theory are given by equations (\ref{eqn:RS1planck}) and (\ref{eqn:Higgsmass}). To
ensure that $m_{H} \sim 10^3 ~\textrm{GeV}$ and $m_{pl} \sim 10^{19} ~\textrm{GeV}$  we require
that $e^{kz_c} \sim 10^{15}$. The presence of the exponential here is
crucial because all we really need is $k z_c \sim 50$.  We
see that we  have solved the  hierarchy problem  without introducing a
second hierarchy involving  the compactification scale, $1/z_c$ or the
AdS length, $1/k$. We should emphasize here that this is only true if
the radion is stabilised. If not, its fluctuations appear in the
exponential, spoiling the solution to the problem.

At this point we should note that we have set the fundamental mass
scale to be around $10^{19}$ GeV. We could easily have chosen the
fundamental scale to be as low as a few TeV because what really matters
is the ratio between the physical masses, as this is a
dimensionless quantity. We can see this explicitly if we
change coordinates $x^{\mu} \to e^{kz_c}x^{\mu}$. The warp factor at
$z=z_c$ is unity, whereas at $z=0$ it is exponentially large,
$e^{2kz_c}$. This time, the Higgs mass does not get rescaled,
$m_{H} \sim m_0$, unlike the Planck mass which behaves like $m_{pl}^2 \sim e^{2kz_c}
\frac{M^3}{k}$. If both $M$ and $m_0$ are around a few TeV, we again
only need    $k z_c \sim 50$ to recover the correct physical masses in
the effective theory.

To summarise, even though all scales in the fundamental theory are near the
weak scale, the extra dimension ensures
that $m_{pl}$ is close to the large value we observe
in Nature. What  is more, this is achieved   without the need for  the
extra dimension to be disturbingly large. From a phenomenological point
of view this is particularly exciting. If the fundamental scale of
gravity is indeed as low as a few TeV then we would expect quantum
gravity effects to start showing up in forthcoming collider
experiments. The path to a ``theory of everything'' could be
dictated by experiment rather than the imagination. 
 
\section{Randall-Sundrum II (RS2)} \label{section:RS2}
When we introduced braneworlds at the start of this chapter we stated that the
Standard Model fields are localised on the
brane~\cite{Rubakov:domain, Akama:braneworld} in contrast to gravity
which can propagate into the fifth dimension. This should worry a braneworld observer because Newton's $1/r^2$ law for
gravitational force is a property of four-dimensional gravity and is
experimentally verified as low as $r \sim 0.2$~mm. The problem is
solved if the extra
dimension is small and compact owing to the large mass gap between the
graviton zero mode and the first heavy Kaluza-Klein mode. This
ensures that gravity behaves four dimensionally, except at very high
energies near the heavy mode masses. In braneworld models we have seen how the extra
dimension can be of order one or larger so we would naively expect
gravity to look five dimensional even at fairly low energies. This would violate Newton's law and be
unacceptable. The RS2 model is more subtle than this. Even though it has an infinite
extra dimension it still manages
to reproduce Newton's law on the brane. This is because we have a
negative cosmological constant in the bulk. RS2 does not solve the
hierarchy problem in the way that RS1 does, and is of interest from a
purely gravitational point of view.

\subsection{The model}

\begin{figure}
\begin{center}
\includegraphics[width=8cm]{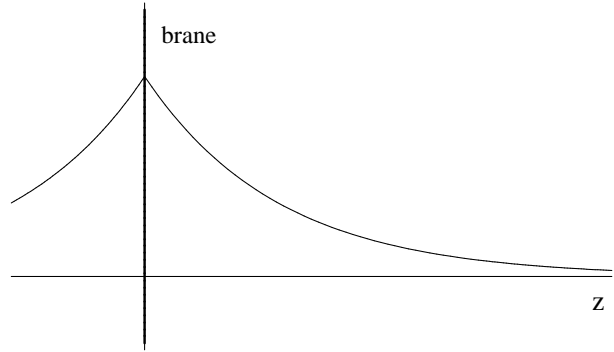}
\vskip 5mm
\caption{The behaviour of the warp factor in the RS2 model} \label{fig:RS2}
\end{center}
\end{figure}

To arrive at the RS2 model we start with RS1, and extend the brane separation to infinity so that we are left with a single brane of positive tension. The old negative tension brane will act as
a regulator in the subsequent analysis. The geometry of this new
set-up is again described by the metric (\ref{eqn:RS1metric}) with $z_c \to \infty$.  We can see the behaviour of the
warp factor in figure~\ref{fig:RS2}. 
It has a peak at $z=0$ indicating that the brane there has
positive tension. Note also the $\mathbb{Z}_2$ symmetry about $z=0$
which is, of course, explicit in the metric.

\subsection{Localisation of gravity} \label{section:local}
In the absence of any additional matter, we have a single brane with tension $\sigma=12M^3k$
embedded in five-dimensional anti-de Sitter space with cosmological
constant $\Lambda=-6k^2$. In order to investigate whether gravity is
localised on the brane, we will consider small gravitational
perturbations about the background metric 
\begin{equation} \label{eqn:RS2metric}
ds^2=\tilde
g_{ab}dx^adx^b=e^{-2k|z|}\eta_{\mu\nu}dx^{\mu}dx^{\nu}+dz^2
\end{equation}
This may be achieved by placing a
point mass on the brane, and solving the relevant perturbation equations. In the event of gravity localisation we would hope to see the
graviton zero mode dominating at large enough distances. This would
reproduce observed phenomena such as Newton's inverse square law and
gravitational light bending. In the remainder of this section we will
adopt Garriga and Tanaka's 
delightful approach to gravity in the
Randall-Sundrum model~\cite{Garriga:gravity}.
\subsubsection{The Newtonian potential on the brane} \label{sec:Newton}
We begin by deriving the Newtonian
potential due to a point mass, $m_0$, bound to the brane. If we denote the perturbed
metric by $g_{ab}=\tilde g_{ab}+h_{ab}$, the
Randall-Sundrum gauge~\cite{Randall:compactification} is given by
\begin{equation}\label{eqn:RSgauge}
h_{zz}=h_{\mu z}=0, \quad h_{\mu}{}^{\nu}{}_{,\nu}=0, 
\quad h^{\mu}{}_{\mu}=0. 
\end{equation}
Since we have no additional
matter in the bulk, the bulk equations of motion for $h_{ab}$ are
given by
\begin{equation} \label{eqn:Lich}
0=\delta R_{ab}=-\frac{1}{2}\Delta_L h_{ab}
\end{equation}
where $\Delta_L$ is the Lichnerowicz operator\footnote{The
Lichnerowicz operator is defined by $\Delta_{L} h_{ab}=\tilde \Box h_{ab} -2
\tilde \nabla_{(a}\tilde \nabla_{|c|} \bar{h}_{b)}^{c} -2\tilde R_{c(a}h^{c}_{b)}+2\tilde R_{acbd}h^{cd}$
where  $\bar{h}_{ab}=h_{ab}-\frac{1}{2}h\tilde g_{ab}$ and the covariant
derivative and Riemann tensor are constructed out of the
unperturbed metric $\tilde g_{ab}$.}. We are free to take the RS gauge (\ref{eqn:RSgauge}) everywhere in the bulk~\cite{Randall:compactification}
so that equation (\ref{eqn:Lich}) is reduced to
\begin{equation} \label{eqn:RSeq}
\left[e^{2k|z|}\Box^{(4)} + \partial_z^2 -4k^2\right]h_{\mu\nu}=0.
\end{equation}
Boundary conditions for this equation are given by the jump conditions
at the brane. However, if we take the RS gauge in the bulk
then additional matter causes the brane to bend
and we can no longer say that it lies at
$z=0$. For this reason, we will temporarily relax our choice
of gauge and work in Gaussian
normal (GN) coordinates, denoted by $(\hat x^{\mu}, \hat z)$. By definition, we now
have $\hat h_{zz}=\hat h_{\mu z}=0$ and can set the brane to be
located at $\hat z=0$. By using the Israel junction
conditions~\cite{Israel:junction} we can relate the jump in extrinsic
curvature\footnote{$\Delta K_{ab}=K_{ab}^+-K_{ab}^-$
where $K_{ab}^-=\hat g_{0~a}^{~c}\hat g_{0~b}^{~d}\nabla_{(c}n_{d)}$
and $n^a$
is the unit normal to the brane pointing in the direction of
increasing $z$, and $\hat g_{0ab}$ is the induced metric on the brane.}, $\Delta K_{ab}$, across the brane to the energy-momentum
tensor, $S_{ab}$ on the brane.
\begin{equation} \label{eqn:Israel}
\Delta K_{ab}=-8\pi
G_5\left(S_{ab}-\frac{1}{3}S\hat g_{0ab} \right).
\end{equation}
Here, $\hat g_{0ab}=\hat g_{ab}(\hat z=0)$ is the induced metric on the brane and
$G_5=1/16\pi M^3$ is the five-dimensional Newton's constant. Note that
the energy momentum tensor is dominated by the brane tension, $\sigma$
with a small additional contribution coming from the point mass,
$\mathcal{T}_{ab}$. Explicitly
\begin{equation}
S_{ab}=-\sigma \hat g_{0ab} +\mathcal{T}_{ab}.
\end{equation}
By imposing $\mathbb{Z}_2$ symmetry across
the brane we arrive at
\begin{equation} \label{eqn:GNjc}
\left( \partial_z+2k \right)\Big\vert_{z=0^+} \hat h_{\mu\nu}=-8\pi
G_5\left(\mathcal{T}_{\mu\nu}-\frac{1}{3}\mathcal{T}\eta_{\mu\nu} \right)
\end{equation}
where we have used the fine-tuning conditions (\ref{eqn:finetune}) and
have ignored all terms non-linear in $\hat h_{\mu\nu}$ and
$\mathcal{T}_{\mu\nu}$. Note that there are no $\mu z$ or $zz$ components
of equation (\ref{eqn:Israel}) because we chose a GN coordinate
system.
We will now attempt to construct the junction condition
(\ref{eqn:GNjc}) in the RS gauge. The most general transformation
between GN and RS gauge is given by
\begin{equation} \label{eqn:GNtoRS}
\xi^z = f(x^{\rho}),\qquad
\xi^{\mu} = -\frac{1}{2k}e^{2k|z|}\eta^{\mu\nu}\partial_{\nu}f +
F^{\mu}(x^{\rho})
\end{equation}
where $f$ and $F^{\mu}$ are independent of $z$. The perturbation in
the RS gauge, $h_{\mu\nu}$, is related to its GN counterpart by
\begin{equation} \label{eqn:gauget}
h_{\mu\nu} = \hat h_{\mu\nu} - \frac{1}{k} f_{,\mu\nu}
 -2 ke^{-2k|z|}\eta_{\mu\nu}f + e^{-2k|z|}\eta_{\rho(\nu}F^{\rho}{}_{,\nu)}, 
\end{equation}
Inserting this back into (\ref{eqn:GNjc}) we derive the junction
condition in
the RS gauge
\begin{equation} \label{eqn:RSjc}
\left( \partial_z+2k \right)\Big\vert_{z=0^+}
h_{\mu\nu}=-\Sigma_{\mu\nu}
\end{equation}
where
\begin{equation} \label{eqn:Sigma}
\Sigma_{\mu\nu}=8\pi
G_5\left(\mathcal{T}_{\mu\nu}-\frac{1}{3}\mathcal{T}\eta_{\mu\nu} \right)+2 f_{,\mu\nu}.
\end{equation}
Equations (\ref{eqn:RSeq}) and (\ref{eqn:RSjc}) fully define the bulk
equations of motion with boundary conditions at the brane. Given that
a solution must be 
$\mathbb{Z}_2$ symmetric about $z=0$, we see that $\partial_z
h_{\mu\nu}$  must be discontinuous there. Both (\ref{eqn:RSeq}) and
(\ref{eqn:RSjc}) can be contained in a single equation if we include
delta functions at the discontinuity.
\begin{equation} \label{eqn:fullRSeq}
\left[e^{2k|z|}\Box^{(4)} + \partial_z^2
-4k^2+4k\delta(z)\right]h_{\mu\nu}=-2\delta (z)\Sigma_{\mu\nu}
\end{equation}
Before we can solve  equation
(\ref{eqn:fullRSeq}) we need to identify $f(x)$. Nevertheless, we shall proceed blindly and
define  $G_R(x,z;x^{\prime}, z^{\prime})$ to be the five-dimensional
retarded Green's function satisfying
\begin{equation} \label{eqn:Greens}
\left[e^{2k|z|}\Box^{(4)} + \partial_z^2
-4k^2+4k\delta(z)\right]G_R(x,z; x^{\prime}, z^{\prime})=\delta^{(4)}(x-x^{\prime}) \delta(z-z^{\prime}).
\end{equation}
The solution to the perturbation equation (\ref{eqn:fullRSeq}) is then
given by
\begin{equation} \label{eqn:pert}
h_{\mu\nu}(x,z)=-2\int d^4x^{\prime}G_R(x,z; x^{\prime},
0)\Sigma_{\mu\nu}(x^{\prime})
\end{equation}
where we have integrated across the surface $z^{\prime}=0$. Since we
are in the RS gauge, $h^{\mu}_{\mu}=0$  and so 
\begin{equation} \label{eqn:feq}
\Sigma^{\mu}_{\mu}=0 \qquad \Rightarrow \qquad \Box^{(4)}f=\frac{4 \pi G_5}{3}\mathcal{T}.
\end{equation}
$f(x)$ represents the
brane position in RS gauge and in principle we can calculate it by
solving equation (\ref{eqn:feq}). Here we see explicitly that the brane is bent by the
presence of additional matter because $\mathcal{T}$ acts as a source
for $f(x)$.

In order to evaluate the full Green's function we will use techniques
from Sturm Liouville theory. We will
simply state the result here although a detailed derivation can be
found in appendix~\ref{app:Greens}.  
\begin{equation} \label{eqn:Greenssoln}
G_R(x,z;x',z')=-\int \frac{d^4 p}{(2\pi)^4}
e^{ip_{\mu}(x^{\mu}-x'{}^{\mu})}\Biggl[
\frac{e^{-2k(|z|+|z'|)} k}{ {\bf p}^2-(\omega+i\epsilon)^2}
+\int_0^{\infty} dm\,
 \frac{v_m(z) v_m(z')}{ m^2+{\bf p}^2-(\omega+i\epsilon)^2}\Biggr],
\end{equation}
where
\begin{equation}
v_m(z)=\frac{\sqrt{m/2k} \left[J_1(m/k)Y_2(me^{k|z|}/k) - 
Y_1(m/k) J_2(me^{k|z|}/k)\right]}{\sqrt{J_1(m/k)^2+Y_1(m/k)^2}}. 
\end{equation}
and $J_n, ~Y_n$ are Bessel's functions of integer order $n$.

If we return to GN coordinates, we can define the stationary point
mass $m_0$ to be located at $(t, {\bf x}, z)=(t, {\bf 0}, 0)$ so that
its energy momentum tensor on the brane is given by
\begin{equation} \label{eqn:src}
\mathcal{T}_{ab}=m_0\delta^{(3)}({\bf x})\textrm{diag}(1,0,0,0,0)
\end{equation}
Combining equation (\ref{eqn:gauget})
 with equation (\ref{eqn:pert}) we obtain an expression for the
 gravitational perturbation in this gauge.
\begin{equation}
\hat h_{\mu\nu}(x,z)=h_{\mu\nu}^{(m)}+h^{(f)}_{\mu\nu} +\frac{1}{k} f_{,\mu\nu}
 +2 ke^{-2k|z|}\eta_{\mu\nu}f -e^{-2k|z|}\eta_{\rho(\nu}F^{\rho}{}_{,\nu)}, 
\end{equation}
where the matter part and the brane bending part are given by
\begin{equation} \label{eqn:matterpert}
h_{\mu\nu}^{(m)}=-16\pi
G_5\int d^4x'
~G_R(x,z;x',0)\left(\mathcal{T}_{\mu\nu}-\frac{1}{3}\mathcal{T}
\eta_{\mu\nu} \right)
\end{equation}
\begin{equation}
h^{(f)}_{\mu\nu}=-4 \int d^4x'
~G_R(x,z;x',0)f_{, \mu\nu}
\end{equation}
Since we are only interested in the perturbation on the brane, we set $z=0$, and can choose $F^{\mu}$ appropriately so that
\begin{equation} \label{eqn:fullpert}
\hat h_{\mu\nu}(x,0)=2k\eta_{\mu\nu}f-16\pi
G_5\int d^4x'~
G_R(x,0;x',0)\left(\mathcal{T}_{\mu\nu}-\frac{1}{3}\mathcal{T}
\eta_{\mu\nu} \right)
\end{equation}
To evaluate $f(x)$,  we solve equation (\ref{eqn:feq}) with
$\mathcal{T}=m_0\delta^{(3)}({\bf x})$. Note that our source is
stationary so we look for time independent solutions. With this
ansatz, the differential operator in
equation (\ref{eqn:feq}) is reduced to the Laplacian so that
\begin{equation} \label{eqn:fsoln}
f(x)=\frac{G_5m_0}{3r}
\end{equation}
where $r=|{\bf x}|$. We now evaluate the matter part of the
perturbation $h_{\mu\nu}^{(m)}(x,0)$ when we insert the energy
momentum tensor
(\ref{eqn:src}).
\begin{equation}
h_{\mu\nu}^{(m)}(x,0) =-\frac{16 \pi G_5m_0}{3}\textrm{diag}(2,1,1,1)\int
dt'~ G_R(t, {\bf x}, 0;t',{\bf 0},0) 
\end{equation}
where 
\begin{equation}
\int dt'~ G_R(t, {\bf x}, 0;t',{\bf 0},0) = -\frac{k}{4 \pi
r}-\int_0^{\infty} dm ~\frac{e^{-mr}}{4 \pi r}[v_m(0)]^2 
\end{equation}
The integration over $m$ is exponentially suppressed for $m>1/r$. For
small $m$,
\begin{equation}
[v_m(0)]^2 = \frac{m}{2k} + \mathcal{O}(m/k)^2
\end{equation}
where we have used the fact that
\begin{equation}
J_n(m/k) \sim \frac{1}{n!}\left(m/2k\right)^n, \qquad Y_n(m/k) \sim \frac{(n-1)!}{\pi}\left(m/2k\right)^{-n}
\end{equation}
in this limit. The matter part of the perturbation is therefore given
by
\begin{equation}
h_{\mu\nu}^{(m)}(x,0) =\frac{2 
G_5km_0}{3r}\textrm{diag}(2,1,1,1)\left[2+\frac{1}{k^2r^2 }+\mathcal{O}(1/r^3) \right]
\end{equation}
Inserting the solution (\ref{eqn:fsoln}) for $f$ into equation
(\ref{eqn:fullpert}) yields the full metric perturbation
\begin{equation}
\hat h_{\mu\nu}(x, 0)=\frac{2 
G_5km_0}{r}\left[\textrm{diag}(1,1,1,1)+\frac{1}{3k^2r^2 }\textrm{diag}(2,1,1,1)+\mathcal{O}(1/r^3) \right]
\end{equation}
We are ready to read off the Newtonian potential, $\phi(r)$, measured by a
braneworld observer distance $r$ away from the source. This is given by
\begin{equation}
\phi(r)=\frac{1}{2} \hat h_{00}=\frac{ 
G_5km_0}{r}\left[1+\frac{2}{3k^2r^2 }+\mathcal{O}(1/r^3) \right]
\end{equation}
This is the Newtonian potential of four-dimensional gravity, with Yukawa type
corrections at short distances ($r<1/k$). Note that the four-dimensional Newton's
constant on the brane, $G_4=G_5k$. We conclude that this model does
not contradict  experimental tests
of Newton's inverse square law for the
force of gravitational attraction.

\subsubsection{The graviton propagator}
In the previous section we were careful to include the scalar field
$f$ corresponding to brane bending. This appeared because additional
matter on the brane acted as a source for the field. However, consider
what would have happened had we naively ignored it and worked in the RS
gauge throughout, with the brane at a fixed position. The Newtonian potential would still have behaved
like $1/r$ to leading order. We would have been conned
into thinking we had derived four-dimensional gravity.

However, the Newtonian potential is not the only property of four-dimensional gravity
that we can consider. There is also the form of the massless graviton
propagator. In a five-dimensional theory, there is an extra
polarization state that alters the tensor
structure of the propagator. This extra degree of freedom must be removed from the effective theory so that
the massless propagator on the brane looks
four-dimensional. If this didn't happen, the bending of light, for
example, would be $\frac{3}{4}$ of the value accurately predicted by
General Relativity~\cite{Dvali:metastable}.

In RS2 we also have massive KK gravitons. Even in the small mass limit
the tensor structure of their propagator is five
dimensional~\cite{vanDam:discont,Zakharov:discont,Dvali:metastable,Dvali:branebending}.
Since these are only important at high energies we will ignore them in
our effective theory
and focus on the massless graviton bound state.

From equation (\ref{eqn:matterpert}), the matter part of the metric
perturbation on the brane is given by
\begin{equation} \label{eqn:branematterpert}
h_{\mu\nu}^{(m)}=-16\pi
G_5\int d^4x'
~G_R(x,0;x',0)\left(\mathcal{T}_{\mu\nu}-\frac{1}{3}~\mathcal{T}
\eta_{\mu\nu} \right)
\end{equation}
If we ignore the massive modes then the Green's function takes the
following truncated form
\begin{equation} \label{eqn:truncGreens}
G_R(x,0;x',0)=\frac{k}{~\Box^{(4)}}
\end{equation}
where 
\begin{equation} \label{eqn:4DGreens}
\frac{1}{~\Box^{(4)}}=-\int \frac{d^4 p}{(2\pi)^4}
~\frac{e^{ip_{\mu}(x^{\mu}-x'{}^{\mu})}}{ {\bf p}^2-(\omega+i\epsilon)^2} 
\end{equation}
is the massless scalar Green's function for four-dimensional Minkowski
space~\cite{Giddings:linearized,Csaki:lightbending}. If we insert the
truncated Green's function (\ref{eqn:truncGreens}) into equation
(\ref{eqn:branematterpert}) we see that we do not have the usual
propagator for a massless four-dimensional graviton. We need the
factor of $\frac{1}{3}$ to be replaced by $\frac{1}{2}$. This task is
carried out by the brane bending term as we shall now demonstrate.

The {\it full} metric perturbation (\ref{eqn:fullpert}) contains a term
proportional to $f$. We can express $f$ in terms of the
four-dimensional Green's function using equation (\ref{eqn:feq})
\begin{equation}
f(x)=\frac{4 \pi G_5}{3}\int d^4x' ~\frac{1}{~\Box^{(4)}} \mathcal{T}.
\end{equation}
When this is introduced into equation (\ref{eqn:fullpert}) we find
that the (massless) metric perturbation is given by
\begin{equation} \label{eqn:truncpert}
h_{\mu\nu}=-16\pi
G_5k\int d^4x'
~\frac{1}{~\Box^{(4)}}\left(\mathcal{T}_{\mu\nu}-\frac{1}{2}\mathcal{T}
\eta_{\mu\nu} \right)
\end{equation}
This has the correct tensor structure for a four-dimensional massless
graviton. The extra degree of freedom in the five-dimensional
propagator has been compensated for by the brane bending scalar field
$f$.

The two results derived in this section are good evidence that
braneworld gravity agrees with General Relativity, at least for small
perturbations about the background metric. The warped geometry of the
bulk causes these perturbations to be damped away from the brane, so
that gravity is localised. The fact that the brane has positive
tension is crucial as the warp factor is a maximum there. In RS1, we
chose to live on the negative tension brane which is at a minimum of
the warp factor. We would not therefore expect gravity to be localised
on this type of braneworld, which makes its solution to the hierarchy
problem a little pointless. However, the ideas of both models can be
combined such they solve the hierarchy problem and exhibit localisation
of gravity~\cite{Lykken:gravity}. In this case there are two positive
tension branes, the Planck brane and the TeV brane. The Planck brane
has a much larger tension than the TeV brane, which in some sense is
regarded as a probe.  The hierarchy problem is solved in exactly the same way
as in RS1 provided we live on the TeV brane. In a similar way to RS2,
we find that gravity looks four-dimensional
at least
up to a few TeV on both branes.

\subsection{Non-critical braneworlds} \label{sec:noncrit}
Although the RS2 model agrees with Newton's Law and other properties of
four-dimensional gravity, it certainly contradicts one recent
experimental observation. The study of supernovae suggest that the
universe contains a small positive cosmological
constant~\cite{Perlmutter:astro, Riess:astro}. In RS2, we
have Minkowski space on the brane which has a vanishing cosmological
constant. In this section we shall show how to extend the model
to allow for de Sitter or anti-de Sitter braneworlds.

Recall that we have so far demanded that our braneworlds should exhibit
four-dimensional Poincar\'e invariance. This led to the ansatz
(\ref{eqn:RS1ansatz}) which has Minkowski spacetime induced on the
brane. We found that we then had to fine tune the brane tension,
$\sigma$ against the bulk cosmological constant, $\Lambda$, in the
following way
\begin{equation} \label{eqn:5Dcritcond}
\frac{4 \pi G_5\sigma}{3}=k, \qquad \Lambda=-6k^2
\end{equation}
This is the {\it criticality condition} and as such the flat
braneworlds that satisfy it are known as {\it critical}. We now generalise
the ansatz (\ref{eqn:RS1ansatz}) to allow for dS and AdS branes.
\begin{equation} \label{eqn:noncritansatz}
ds^2=a^2(z)g_{\mu\nu}dx^{\mu}dx^{\nu}+dz^2
\end{equation}
where $g_{\mu\nu}$ can be Minkowski, de Sitter or anti-de Sitter. The
solutions to the bulk equations of motion with appropriate boundary
conditions are derived in~\cite{Kaloper:bent, Kim:inflation,
Nihei:inflation} although a review may be found in appendix
\ref{app:noncrit}. In this section we will proceed as in~\cite{Karch:locally} and simply quote the results.
\begin{align} 
\textrm{de Sitter}&:&a(z)&=
\frac{1}{k}\sqrt{\frac{\lambda}{3}}\sinh(c-k|z|)
&k&=\sqrt{\frac{\lambda}{3}}\sinh c, \label{eqn:DSsoln}\\
\textrm{Minkowski}&:&a(z)&= e^{-k|z|}, \label{eqn:flatsoln}\\
\textrm{anti-de Sitter}&:&a(z)&=
\frac{1}{k}\sqrt{-\frac{\lambda}{3}}\cosh(c-k|z|),
&k&=\sqrt{-\frac{\lambda}{3}}\cosh c, \label{eqn:ADSsoln}
\end{align}
where the cosmological constant on the brane is given by 
\begin{equation} \label{eqn:branecc}
\lambda=3(\tilde \sigma^2-k^2), \qquad \tilde \sigma=\frac{4 \pi
G_5 \sigma}{3}.
\end{equation}
When $\sigma$ takes its critical value we have $\tilde \sigma=k$, and
the cosmological constant on the brane vanishes. For de Sitter branes,
$\sigma$ exceeds its critical value $(\tilde \sigma>k)$ where as the opposite
is true for anti-de Sitter branes. For this reason we refer to dS and AdS branes as {\it
supercritical} and {\it subcritical} branes respectively.

\begin{figure}[ht]
\begin{center}
\includegraphics[width=8cm]{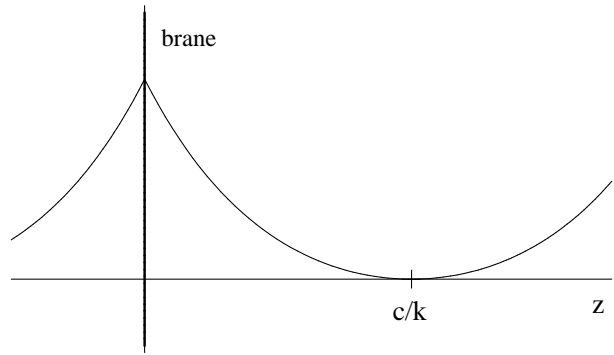}
\vskip 5mm
\caption{The behaviour of the warp factor around a supercritical (ie
de Sitter) brane.} \label{fig:dS}
\end{center}
\end{figure}

\begin{figure}[ht]
\begin{center}
\includegraphics[width=8cm]{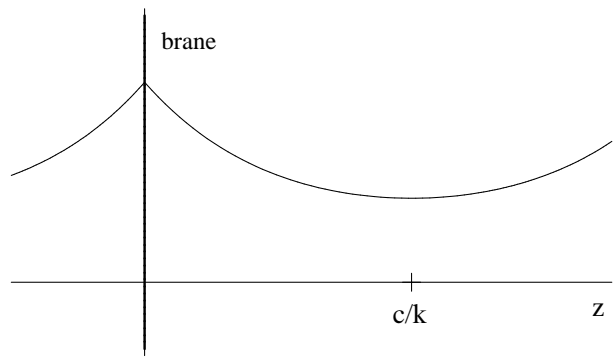}
\vskip 5mm
\caption{The behaviour of the warp factor around a subcritical (ie
anti-de Sitter) brane.} \label{fig:AdS}
\end{center}
\end{figure}

In section \ref{section:local} we saw how gravity was localised on
critical braneworlds. This was due to the behaviour of the warp
factor, which damped gravitational perturbations as they went
further into the bulk. We can ask whether the same is true for
supercritical and subcritical braneworlds. Without performing a
detailed analysis we can see the behaviour of the warp factors in
figures~\ref{fig:dS} and~\ref{fig:AdS}.
In each case, there is a turnaround in the warp factor. For the de
Sitter brane this corresponds to the de Sitter horizon where the
warp factor vanishes altogether, and the spacetime ends. It is clear that de Sitter branes are
even more likely to exhibit four-dimensional gravity than flat branes,
because the damping is greater. This is argued
in~\cite{Karch:locally} and proven in~\cite{Gen:dSbrane,Brevik:local}. Unlike in
RS2, there is a mass gap between the zero mode and the heavy modes in
the metric perturbations. We further note that the Newton's constant on the brane is found to
be proportional to the brane tension, $\sigma$, as opposed to the
bulk quantity $k$.

The situation for the anti-de Sitter brane is less clear. Near the brane the fluctuations in the metric behave in the same way
as for de Sitter and flat branes. However, the warp factor does not
vanish at the turnaround point, and beyond this the metric
perturbations start to grow. If we assume that this point lies far
from the brane we might yet believe that gravity is localised at low
enough energies. At finite temperature we could even hide the
point behind a black hole horizon. Despite the absence of a
normalisable zero mode the case for localisation is
presented in~\cite{Karch:locally}.

Finally, in this section we have seen how braneworld models can
exhibit four-dimensional gravity in line with experimental
observations. They also provide an unusual resolution of the hierarchy
problem, without the need for an unacceptably large (but finite) extra
dimension. Given our extension to non-critical branes, we could also
rephrase the cosmological constant problem. This is now a question of
balancing the tension and other matter fields on the brane against the bulk
cosmological constant~\cite{Csaki:constant, Csaki:essay}.

\chapter{Brane Cosmology} \label{chap:bwcosmo}
\setcounter{equation}{0}
\section{Introduction}
We have seen how Randall-Sundrum braneworlds provide a radical new way of
thinking about our universe and the extra dimensions that might
exist. If this extra dimension is warped anti-de Sitter space then it
can be infinitely large and still exhibit localisation of gravity on
the brane. We have also seen how to generalise the RS2 model to include
super/subcritical braneworlds which have a positive/negative
cosmological constant in four dimensions. 

To better understand these models we can and should generalise further. We note
that in the last section we always assumed a five-dimensional bulk which was
$\mathbb{Z}_2$ symmetric about a brane of codimension one. In this
section we will
consider bulk spacetimes which are $n$-dimensional and in some cases
relax the  $\mathbb{Z}_2$ symmetry. We will not generalise to branes of higher codimension although they have been studied (see for
example~\cite{Gherghetta:codimension2,Gherghetta:hegdehog,Ponton:codimension2}).

Another very important assumption of the last section was the fact
that perturbations about the background spacetime were small: the
energy-momentum due to additional matter on the brane was far less
than the brane tension.
\begin{equation}
\mathcal{T}_{00} \ll \sigma
\end{equation}
Unfortunately, life is not so easy as to be fully
described by perturbative physics. We will begin a study of
non-perturbative physics on the brane by examining their
cosmology. There are two main approaches: the {\it brane based}
approach and the {\it bulk based} approach, although we will show that
these are in fact equivalent. Each approach has its advantages and
disadvantages. For example, if we wished to examine
non-$\mathbb{Z}_2$ symmetric theories it would be much easier to use
the latter. However, we begin with a review of  the
brane based approach of Shiromizu {\it et al}~\cite{Shiromizu:einstein}, and although we
will retain $\mathbb{Z}_2$ symmetry we will generalise their work to $n$-dimensions.
\section{Brane based braneworld cosmology}
\begin{figure}[ht]
\begin{center}
\includegraphics[width=10cm, angle=-90]{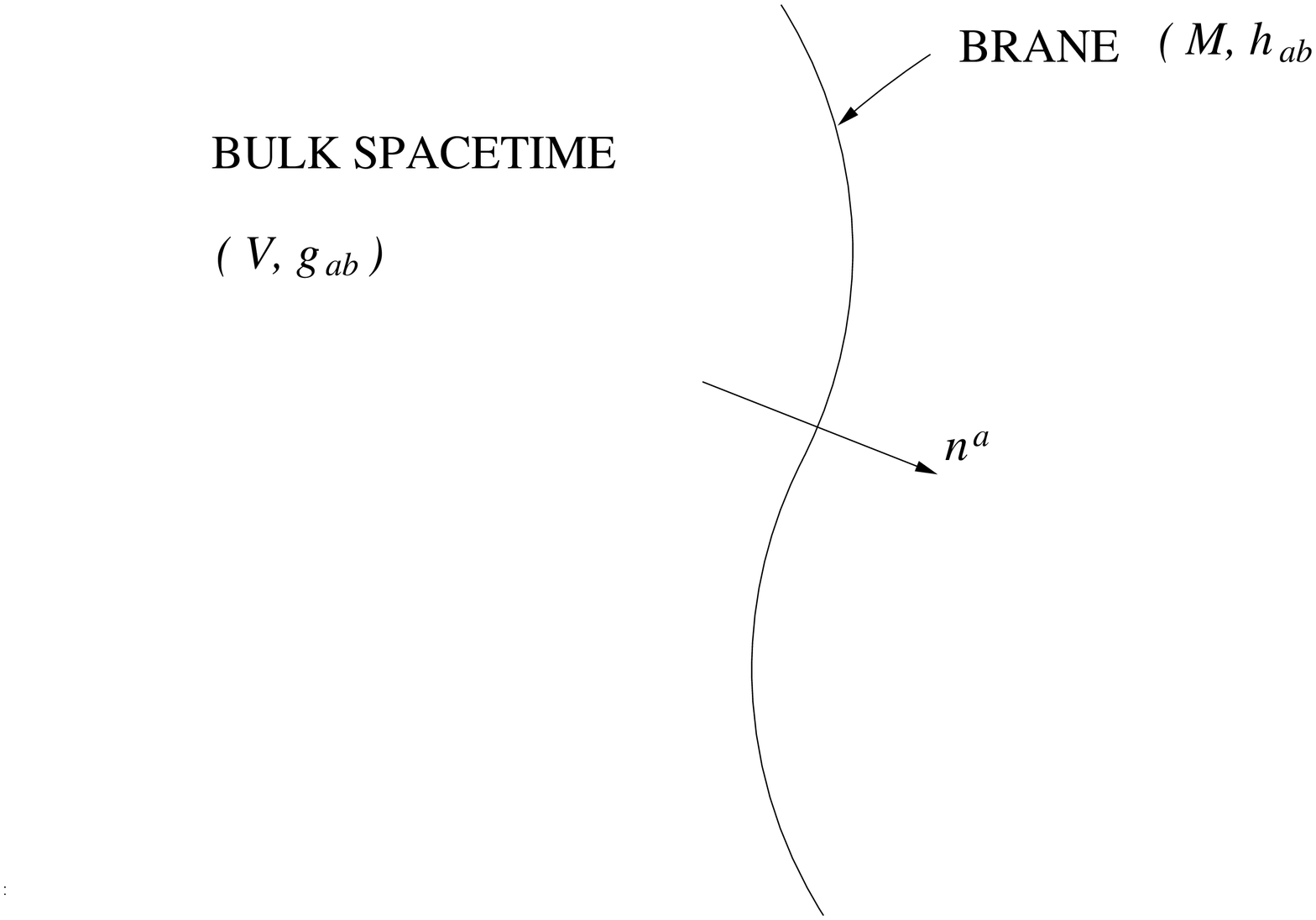}
\vskip 5mm
\caption{$(n-2)$-brane embedded in an $n$-dimensional bulk.} \label{fig:brane}
\end{center}
\end{figure}

Consider a timelike $(n-2)$-brane, $(M, h_{ab})$, in an $n$-dimensional bulk
spacetime $(V, g_{ab})$. The induced metric on $M$ is given by
\begin{equation} \label{eqn:indmetric}
h_{ab}=g_{ab}-n_an_b
\end{equation}
where $n^{a}$ is the unit normal to $M$ (see figure \ref{fig:brane}).
By using the Gauss-Codazzi equations~\cite{Wald:Gauss-Codacci} we can relate the
$(n-1)$-dimensional geometry on $M$ to its extrinsic curvature
$K_{ab}=h_a^ch_b^d\nabla_{(c}n_{d)}$ in $V$ and the bulk geometry. If we
label curvature tensors with an $n$ or $(n-1)$ depending on whether
they correspond to the bulk or the brane respectively, we have
\bml\begin{eqnarray}
^{(n-1)}R_{abcd}&=&
^{(n)}R_{pqrs}h_a^ph_b^qh_c^rh_d^s+K_{ac}K_{bd}-K_{ad}K_{bc}
\label{eqn:Gauss} \\
D_b (K_a^b-Kh_a^b) &=& ^{(n)}R_{cd}n^ch_a^d \label{eqn:Codazzi} \\
-2{}^{(n)}G_{ab}n^an^b &=& {}^{(n-1)}R-K^2+K_{ab}K^{ab} \label{eqn:trace} 
\end{eqnarray}\eml
where $D_{a}$ is the covariant derivative made out of $h_{ab}$. When
there is no $\mathbb{Z}_2$ symmetry, we label the ``left hand'' bulk
with a ``$-$''
and the ``right hand'' bulk with a ``$+$''. There is a version of equations
(\ref{eqn:Gauss}) to (\ref{eqn:trace}) for both ``$+$'' and ``$-$'', so in
principle we should label each of the bulk quantities $(^{(n)}R_{abcd}$
and $K_{ab})$ with the appropriate sign. However, for now we shall
assume $\mathbb{Z}_2$ symmetry so we drop the labels.

From equation (\ref{eqn:Gauss}) we are able to construct the Einstein
tensor on the brane
\begin{eqnarray} \label{eqn:Etensor1}
^{(n-1)}G_{ab} &=&
{}^{(n)}G_{cd}h_a^ch_b^d-{}^{(n)}R_{pqrs}n^pn^rh_a^qh_b^s+{}^{(n)}R_{cd}n^cn^dh_{ab}\nonumber\\
&& \qquad +KK_{ab}-K^c_aK_{bc}-\frac{1}{2}h_{ab}\left(K^2-K^{cd}K_{cd}\right)
\end{eqnarray}
We now use the bulk equations of motion
\begin{equation} \label{eqn:Einstein1}
^{(n)}G_{ab}={}^{(n)}R_{ab}-\frac{1}{2}{}^{(n)}Rg_{ab}=-\Lambda_{n}g_{ab}+8\pi
G_nT_{ab}
\end{equation}
where $\Lambda_{n}$ is the bulk cosmological constant,  $G_n$ is
the Newton's constant in $n$-dimensions, and $T_{ab}$ is
the energy-momentum tensor due to any additional bulk fields.
We can also express the bulk Riemann tensor in terms of the Weyl and
Ricci tensors.
\begin{eqnarray} \label{eqn:Weyl1}
^{(n)}R_{abcd} &=&^{(n)}C_{abcd}+\frac{1}{n-2}\left({}^{(n)}R_{ac}g_{bd}-{}^{(n)}R_{ad}g_{bc}+
{}^{(n)}R_{bd}g_{ac}- {}^{(n)}R_{bc}g_{ad}\right) \nonumber \\
&& \qquad-\frac{1}{(n-1)(n-2)}{}^{(n)}R\left(g_{ac}g_{bd}-g_{ad}g_{bc}\right)
\end{eqnarray}
Inserting equations (\ref{eqn:Einstein1}) and (\ref{eqn:Weyl1}) into
equation (\ref{eqn:Etensor1}) we find
\begin{equation} \label{eqn:Etensor2}
^{(n-1)}G_{ab}=-\Lambda_n
\left(\frac{n-3}{n-1}\right)h_{ab}-E_{ab}+KK_{ab}-K^c_aK_{bc}-\frac{1}{2}h_{ab}\left(K^2-K^{cd}K_{cd}\right)
\end{equation}
where
\begin{equation}
E_{ab}=C_{pqrs}n^pn^rh_a^qh_b^s-\left(\frac{n-3}{n-2}\right)\left[h_a^ch_b^d+n^cn^dh_{ab}-\frac{1}{n-1}g^{cd}h_{ab}\right]8\pi
G_n 
T_{cd}
\end{equation}
This term is often described as the ``electric'' part of the Weyl
tensor although this is only the case when there are no extra
bulk fields and $T_{ab} \equiv 0$. We can make sense of the extrinsic
curvature terms by using the Israel equations~\cite{Israel:junction} at
the brane
\begin{equation} \label{eqn:Israelinn}
\Delta K_{ab}=-8\pi G_n\left(S_{ab}-\frac{1}{n-2}Sh_{ab} \right)
\end{equation}
where the energy-momentum tensor for the brane is given by
\begin{equation} \label{eqn:braneEM}
S_{ab}=-\sigma h_{ab}+\mathcal{T}_{ab}
\end{equation}
with $\mathcal{T}_{ab}n^b=0$. Here we understand $\sigma$ to correspond to brane tension and
$\mathcal{T}_{ab}$ to additional matter, although it not obvious that we
should do this. In section \ref{sec:Newton} we assumed the additional
matter $\mathcal{T}_{ab}$ was much smaller
than the brane tension. This meant that the split between tension and
extra matter in equation
(\ref{eqn:braneEM}) was natural. However, we are now allowing for
larger values of $\mathcal{T}_{ab}$ which makes the split an arbitrary
one. It is not clear why we should have tension $\sigma$ rather than (say)
$\sigma/2$ because we could always redefine $\mathcal{T}_{ab}$ to absorb the
left over terms. However,  we shall see in
chapter \ref{chap:exact} some evidence that we are in fact interpreting
equation (\ref{eqn:braneEM}) in the right way.

At this stage we are assuming $\mathbb{Z}_2$ symmetry across the brane
so we have $\Delta K_{ab}=2K_{ab}$. Using the Israel equation
({\ref{eqn:Israelinn}) we can replace the extrinsic curvature terms in
equation (\ref{eqn:Etensor2}) with terms involving $\sigma$ and
$\mathcal{T}_{ab}$.
\begin{equation} \label{eqn:Etensor3}
^{(n-1)}G_{ab} = -\Lambda_{n-1}h_{ab}
+8\pi G_{n-1}\mathcal{T}_{ab} +(4 \pi G_n)^2\Pi_{ab}-E_{ab}
\end{equation}
where 
\begin{eqnarray}
\Lambda_{n-1} &=& \frac{1}{2}(n-2)(n-3)\left[ \sigma_n^2+
\frac{2}{(n-1)(n-2)}\Lambda_n\right] \\
G_{n-1} &=&\frac{G_n\sigma_n(n-3)}{2} \label{eqn:NC}\\
\Pi_{ab} &=& -\mathcal{T}^c_a\mathcal{T}_{bc}+\frac{1}{n-2}\mathcal{T}~\mathcal{T}_{ab}+\frac{1}{2}\mathcal{T}^{cd}\mathcal{T}_{cd}h_{ab}-\frac{1}{2n-4}\mathcal{T}^2h_{ab}
\end{eqnarray}
and
\begin{equation} \label{eqn:sigman}
\sigma_{n}=\frac{4\pi G_n \sigma}{n-2}
\end{equation}
The most striking feature of equation (\ref{eqn:Etensor3}) is the
presence of the quadratic matter terms contained in $\Pi_{ab}$. We
will discuss these in more detail later on. Meanwhile, we see that we
should interpret $\Lambda_{n-1}$ and $G_{n-1}$ as the braneworld
cosmological constant and Newton's constant respectively. As we hinted
at the end of section \ref{sec:noncrit}, $G_{n-1}$ is proportional to
the brane tension, rather than $\sqrt{|\Lambda_n|}$. This is highly
relevant to non-critical branes, although it is often ignored.

The other term in equation (\ref{eqn:Etensor3}) is of course the
``Weyl tensor'' term, $E_{ab}$. It contains
information about the bulk but is constrained by the matter on the
brane. We might hope to fully determine $E_{ab}$ from knowledge of
this matter, but this turns out not to be the case. In general
we need to solve the bulk equations of motion to derive $E_{ab}$ and
then insert it into the braneworld Einstein equation. We will discuss
this mysterious term from a holographic point of view in chapters~\ref{chap:holography} and~\ref{chap:exact}.
\subsection{A Friedmann-Robertson-Walker brane} \label{sec:FRWbrane}
We will now simplify the discussion further by assuming that the bulk
spacetime has negative cosmological constant with no additional
fields, that is
\begin{equation}
\Lambda_n=-\frac{1}{2}(n-1)(n-2)k_n^2, \qquad T_{ab} \equiv 0
\end{equation}
where $k_n$ is the inverse AdS length in $n$-dimensions. The
cosmological constant on the brane is now given by
\begin{equation} \label{eqn:braneccinn}
\Lambda_{n-1} = \frac{1}{2}(n-2)(n-3)\left[ \sigma_n^2-k_n^2\right]
\end{equation}
Note that equations (\ref{eqn:sigman}) and (\ref{eqn:braneccinn}) are
the $n$-dimensional analogue of equation (\ref{eqn:branecc}). Critical
branes are now defined as those satisfying the $n$-dimensional
criticality condition $\sigma_n=k_n$. Super/subcritical branes now
have $\sigma_n>k_n/\sigma_n<k_n$ respectively.
For a study of cosmology it is important to examine the behaviour of
a homogeneous and isotropic braneworld described by a
Friedmann-Robertson-Walker (FRW) metric.
\begin{equation}
ds_{n-1}^2=h_{ab}dx^adx^b=-d\tau^2+Z^2(\tau)d{\bf x}_{\kappa}^2
\end{equation}
where $d{\bf x}_{\kappa}^2$ is the metric on an $(n-2)$-dimensional Euclidean
space, $X$ of
constant curvature, $\kappa=0, \pm 1$. 
\begin{equation}
X= \begin{cases}  S^{n-2} & \textrm{for }\kappa=1 \\
 \mathbb{R}^{n-2} &  \textrm{for }\kappa=0 \\
H^{n-2} & \textrm{for }\kappa=-1
\end{cases}
\end{equation}
where $S^{n-2}, \mathbb{R}^{n-2}, H^{n-2}$ are the unit sphere,
plane, and hyperboloid respectively. $Z(\tau)$ represents
the scale factor for our braneworld.
We will assume the matter on the brane is given by a homogeneous perfect
fluid of density $\rho(\tau)$ and pressure $p(\tau)$ so that
\begin{equation} \label{eqn:perfectfluid}
\mathcal{T}_{ab}=\rho \tau_a\tau_b+p(h_{ab}+\tau_a\tau_b)
\end{equation}
where $\tau^a$ are the components of $\frac{\partial}{\partial
\tau}$. Finally, we avoid difficulties with $E_{ab}$ by setting it to zero,
which corresponds to {\it pure} anti-de Sitter space in the bulk.
We now use the braneworld Einstein equation (\ref{eqn:Etensor3}) to
derive the FRW equations for the cosmological evolution of the
brane. Defining the Hubble parameter, $H=\dot Z/Z$, where dot
denotes differentiation with respect to $\tau$, we find
\bml\begin{eqnarray}
H^2 &=& a -\frac{\kappa}{Z^2}+\frac{16\pi G_{n-1}}{(n-2)(n-3)}\rho
+\left(\frac{4 \pi G_n}{n-2}\right)^2\rho^2
\label{eqn:evolADS1}\\
\dot H &=& \frac{\kappa}{Z^2}-\frac{8\pi
G_{n-1}}{(n-3)}(\rho+p)-(n-2)\left(\frac{4 \pi G_n}{n-2}\right)^2
\rho(\rho+p) \label{eqn:evolADS2}
\end{eqnarray}\eml
where $a=\sigma_n^2-k_n^2$. These are not the standard FRW equations because they contain terms
quadratic in $\rho$ and $p$. Braneworld cosmology is therefore
different to the standard cosmology. This unconventional behaviour was
first discovered in five dimensions by Binetruy {\it et
al}~\cite{Binetruy:branecos1}. Notice that we
recover the standard cosmology for large values of the scale factor,
because we can ignore the non-linear density terms. 
\section{Bulk based braneworld cosmology} \label{section:bulkbranecos}
In the last section we saw a number of the limitations of the brane
based approach to braneworld cosmology. We chose to impose
$\mathbb{Z}_2$ symmetry across the brane and ignored the possibility
of non-zero Weyl terms. These were difficult to get a handle on
because we were working with a static brane in a dynamic bulk. The
bulk based approach turns everything around by having a dynamic brane
in a static bulk. This allows us to include non-$\mathbb{Z}_2$
symmetric branes and non-vanishing Weyl terms. The disadvantage now is
that we will only be considering FRW branes, and will not have the
generalisation provided by equation (\ref{eqn:Etensor3}).
\subsection{Generalised Birkhoff's Theorem}
Since the bulk based approach works on the premise of there being a
static bulk spacetime, we immediately think of Birkhoff's
Theorem~\cite{Birkhoff:theorem, Bronnikov:Birkhoff}. This
states that {\it if the geometry of a given region of spacetime is
spherically symmetric and a solution to the vacuum Einstein
equations, then it is necessarily a piece of the Schwarzschild
geometry.} In order to bridge the gap between the brane based approach
to braneworld cosmology and the bulk based approach, we will prove a
generalised version of this theorem. This was first shown by Bowcock
{\it et al}~\cite{Bowcock:branecos} in five dimensions, but once again we will extend the ideas to $n$-dimensions.

We start by assuming that our spacetime contains a codimension two Euclidean surface
of constant curvature. This will ultimately provide us with
spatial homogeneity on our braneworld. The most general metric admitting this
symmetry is given by~\cite{Gregory:nested, Gregory:instantons}
\begin{equation} \label{eqn:Birkhoffmetric}
ds^2 = A^\frac{2}{n-2} d{\bf x}_\kappa^2 +e^{2\nu} A^{-\left(\frac{n-3}{n-2}\right)} 
(-dt^2 + dz^2)
\end{equation}
where $A$ and $\nu$ are functions of $t$ and $z$ to be determined by
the bulk Einstein equations, as well as the jump conditions across the
brane. Again,  $d{\bf x}_\kappa^2$ represents the metric on the
Euclidean surface of constant curvature, $\kappa=0, \pm1$. Here we have used the fact that the rest of the metric is two
dimensional and therefore conformally flat. Without loss of generality,
we can say that the brane sits at $z=0$\footnote{If the brane sits at
$z'=\zeta(t')$ we use the conformal transformation $t' \pm z'=t \pm z
\pm \zeta(t \pm z)$ to shift the wall back to $z=0$ without spoiling
the form of the metric (\ref{eqn:Birkhoffmetric})~\cite{Bowcock:branecos}.}.

We will assume that the bulk spacetime contains no additional matter
$(T_{ab} \equiv 0)$. When we insert our metric ansatz into the bulk
Einstein equations (\ref{eqn:Einstein1}) we arrive at the following set of
differential equations
\bml
\begin{eqnarray}
A,_{tt}-A,_{zz} &=& \left[2\Lambda_nA^{\frac{1}{n-2}}-(n-2)(n-3)\kappa
A^{-\frac{1}{n-2}}\right]e^{2\nu} \\
\nu,_{tt}-\nu,_{zz} &=&
\left[\frac{\Lambda_n}{n-2}A^{-\left(\frac{n-3}{n-2}\right)}+\frac{n-3}{2}\kappa
A^{-\left(\frac{n-1}{n-2}\right)}
\right]e^{2\nu} \\
A,_{tt}+A,_{zz} &=& 2\nu,_{z}A,_{z}+2\nu,_{t}A,_{t} \\
A,_{tz} &=& \nu,_{z}A,_{t}+\nu,_{t}A,_{z}
\end{eqnarray}
\eml
It is convenient to change to lightcone coordinates
\begin{equation}
u=\frac{t-z}{2}, \qquad v=\frac{t+z}{2}
\end{equation}
so that we now have
\bml
\begin{eqnarray}
A,_{uv} &=&  \left[2\Lambda_nA^{\frac{1}{n-2}}-(n-2)(n-3)\kappa
A^{-\frac{1}{n-2}}\right]e^{2\nu} \label{eqn:birkhoff1}\\
\nu,_{uv} &=&
\left[\frac{\Lambda_n}{n-2}A^{-\left(\frac{n-3}{n-2}\right)}+\frac{n-3}{2}\kappa
A^{-\left(\frac{n-1}{n-2}\right)}
\right]e^{2\nu}\label{eqn:birkhoff2} \\
2\nu,_uA,_u &=& A,_{u}\left[\ln(A,_u)\right],_u \label{eqn:birkhoff3}\\
2\nu,_vA,_v &=& A,_{v}\left[\ln(A,_v)\right],_v \label{eqn:birkhoff4}
\end{eqnarray}
\eml
We can easily integrate equations (\ref{eqn:birkhoff3}) and
(\ref{eqn:birkhoff4}) to give
\begin{align}
\textrm{Case I} &:~ A ~\textrm{is constant}&   \nonumber\\
\textrm{Case II} &:~ A=A(u),&  e^{2\nu}&=A'(u)V'(v) \nonumber \\
\textrm{Case III} &:~ A=A(v),&  e^{2\nu}&=A'(v)U'(u) \nonumber\\
\textrm{Case IV} &:~ A=A(u, v),& e^{2\nu}&=V'(v)A,_u=U'(u)A,_v \nonumber
\end{align}
where $U'(u)$ and $V'(v)$ are arbitrary non-zero functions of $u$ and $v$
respectively. Note that {\it prime} denotes differentiation with
respect to the unique argument of the function. Cases I to III imply
that $\Lambda_n=\kappa=0$, which is not relevant here
(see~\cite{Taub:empty, Bowcock:branecos} for some discussion). We will
focus on case IV, for which it is easy to see that
\begin{equation}
A=A(U(u)+V(v)), \qquad e^{2\nu}=A'U'V'
\end{equation}
so that equation (\ref{eqn:birkhoff1}) is reduced to an ODE
\begin{eqnarray}
&& A''-\left[2\Lambda_nA^{\frac{1}{n-2}}-(n-2)(n-3)\kappa
A^{-\frac{1}{n-2}}\right]A'=0 \\
&&\Rightarrow A'-2\left(\frac{n-2}{n-1}\right)\Lambda_nA^{\left(\frac{n-1}{n-2}\right)}+(n-2)^2\kappa
A^{\left(\frac{n-3}{n-2}\right)}=(n-2)^2c  \label{eqn:Aprime}
\end{eqnarray}
where $c$ is a constant of integration. Notice that equation
(\ref{eqn:birkhoff2}) just gives the derivative of the ODE, and is
satisfied automatically.
We are now ready to impose the jump conditions on the brane. Once
again we will assume that the matter on the brane is homogeneous and
isotropic so that
\begin{equation}
S_{ab}=-\sigma h_{ab}+\mathcal{T}_{ab}, \qquad \mathcal{T}_{ab}=\rho
\tau_a\tau_b+p(h_{ab}+\tau_a\tau_b) \nonumber
\end{equation}
where $\tau^a$ is the unit timelike vector parallel to
$\frac{\partial}{\partial t}$. When there is $\mathbb{Z}_2$ symmetry across the brane at $z=0$, the Israel
equations (\ref{eqn:Israelinn}) give
\begin{gather}
4\pi G_n(\sigma+\rho)
=-e^{-\nu}A^{-\frac{1}{2}\left(\frac{n-1}{n-2}\right)}A,_z =
\frac{1}{2}e^{-\nu}A^{-\frac{1}{2}\left(\frac{n-1}{n-2}\right)}[U'-V']A' \\
\begin{split}
&4 \pi G_n\left[\frac{n-3}{n-2}\left(\sigma+\rho\right)-\sigma+p\right]
=-\partial_z\left[e^{-\nu}A^{\frac{1}{2}\left(\frac{n-3}{n-2}\right)}\right]\\
&=\frac{1}{4}e^{-\nu}A^{\frac{1}{2}\left(\frac{n-3}{n-2}\right)}\left[(V'-U')\left(\frac{A''}{A}-\left(\frac{n-3}{n-2}\right)\frac{A'}{A}\right) +\frac{V''}{V}-\frac{U''}{U}\right]
\end{split}
\end{gather}
Note that we could use equation (\ref{eqn:Aprime}) to
eliminate $A'$ and $A''$. If we make the following coordinate
transformation
\begin{equation}
u \to f(u), \qquad v \to f(v)
\end{equation}
then the boundary conditions at the brane are unchanged\footnote{This
is seen if we note that the brane is given by $u=v$, where the
coordinate change gives $U' \to  f'(u)U'$, $V' \to  f'(u)V'$ and
$e^{-\nu}=1/\sqrt{A'U'V'} \to e^{-\nu}/f'(u)$. $A'$ and $A''$ are
unchanged}. This symmetry is related to the conformal symmetry on the
$t-z$ plane. To eliminate this unphysical gauge freedom we choose
$f=V$ , thereby setting $V=v$. We are now left with only one physical
degree of freedom, $U(u)$. Setting
\begin{equation}
Z=A^{\frac{1}{n-2}}, \qquad T=(n-2)(v-U)
\end{equation}
we see that the bulk metric can locally be written in the explicitly static form
\begin{equation} \label{eqn:schmetric}
ds_n^2=-h(Z)dT^2+\frac{dZ^2}{h(Z)}+Z^2d{\bf x}_{\kappa}^2
\end{equation}
where
\begin{equation}
h(Z)=-\frac{Z'}{n-2}=-\frac{A'A^{-\left(\frac{n-3}{n-2}\right)}}{(n-2)^2}
\end{equation}
From equation (\ref{eqn:Aprime})
\begin{equation}
h(Z)=-\frac{2\Lambda_n}{(n-1)(n-2)}Z^2+\kappa-\frac{c}{Z^{n-3}}
\end{equation}
For $c>0$, the metric (\ref{eqn:schmetric}) clearly takes the form of
the Schwarzschild black hole in de Sitter, flat or anti-de Sitter
space, depending on the value of $\Lambda_n$. Given that our starting
point was that our braneworld contained spatial geometry of constant
curvature, we conclude that we have indeed proved a generalised
version of Birkhoff's theorem. In this work we assumed our bulk physics
was described by pure Einstein gravity with a
cosmological constant. Similar proofs have been carried out for
Einstein-Maxwell gravity~\cite{Cai:topologicalBHs} and Gauss-Bonnet
gravity~\cite{Charmousis:gauss-bonnet}.

Although this generalisation of Birkhoff's Theorem is of interest from
a mathematical point of view, our focus is on braneworld physics. We
have shown that we can express the bulk geometry in the static form
given by equation (\ref{eqn:schmetric}), although in doing so we can
no longer say that we have a static brane sitting quietly at $z=0$. On the contrary, we now have a
dynamic brane, whose trajectory in the new coordinates is far more
complicated. Braneworld cosmology from this perspective was first
studied by Ida~\cite{Ida:branecos},  although moving branes in a static
anti-de Sitter bulk were
considered earlier by Kraus~\cite{Kraus:dynamics}.  
\subsection{A dynamic brane in a static bulk} 
Having bridged the gap from the brane based approach to braneworld
cosmology we are ready to give a generalisation of Ida's bulk based
approach. We will see that by transferring the dynamics of the system from the bulk to the brane
we allow ourselves more flexibility regarding the structure of the
bulk spacetime. We will no longer assume $\mathbb{Z}_2$ symmetry across
the brane and will even allow the cosmological constant on either side to
differ.

We start by taking the general static solution (\ref{eqn:schmetric})
to the Einstein equations with cosmological constant, $\Lambda_n$. To
construct the brane solution, we treat the brane as the boundary 
\begin{equation}
X^a=({\bf x}^{\mu}, t(\tau), Z(\tau))
\end{equation}
of the bulk (\ref{eqn:schmetric}). We now patch this bulk spacetime
(labelled with a``$-$'') onto
another appropriate bulk (labelled with a ``$+$'') with the same
boundary value $Z(\tau)$. Note that we have reintroduced the ``$\pm$'' notation to indicate which side of the brane a given quantity resides\footnote{For
example, $g_{ab}^+$
and $\Lambda_n^+$ are the bulk metric and cosmological constant on the
``$+$'' side of the brane.}. We set the parameter $\tau$ to correspond
to the proper time with respect to an observer comoving with the
brane. This imposes the conditions
\begin{equation}
-h^{\pm}\dot t_{\pm} ^2+\frac{\dot Z ^2}{h^{\pm}}=-1
\end{equation}
so that whichever side of the brane you look from, the induced metric
on the brane takes the standard FRW form
\begin{equation} \label{eqn:FRWbranemetric}
ds_{n-1}^2=h_{ab}dx^adx^b=-d\tau^2+Z^2(\tau)d{\bf x}_{\kappa}^2
\end{equation}
and $Z(\tau)$ is understood to be the scale factor of the brane
universe. It is clear that the bulk metric is continuous across the
brane because both $\tau$ and $Z(\tau)$ agree there. Note that $t$ can
be discontinuous at the brane, because neither $g_{ab}$ nor $h_{ab}$
depend on it explicitly. 

In order to produce the type of brane required, it is important we
patch together the two bulk spacetimes in such a way that the Israel
equations (\ref{eqn:Israelinn}) are satisfied. We take the energy
momentum tensor of the brane to be given by a tension $\sigma$ and a
perfect fluid of energy density $\rho$ and pressure $p$ (that is,
equation (\ref{eqn:braneEM}) with $\mathcal{T}_{ab}$ given by
(\ref{eqn:perfectfluid})). In defining the extrinsic curvature of the
brane on either side, we need some knowledge of the outward
normal.
\begin{equation}
n^{\pm}_a=\epsilon_{\pm}({\bf 0}, -\dot Z(\tau), \dot t_{\pm}(\tau))
\end{equation}
where $\epsilon^{\pm}=\pm 1$ depending on which part of the spacetime
is kept\footnote{If we wished to keep (say) $Z<Z(\tau)$ on the ``$-$''
side we would choose $\epsilon_-=1$, assuming of course that $\dot
t_->0$.}.  With reference to appendix \ref{app:extrinsic}, the Israel equations now yield the following 
\begin{equation} \label{eqn:ext1}
\frac{1}{Z}\overline{[\epsilon h \dot t]}=\frac{4 \pi G_n}{n-2}(\sigma+\rho)
\end{equation}
\begin{equation} \label{eqn:ext2}
\overline{\left[\frac{\ddot Z+\frac{1}{2}h'}{\epsilon h \dot
t}\right]}=\frac{4 \pi G_n}{n-2}\left[\sigma-(n-3)\rho-(n-2)p \right]
\end{equation}
where $\overline{Q}=\frac{Q_++Q_-}{2}$ for a given quantity $Q$. Note that
while $K_{ab}^-=h_a^ch_b^d\nabla_{(c}n_{d)}^-$, the process of gluing
together spacetimes causes the ``$+$'' side to flip orientation so that we
must define $K_{ab}^+=-h_a^ch_b^d\nabla_{(c}n_{d)}^+$.
We now refer back to the third Gauss-Codazzi equation
(\ref{eqn:trace}), with the understanding that it is valid on both
sides of the brane, and $G^{\pm}_{ab}=-\Lambda^{\pm}_n
g^{\pm}_{ab}$. If we now take the difference between the ``$+$''
equation and the ``$-$'' equation we find that
\begin{equation}
-\Delta \Lambda_n={\overline K} \Delta K -{\overline K_{ab}} \Delta K^{ab}. 
\end{equation}
Inserting the values of the extrinsic curvature found in
appendix {\ref{app:extrinsic} we obtain
\begin{equation} \label{eqn:ext3}
-\Delta \Lambda_n=4 \pi G_n (n-2)(\sigma-p)\frac{\Delta [\epsilon h
\dot t]}{Z}+4 \pi G_n(\sigma+\rho)\Delta \left[\frac{
\ddot Z+\frac{1}{2}h'}{\epsilon h \dot t}\right].
\end{equation}
After careful and tedious manipulations of equations (\ref{eqn:ext1}),
(\ref{eqn:ext2}) and (\ref{eqn:ext3}) we arrive at the following
expressions for derivatives of the scale factor
\begin{equation} \label{eqn:dotZ}
\dot Z ^2 = -\overline{h}+\left[\frac{4 \pi G_n}{n-2}(\sigma+\rho)Z\right]^2
+ \left[\frac{(n-2)\Delta h}{16 \pi G_n(\sigma+\rho)Z}\right]^2
\end{equation}
and
\begin{eqnarray} \label{eqn:ddotZ}
\ddot Z &=&-\frac{1}{2}\overline{h'}-\left(\frac{4 \pi
G_n}{n-2}\right)^2(\sigma+\rho)\left[(n-3)\rho+(n-2)p -\sigma
\right]Z \nonumber \\
&&+\left(\frac{(n-2)\Delta h}{16 \pi G_n (\sigma+
\rho)Z}\right)^2\left[\frac{(n-3)\rho+(n-2)p
-\sigma}{(\sigma+\rho)Z}\right] \nonumber \\
&&+\left(\frac{n-2}{16 \pi G_n (\sigma+
\rho)Z}\right)^2\Delta h \Delta h'.
\end{eqnarray}
Note that for equations (\ref{eqn:dotZ}) and (\ref{eqn:ddotZ}) to be
consistent, we require that the conservation of energy equation holds on
the brane
\begin{equation}
\dot \rho =-(n-2)\frac{\dot Z}{Z}(\rho +p).
\end{equation}
Here we have seen the beauty of the bulk based approach to braneworld
cosmology. We have found the cosmological evolutions equations
(\ref{eqn:dotZ}) and (\ref{eqn:ddotZ}) for the brane without assuming
$\mathbb{Z}_2$ symmetry. This is particularly important when studying
braneworld models that have differing cosmological constants on either
side of the brane (eg.~\cite{Gregory:ultralarge,Lykken:gravity}). Furthermore, by considering general values of
$h$, we have allowed the bulk Weyl tensor
on either side to be non-zero. Recall that in the brane based approach
the Weyl tensor contribution was just hidden away behind the mysterious $E_{ab}$ term,
without any real understanding of its effects. That is not the case
here.
\subsubsection{A $\mathbb{Z}_2$ symmetric brane in AdS-Schwarzschild} \label{sec:unconventional}
As a consistency check, we will now examine the evolution equations
when we do indeed have $\mathbb{Z}_{2}$ symmetry across the brane. This has the
effect that for a given quantity $Q$, $\overline{Q} \to Q$ and $\Delta
Q \to 0$.
We will also assume that the bulk cosmological constant is negative,
and set
\begin{equation}
\Lambda_n=-\frac{1}{2}(n-1)(n-2)k_n^2.
\end{equation}
Our bulk solution is therefore given by equation (\ref{eqn:schmetric})
with
\begin{equation}
h(Z)=k_n^2 Z^2+\kappa-\frac{c}{Z^{n-3}}
\end{equation}
Note that the integration constant $c$ gives the Weyl tensor
contribution. For $c=0$,  (\ref{eqn:schmetric}) represents pure AdS
space with the appropriate slicing (depending on $\kappa$). For $c>0$
we have the AdS-Schwarzschild metric, with its horizon at the point
where $h$ vanishes. In the spirit of Randall-Sundrum, we will
construct the brane by cutting away the AdS boundary in each bulk, and
then gluing together. This imposes the choice $\epsilon=1$.
Again, defining $H=\dot Z/Z$, we find that the cosmological evolution
equations now simplify somewhat
\bml
\begin{equation} \label{eqn:tdot}
\frac{h \dot t}{Z}=\frac{4 \pi G_n}{n-2}(\sigma+\rho)
\end{equation} 
\begin{equation} \label{eqn:evolSADS1}
 H^2=a -\frac{\kappa}{Z^2} +\frac{c}{Z^{n-1}}+\frac{16
\pi G_{n-1}}{(n-2)(n-3)}\rho +\left(\frac{4
\pi G_n}{n-2}\right)^2 \rho^2 
\end{equation}
\begin{equation} \label{eqn:evolSADS2}
\dot H =
\frac{\kappa}{Z^2}-\left(\frac{n-1}{2}\right)\frac{c}{Z^{n-1}}-\frac{8\pi
G_{n-1}}{(n-3)}(\rho+p)-(n-2)\left(\frac{4 \pi G_n}{n-2}\right)^2
\rho(\rho+p)
\end{equation}
\eml
where we recall that $a=\sigma_n^2-k_n^2$ represents the cosmological
constant on the brane, and $\sigma_n$ is defined by equation
(\ref{eqn:sigman}).  We have also used the relationship (\ref{eqn:NC})
to include the $(n-1)$-dimensional Newton's constant.
Notice that equations (\ref{eqn:evolSADS1}) and (\ref{eqn:evolSADS2})
agree with equations (\ref{eqn:evolADS1}) and (\ref{eqn:evolADS2}) derived using the brane
based approach. However we have now been able to explicitly
include the the bulk Weyl term, which we were not able to do
previously.

Although we have come a long way using the bulk based approach, this
is as far as we can go. The main limitation is that we can only
consider FRW branes, but that is fine if we wish to examine
cosmological branes. The brane based approach had the advantage that
we can generalise to more complicated brane geometries.

To conclude this section, we reiterate two interesting features to
arise in brane cosmology. The first is the quadratic
energy-momentum terms. One can generally ignore these if the densities
are small (for example, when the scale factor is very large), although not
otherwise. The second feature is the effect of the bulk Weyl tensor on
these cosmologies. We will see in chapters \ref{chap:holography} and \ref{chap:exact} how this
can be understood from the point of view of AdS/CFT.

\chapter{Bubbles and ribbons on the brane} \label{chap:bubbles}
\setcounter{equation}{0}
\section{Introduction}
In chapter~\ref{chap:RSbw}, we saw why the RS2 model was so
compelling, and why it has been taken as a viable toy model for our
universe. The key feature is that gravity on
the brane is precisely Einstein gravity at low energies, {\it i.e.,}
\begin{equation} \label{4deeq}
R_{ab} - \frac{1}{2} R g_{ab} = 8\pi GT_{ab} 
\end{equation}
This result is of course perturbative~\cite{Randall:compactification,Garriga:gravity}, and does not include
the effect of the short-range KK corrections. Strictly speaking it is only 
valid for a single brane universe -- the presence of a second wall, as
in RS1~\cite{Randall:hierarchy}, introduces 
a radion, representing the distance between the branes and modifying the
Einstein gravity to Brans-Dicke
gravity~\cite{Garriga:gravity,Giddings:linearized,Charmousis:radion,Chiba:scalar-tensor}.
Non-perturbative results however, particularly understanding the effect
of the KK modes, are somewhat sparse. In chapter~\ref{chap:bwcosmo},
we began a study of non-perturbative braneworld gravity by examining
their cosmology. The most notable effect was the deviation from the
standard four-dimensional cosmology via quadratic energy density and
pressure terms in the FRW equations. The most obvious example of
strong brane gravity would be a black hole bound to the
brane. Although this has been well understood for a 2-brane in four
dimensions~\cite{Emparan:BHsonbranes}, we know very little about the
higher dimensional analogue.




In this chapter we will investigate non-perturbative gravity by
considering the effect of a domain wall living entirely on the
brane~\cite{Gregory:nested, Gregory:instantons}. Recall that
braneworld universes are really only domain walls themselves~\cite{Rubakov:domain}, so the
codimension 2 objects (or vortices) we are considering can be regarded as
{\it nested}
domain walls (see figure~\ref{fig:vortex})
\begin{figure}[ht]
\begin{center}
\includegraphics[width=10cm, angle=-90]{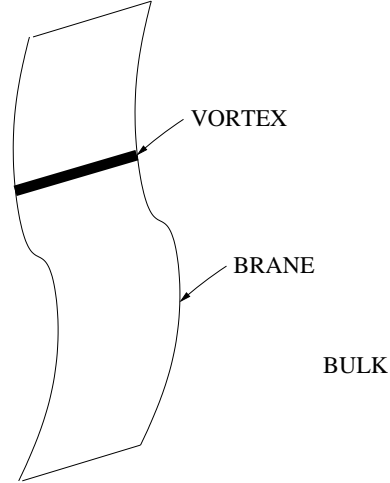}
\vskip 5mm
\caption{A nested domain wall, or vortex on an $(n-1)$-brane.} \label{fig:vortex}
\end{center}
\end{figure}
These kind of objects can arise naturally from domain wall
configurations. For example, suppose we have a $\lambda \phi^4$ kink interacting
with an additional scalar, $\sigma$, via a potential of the form
\be\label{genpotl}
V(\phi,\sigma) = \frac{\lambda}{4} (\phi^2-\eta^2)^2 + \frac{\tilde \lambda}{4}
\sigma^4 + (\phi^2 - m^2) \sigma^2.
\ee
In the true vacuum, $\langle \phi\rangle=\pm\eta$, the state  $\langle
\sigma \rangle=0$ is energetically favoured. However, this is not the
case in the core of the wall. For example, when $\langle \phi
\rangle=0$, the potential is minimised when $\langle
\sigma \rangle=\pm m \sqrt{2/\tilde \lambda}$. We see, then, that we
can generate a kink in the $\sigma$ field within the core of the
domain wall. 
%
Such a configuration is quite well studied in the context
of nested topological defects in field theory~\cite{Morris:nested,
Morris:strings+walls,Bazeia:topdefects,Bazeia:domain,
Edelstein:kinks}, although gravity is absent. This
particular configuration is known as a {\it domain ribbon}~\cite{Morris:nested, Morris:strings+walls}. In this chapter we will show that we can fully derive
the gravitational field associated with these nested defects. This
will not only give us an insight into strong gravity on the brane, it
will also enable us to construct a whole class of new configurations,
including {\it nested braneworlds} and the braneworld analogue of the
{\it Coleman-De Luccia instantons}~\cite{Coleman:vacuumdecay}.
\section{Equations of motion for the domain ribbon}\label{section:dr}
Consider the gravitational field generated by a domain ribbon source.
In general, it will depend on only two
spacetime coordinates, $r$ and $z$ say, with $z$ roughly
representing the direction orthogonal to the brane and $r$, the direction orthogonal to the domain
ribbon (or vortex) within our brane universe.
Schematically, the energy-momentum 
tensor of this source will have the following form:
\be
T_{ab} = -\sigma h_{ab} \frac{\delta(z)}{\sqrt{g_{zz}}}
-\mu \gamma_{ab} \frac{\delta(z) \delta(r)}{\sqrt{g_{zz}g_{rr}}}
\label{emtens}
\ee
where $h_{ab}$ is the induced metric on the brane universe, and $\gamma_{ab}$
the induced metric on the vortex. The symmetries of this energy-momentum
tensor mean that we can treat the vortex as a constant curvature spacetime. 
The most general metric consistent with
these symmetries can, in $n$ dimensions,
be reduced to the form
\be
ds^2_n = A^{\frac{2}{n-2}} d{\bf x}_\kappa^2 +e^{2\nu} A^{-\left(\frac{n-3}{n-2}\right)} 
(dr^2 + dz^2)
\ee
where $d{\bf x}_\kappa^2$ represents the `unit' metric on an
$(n-2)$-dimensional spacetime of constant curvature ($\kappa=0$ corresponds to a Minkowski spacetime, $\kappa = \pm 1$ to de-Sitter and anti-de Sitter spacetimes). $A$ and $\nu$ are functions
of $r$ and $z$ to be determined by the equations of motion. Here, 
the brane universe sits at $z = 0$ with the vortex at $r = z = 0$. 
This is basically an analytic continuation of the cosmological metric (\ref{eqn:Birkhoffmetric})
in section~\ref{section:bulkbranecos}, where it is the time translation symmetry $\partial_t$ which is
broken, rather than $\partial_r$. The key result of that section of relevance
here was to show that the conformal symmetry of the $t,z$ plane meant that
the gravity equations were completely integrable in the bulk, and the 
brane universe was simply a boundary $(\bf x\rm^{\mu}, t(\tau), Z(\tau))$ of that
bulk (identified with another boundary of another general bulk). The 
dynamical equations of the boundary reduced to
pseudo-cosmological equations for $Z(\tau)$. We now briefly review this
argument in the context of the current problem.

First of all, transform the $(r,z)$ coordinates to complex coordinates 
$(\omega, \bar{\omega})$ where $\omega=z+ir$, $\bar{\omega}=z-ir$, in which
the bulk equations of motion reduce to:
\bml \label{complextraj}
\bea 
\partial \bar{\partial} A &=& -\frac{1 }{ 2}\Lambda_nA^{\frac{1 }{ n-2}}e^{2\nu} 
+ \frac{(n- 2)(n-3)}{ 4}\kappa A^{-\frac{1 }{n-2}}e^{2\nu} \label{complextraja} \\
\partial \bar{\partial } \nu &=& -\frac{1 }{ 4(n-2)}\Lambda_nA^{-\left(\frac{n-3}{ n-2}\right)}
e^{2\nu} - \frac{n-3}{ 8}\kappa  A^{-\left(\frac{n-1}{ n-2}\right)}e^{2\nu} \label{complextrajb} \\
\partial A \partial [\ln\partial A] &=& 2\partial \nu \partial A  
\label{complextrajc} \\
\bar{\partial} A \bar{\partial} [\ln\bar{\partial}A] &=& 
2\bar{\partial} \nu\bar{ \partial} A \label{complextrajd}
\eea
\eml 
where $\partial$ and $\bar{\partial}$ denote partial differentiation with
respect to $\omega$ and $\bar{\omega}$ respectively. For non-zero $\Lambda_n$ or
$\kappa$, equations (\ref{complextrajc}) and (\ref{complextrajd}) can be
integrated to give $e^{2\nu}=A^{\prime}f^{\prime}g^{\prime}$, where
$A = A \big( f(\omega)+g(\bar{\omega}) \big)$ with $f$ and $g$ being
arbitrary functions of the complex variables. The remaining equation
(\ref{complextraja}) for $A$ becomes an ODE.

Were the brane not present, we could use the fact that the metric depends only
on the combination $f+g$ to make a coordinate transformation in the bulk which
would give the metric in the familiar simple canonical form
\be\label{canbulk}
ds^2_n = Z^2 d{\bf x}_\kappa^2 +h(Z) dR^2 +\frac{dZ^2 }{ h(Z)} 
\ee
where $d{\bf x}_\kappa^2$ is now a constant curvature Lorentzian spacetime,
and in general the function $h$ is
\be
h(Z) = -\frac{2\Lambda_n}{(n-1)(n-2)} Z^2 + \kappa - \frac{c}{ Z^{(n-3)}}
\ee
%

The addition of the brane, however, requires that the Israel conditions be 
satisfied at $z=0$ in the original coordinates. These turn out to have a scaling symmetry $\omega \to W(\omega)$, 
${\bar\omega}\to W({\bar\omega})$, so we are free to choose $f$ 
or $g$ (but not both) as we wish. The net result is that our brane becomes some boundary 
of the bulk (\ref{canbulk}) identified with the boundary of some other 
general bulk. The vortex (or ribbon), in these coordinates, becomes a kink 
on this boundary as we shall see.  Introducing the affine parameter $\zeta$ 
which parametrizes geodesics on the brane normal to the vortex, 
the brane is now  given by the section $({\bf x}^\mu , R(\zeta), Z(\zeta))$ 
of the general bulk metric. Note that we now have the condition
\be
hR'^2+\frac{Z'^2}{h}=1
\ee 
In an exactly analogous procedure to section~\ref{section:bulkbranecos}, we consider the Israel 
equations  for the jump in extrinsic curvature across the brane, as well 
as the normal component of the Einstein equations, and thus obtain the
equations of motion for the source:
\bml \label{srctraject}
\bea
Z^{\prime 2} &=&\bar{h}-\sigma_n^2Z^2-\left(\frac{ \Delta h  }{4\sigma_nZ}  
\right)^2 \label{srctrajecta} \\
Z^{\prime \prime} &=& \frac{\bar{h^{\prime}} }{ 2}-\sigma_n^2Z 
-\frac{\mu_n }{2}\sigma_nZ\delta(\zeta) 
+ \frac{1 }{ Z}\left(\frac{ \Delta h }{4\sigma_nZ}  \right)^2 \nonumber \\  
&-&\frac{\Delta h^{\prime}\Delta h}{(4 \sigma_n Z)^2}+\frac{\mu_n\delta(\zeta)}{2
\sigma_nZ }\left(\frac{ \Delta h}{4\sigma_nZ}\right)^2 \label{srctrajectb} \\
\overline{\epsilon h R^{\prime}} &=& \sigma_nZ \label{srctrajectc}
\eea
\eml
The brane and vortex tensions now appear in $\sigma_n$ and $\mu_n$
respectively. These are defined as follows
\be
\sigma_n = \frac{4\pi G_n \sigma}{n-2}, \qquad \mu_n = 8\pi G_n\mu.
\ee
As in section~\ref{section:bulkbranecos}, the quantity $\epsilon$ in (\ref{srctrajectc}) is related to the sign
of the the outward normal to the boundary of the bulk spacetime, the boundary of 
course being the brane. In particular, it is given by
\be
n_a^{\pm}=\epsilon_{\pm}({\bf 0}, -Z^{\prime}, R_{\pm}^{\prime})
\ee
where $\epsilon_{\pm}=\pm 1$, depending on which part of the spacetime
is kept. Recall that we should define
the extrinsic curvature on the ``$+$'' with an extra minus sign. This is
to account for
reversing its orientation when we glue it onto the ``$-$'' spacetime.

For simplicity, we will now assume our
brane universe is $\mathbb{Z}_2$ symmetric. This has the effect that
for any intrinsic bulk quantity $Q$, $\overline{Q} \to Q$ and $\Delta
Q \to 0$. We also assume that the integration constant, $c$, vanishes
and that the bulk cosmological constant is given by
\be \label{lamkn}
\Lambda_n = -\frac{1}{2}(n-1)(n-2)k_n^2
\ee
This definition is aimed at studying an anti-de Sitter bulk, which is
of course what we find in RS models. However,
we can easily extend to a Minkowski/de Sitter bulk by allowing
$k_n$ to vanish/take imaginary values, as required. Rewriting equation (\ref{srctraject}) for the trajectory of the source 
we obtain the $\mathbb{Z}_2$ symmetric equations of motion: 
\bml\label{traject}\bea
Z^{\prime2}(\zeta) &=& \left ( k_n^2 - \sigma^2_n \right )
Z^2 + \kappa \label{trajecta} \\
Z''(\zeta) &=& \left ( k_n^2 - \sigma^2_n \right )
Z - \frac{\mu_n}{2} \sigma_n Z \delta(\zeta) \label{trajectb} \\
R'(\zeta) &=& \frac{\sigma_n Z}{(k_n^2 Z^2 + \kappa)} \label{trajectc} 
\eea\eml
Note that we have chosen $\epsilon=1$. This ensures that the brane has
positive tension and that in the spirit of Randall-Sundrum, we retain
the $Z<Z(\zeta)$ part of the bulk.
In fact, the Randall-Sundrum brane (in $n$ dimensions) is given
by setting $\kappa = \mu = 0$ (flat, no vortex) and $\sigma_n
= k_n$.  The bulk metric is then
\be
ds^2_n = Z^2 (-dt^2 +dx^2_i +k^2_n dR^2) +\frac{dZ^2}{k_n^2Z^2} 
\ee
with the brane given by $Z=Z_0$ a constant, and $kR = \zeta/Z_0$.
Letting $Z_0=1$, and $Z = e^{-k_nz}$ gives the usual RS coordinates. 

Before turning to the instanton solutions, we will remark on a few 
straightforward domain ribbon solutions in order to gain an understanding 
of the geometrical effect of the ribbon. In particular, we will discuss the
gravity of nested domain walls from the point of view of observers on the brane.
\section{Domain ribbon solutions}\label{section:solns}
In this section we examine the solutions to (\ref{traject}), exploring their 
qualitative features as well as some useful illustrative special
cases. We begin by integrating the $Z$-equation (\ref{trajecta}) away 
from the vortex:
\be\label{genzsolns}
Z = \begin{cases}
\frac{1}{\sqrt{a}} \cos\left[\pm\sqrt{a}(\zeta - \zeta_0)\right] & a>0, 
~\kappa = 1 ~\textrm{only} \\
Z_0 \pm \kappa \zeta & a=0, ~\kappa = 0,1 \\
 \frac{1}{2\sqrt{|a|}} \left [
e^{\pm\sqrt{|a|}(\zeta-\zeta_0)} - \kappa e^{\mp \sqrt{|a|}(\zeta-\zeta_0)} 
\right] & a<0, ~\kappa = 0,\pm 1
\end{cases}
\ee 
where $a = \sigma^2_n-k_n^2 $. Recall that $a=0$ for a critical brane,
whereas $a>0 ~(a<0)$ for a super~(sub) critical brane respectively.
In the absence of the vortex, a critical brane has a
Minkowski induced metric and corresponds to the original RS
scenario~\cite{Randall:hierarchy,Randall:compactification}. A supercritical brane has a de-Sitter induced metric, and
can be regarded as an inflating
cosmology~\cite{Chamblin:dilatonicdw,Kaloper:bent, Kraus:dynamics, Khoury:inflationary}, whereas
the subcritical brane has an AdS induced metric (see~\cite{Kogan:bigravity,Karch:locally,Schwartz:locgrav} for discussion of its phenomenology).
Staying away from the vortex, we can use (\ref{trajectc}) and
the square root of (\ref{trajecta}) to obtain an ODE for $R(Z)$:
\be
R'(Z)=\pm\frac{\sigma_n Z}{\kappa+k_n^2 Z^2}\left( \kappa-a
Z^2 \right)^{-\half}
\ee
This is easily integrated to give
\be\label{Rtraject}
2k_n (R - R_0) =  \pm \begin{cases} \ln\left (1 + k_n^2 Z^2 \right) &
\kappa = 1, \quad~a=0 \\
\ln\left | \frac{k_n\sqrt{1-aZ^2} - \sigma_n}{k_n\sqrt{1-aZ^2}+\sigma_n}
\right |  & \kappa = 1, \quad~a\neq0 \\
-\frac{2\sigma_n}{k_n \sqrt{|a|} Z} & \kappa = 0, \quad~a<0 \\
2 \tan^{-1}\left ( \frac{k_n\sqrt{1 -aZ^2}}{\sigma_n} \right ) &
\kappa = -1, ~~a<0.
\end{cases}
\ee
where the choice of signs refers to the sign of $Z'(\zeta)$.
Note that these trajectories are invariant under Euclideanization of
the metric, therefore instanton trajectories will also have this form.

In order to see how these trajectories embed into the bulk AdS
spacetime, it is useful to transform into conformal coordinates,
$({\tilde t}, {\tilde{\bf x}}, u)$ in which the metric is conformally
flat:
\be\label{confmet}
ds^2_n = \frac{1}{k_n^2u^2} \left [ -d{\tilde t}^2 +d{\tilde {\bf x}}^2 +du^2
\right ]
\ee
For the $\kappa=1$ spacetimes needed to construct the braneworld
instantons, this requires the bulk coordinate transformation
\bml\label{bulkct}\bea
k_n u &=& e^{k_nR} / \sqrt{1+k_n^2 Z^2} \\
\left ( {\tilde t}, {\tilde {\bf x}}\right ) &=& k_n u Z 
\left ( \sinh t , \cosh t {\bf n}_{n-2} \right )
\eea\eml
where ${\bf n}_{n-2}$ is the unit vector in $(n-2)$ dimensions. 
Under such a transformation the trajectories $R(Z)$ in
(\ref{Rtraject}) generally take the form
\be\label{hyper}
(u \mp u_0)^2 + {\tilde {\bf x}}^2 - t^2 = u_1^2
\ee
for $a\neq 0$, where
\be
u_0 = \frac{\sigma_n }{k_n} u_1 
= \frac{\sigma_n }{k_n} \frac{e^{k_nR_0}}{ |a|^{1/2}}
\ee
This means that braneworlds (\ref{hyper}) have the form of hyperboloids (or
spheres in the Euclidean section) in the conformal metric
(\ref{confmet}). In particular, for {\it subcritical branes} $(a<0)$,
we have $u_0<u_1$, and both branches of
the hyperboloid (\ref{hyper}) are allowed, each intersecting 
the AdS boundary (see figures \ref{fig:walltraj:subfig:a} and \ref{fig:walltraj:subfig:b}). An analysis of the
normals to the braneworld shows that for a positive tension $\mathbb{Z}_2$-symmetric 
braneworld, the upper root $Z'>0$ corresponds to keeping the interior of
the hyperboloid, whereas for $Z'<0$ the exterior is kept. {\it Supercritical
branes} $(a>0)$ on the other hand have only the upper root for $u_0$, and as $u_0>u_1$
in a Euclidean signature they represent spheres which are entirely 
contained within the AdS spacetime. For a supercritical brane of
positive tension  the interior of the hyperboloid (or
sphere) is kept (see figure 
\ref{fig:walltraj:subfig:d}). For a {\it critical brane}, $(a=0)$ there are once
again two possible trajectories, one having the form of (\ref{hyper})
but with $u_0=u_1 = e^{k_nR_0}/2k_n$ (figure \ref{fig:walltraj:subfig:c}), and 
the other having $u=$ const.\ -- the RS braneworld.
\begin{figure}
\centering
\subfigure[Subcritical brane centred on $-u_o$]{
    \label{fig:walltraj:subfig:a}
\begin{minipage}[b]{0.3\textwidth}
   \centering
   \includegraphics[width=6cm]{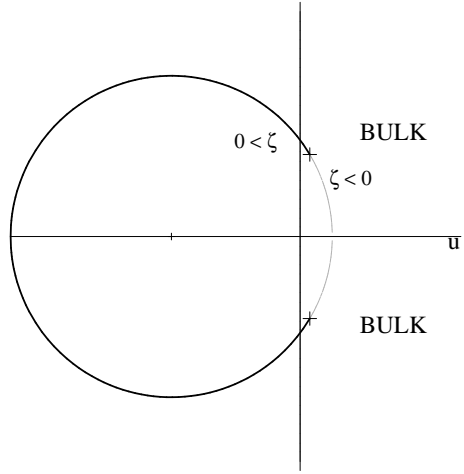}
\end{minipage}}%
\hspace{0.2\textwidth}%
\subfigure[Subcritical brane centred on $+u_o$]{
   \label{fig:walltraj:subfig:b}
\begin{minipage}[b]{0.3\textwidth}
  \centering
    \includegraphics[width=6cm]{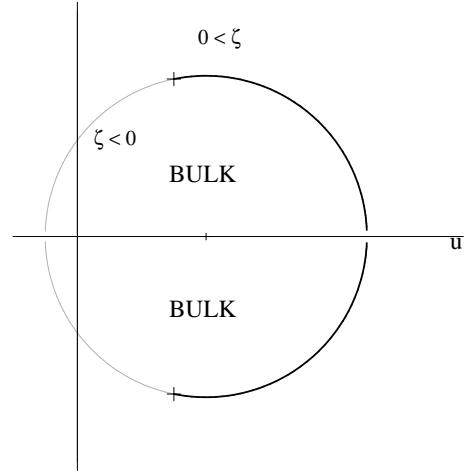} 
\end{minipage}} \\[20pt]
\subfigure[Critical brane]{
    \label{fig:walltraj:subfig:c}
\begin{minipage}[b]{0.3\textwidth}
   \centering
     \includegraphics[width=6cm]{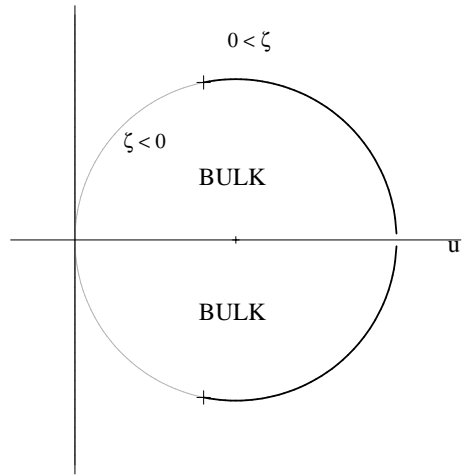}
\end{minipage}}%
\hspace{0.2\textwidth}%
\subfigure[Supercritical brane]{
 \label{fig:walltraj:subfig:d}
\begin{minipage}[b]{0.3\textwidth}
   \centering
     \includegraphics[width=6cm]{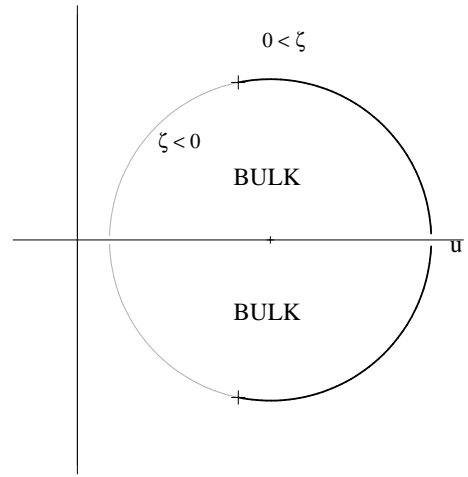}
\end{minipage}}
\caption{Braneworld trajectories given by
equation \ref{hyper}, in Euclidean signature. In each case the location of the 
bulk spacetime is indicated. In
addition we have the simple critical brane trajectory given by
$u=\text{const}$. Here the bulk lies to the right of the brane.} \label{fig:wall
traj:subfig}
\end{figure}

To put a vortex on the braneworld, 
we require solutions with non-zero $\mu_n$, and hence a discontinuity in $Z'$.
To achieve this, we simply patch together different branches of the 
solutions (\ref{genzsolns}) for $\zeta>0$ and $\zeta<0$. 
We immediately see that critical and supercritical branes can
only support a vortex if $\kappa=1$, that is, if the induced metric on the vortex
itself is a de-Sitter universe. A subcritical brane on the other hand can
support all induced geometries on the vortex. 
Defining $k_{n-1}^2 = |a|$, these trajectories are
\be\label{zsolns}
Z = \begin{cases} \frac{1}{2k_{n-1}} \left [
e^{\alpha-k_{n-1}|\zeta|} - \kappa e^{k_{n-1}|\zeta|-\alpha} 
\right] & \text{subcritical brane} \\
Z_0 - |\zeta|& \text{critical brane $(\kappa=1 ~\textrm{only})$ } \\
\frac{1}{k_{n-1}} \cos \left (k_{n-1}|\zeta| + \beta\right) & \text{supercritical brane $(\kappa = 1 ~\textrm{only})$ }
\end{cases}
\ee 
where
\be\label{muvalues}
\mu_n = \begin{cases} \frac{4k_{n-1}}{\sigma_n} \left [
\frac{e^{2\alpha} + \kappa }{
e^{2\alpha} -\kappa } \right ] & \text{subcritical brane} \\
\frac{4}{k_n Z_0} & \text{critical brane} \\
\frac{4k_{n-1}}{\sigma_n} \tan \beta & \text{supercritical brane} 
\end{cases}
\ee
respectively.

\subsection{The domain ribbon in a vacuum bulk} \label{sec:vacuumbulk}
In order to examine the geometry of the ribbon it is useful to consider a vacuum bulk
spacetime. This will
obviously represent a vortex living on a supercritical braneworld.
There is no warping of the bulk due to the
cosmological constant  so we can clearly compare the ribbon spacetime to
that of an isolated vortex $(\sigma_n=0)$ or a pure de-Sitter domain
wall $(\mu_n=0)$. Since the bulk cosmological constant vanishes, we have
\begin{equation} 
k_n=0  ~\Rightarrow ~ k_{n-1} = \sigma_n
\end{equation} 
Note that the pure domain wall universe is a 
hyperboloid  in Minkowski spacetime~\cite{Gibbons:sugradw,Cvetic:dw}. Specifically it is an accelerating bubble of proper 
radius $\sigma_n^{-1}$, with $\kappa=1$.  Meanwhile, we also note that the pure $\delta$-function
isolated vortex solution has a conical deficit metric
\be\label{conic}
ds^2_n = -dt^2 +d{\bf x}^2 +d\rho^2 +
\left (1-\frac{\Delta\theta }{2\pi}\right)^2 \rho^2d\theta^2
\ee
where $\Delta\theta \simeq \mu_n$ for small $\mu_n$~\cite{Vilenkin:vacuumdw}.

We can read off the domain ribbon trajectory from equations
(\ref{zsolns}) and (\ref{muvalues}). In $(R,Z)$ space, this gives
\bml\label{fwtraj}\bea
Z &=& \frac{1}{ \sigma_n \sqrt{16+\mu_n^2}} \left [ 4 \cos(\sigma_n\zeta) 
- \mu_n \sin(\sigma_n|\zeta|)\right ] \\
R &=& \frac{1}{ \sigma_n \sqrt{16+\mu_n^2}} \left [ 4 \sin(\sigma_n\zeta)\pm
\mu_n [ \cos(\sigma_n\zeta) -1] \right]
\eea\eml
where we preserve the region $Z<Z(\zeta)$ of the bulk:
\be
ds_n^2 = Z^2 \left[ -dt^2 +\cosh^2\! t\ d\Omega_{n-3}^2 \right]
+dZ^2 +dR^2
\ee
where $d\Omega_{n-3}^2$ is the metric on a unit $(n-3)$-sphere. This is of course simply a coordinate transformation of 
Minkowski spacetime, with the appropriate limit of (\ref{bulkct}) being
$({\tilde t}, {\tilde{\bf x}}) = 
(Z\sinh t, Z{\bf n}\cosh t)$.
Transforming into Minkowski coordinates, we find that
the vacuum braneworld domain ribbon is given by two copies of the interior of 
the sliced hyperboloid
\be
{\tilde{\bf x}}^2 - {\tilde t}^2 + \left (|R|+\frac{\mu_n}{\sigma_n 
\sqrt{16+\mu_n^2}} \right )^2 = \frac{1}{\sigma_n^2}
\ee
If $\mu_n=0$ this is clearly the standard domain wall hyperboloid. However,
when $\mu_n>0$, this represents a hyperboloid which has had a slice
of width $2\mu_n/\sigma_n \sqrt{16+\mu_n^2}$ removed from it
(see figure \ref{fig:vacdw1}). This corresponds rather well with the intuitive notion
that walls are obtained by slicing and gluing spacetimes.
\begin{figure}[ht]
\begin{center}
\includegraphics[width=8cm]{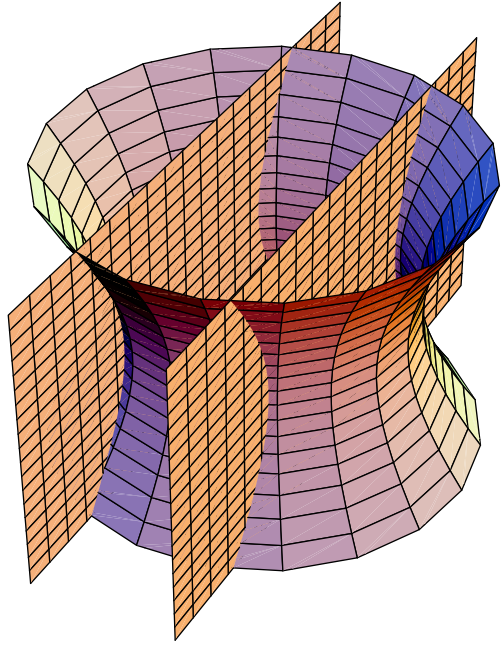}
\vskip 5mm
\caption{Constructing a domain ribbon on a vacuum domain wall. The hyperboloid 
interior has a slice of thickness $2\mu_n/\sigma_n \sqrt{16+\mu_n^2}$ 
removed from it, and is re-identified. The full spacetime consists of
a second copy identified across the hyperboloid.}\label{fig:vacdw1}
\end{center}
\end{figure}

Looking at a constant time slice (figure \ref{fig:vacdw2}) we also see 
how the domain ribbon 
\begin{figure}[ht]
\begin{center}
\includegraphics[width=10cm]{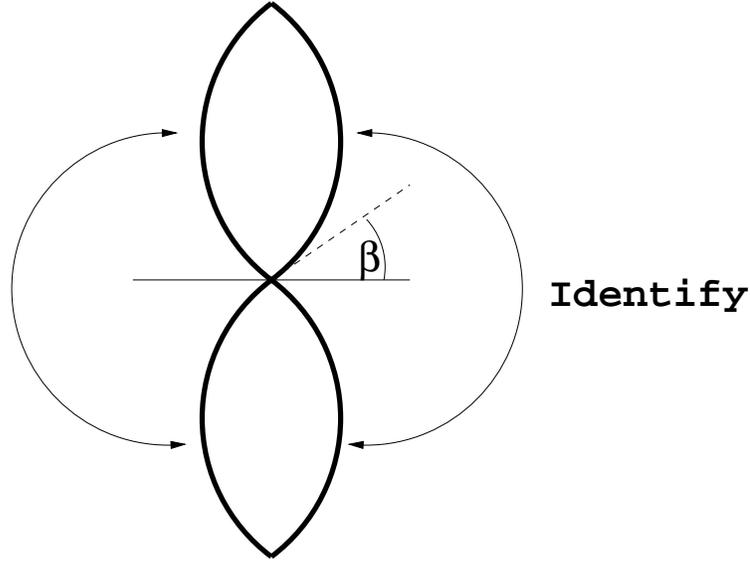}
\vskip 5mm
\caption{Taking a constant time slice through the vacuum domain wall
plus vortex spacetime shows how the deficit angle is built up.}
\label{fig:vacdw2}
\end{center}
\end{figure}
looks like a vortex, with the identifications giving rise to a conical
deficit angle in terms of the overall $n$-dimensional spacetime. We
find that
\be
\Delta \theta = 4 \tan^{-1} \mu_n/4
\ee
Note that for small $\mu$ we have $\Delta \theta \simeq \mu_n$, which
 agrees with the case of the isolated vortex. A
crucial difference however, appears to be that for the ribbon
 spacetime, the vortex can have an arbitrarily
large energy per unit length, as we simply cut out more
and more of the hyperboloid. Indeed, the deficit angle approaches $2\pi$ only as $\mu_n$ approaches
infinity! Contrast this with the spacetime of a pure vortex, (\ref{conic}), 
in which the
deficit angle approaches $2\pi$ as $\mu_n \simeq 1$~\cite{Laguna:cosmicstrings}. The 
ribbon is clearly not behaving as a vortex for large $\mu$. On the other hand, a 
domain wall has the effect of compactifying its spatial sections (the interior
of the hyperboloid) and the transverse dimension only shrinks to zero size as
the tension of the wall becomes infinite. Therefore in this sense, the ribbon
spacetime really does behave as a domain wall.

Finally, we note that the induced metric on the brane is given by
\be \label{eqn:vacmetric}
ds_{n-1}^2 = \frac{[4\cos\sigma_n\zeta - \mu_n \sin \sigma_n |\zeta|]^2}{
\sigma_n^2 (16+\mu_n^2)} \left [ -dt^2 + \cosh^2t d\Omega^2_{n-3} \right]
+d\zeta^2
\ee
This is the metric of an $(n-2)$-dimensional domain wall in an
$(n-1)$-dimensional de-Sitter universe of tension, $\mu$. We see this
by examining the Israel equations in $(n-1)$
dimensions, at $\zeta=0$. For a wall of tension $T$,
\be \label{eqn:tensionT}
\Delta K_{ab} = -\frac{8\pi G_{n-1}T}{n-3} \gamma_{ab}
\ee
Meanwhile, from (\ref{eqn:vacmetric}), the jump in extrinsic curvature
at $\zeta=0$ is given by
\be
\Delta K_{ab}=-\frac {\sigma_n\mu_n}{2} \gamma_{ab}
\ee
With the identification (\ref{eqn:NC}) we conclude that $T\equiv \mu$. In
this sense the geometry of the braneworld seems to know nothing about
the extra dimension. Gravity on the brane appears  $(n-1)$-dimensional
even in this non-perturbative regime.
\subsection{The domain ribbon on a critical RS brane} 
Having constructed this symmetric vacuum domain ribbon spacetime,
we now see the general principle involved in having a domain ribbon.
Whereas a braneworld without a vortex consists of two segments of AdS (or vacuum/dS)
spacetime glued across a boundary, the domain ribbon consists of two
copies of an AdS spacetime with a kinked boundary identified
together. The kink itself could
be viewed as two copies of an AdS bulk glued together across a tensionless
boundary.  Recall that our original motivation was to investigate the
behaviour of domain walls on branes, and in particular the
critical RS brane. With our current insight, we would expect a domain ribbon
on a critical RS brane, in conformal coordinates, to be the critical
hyperboloid sliced by a critical flat RS wall (see figure
\ref{fig:rsdw}). We will investigate this presently.
\begin{figure}[ht]
\begin{center}
\includegraphics[width=8cm]{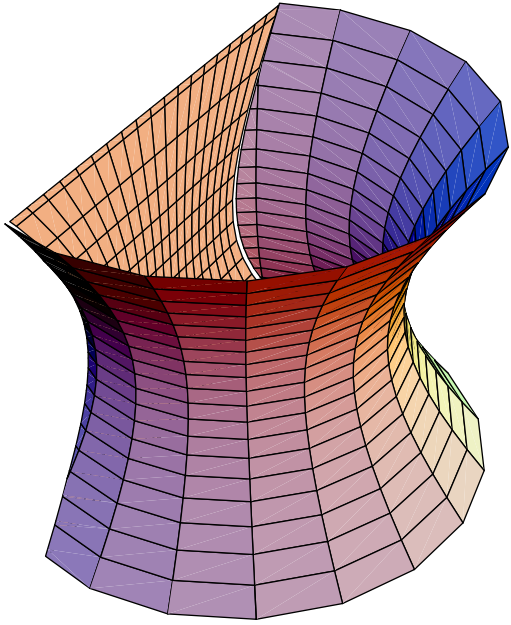}
\vskip 5mm
\caption{A representation of the domain ribbon on a critical RS
brane. }
\label{fig:rsdw}
\end{center}
\end{figure}

The tension of the critical RS brane satisfies the relation
$\sigma_n=k_n$. Here we have pure AdS space in the bulk so
$k_n>0$. Since $a=0$, a domain
ribbon on this brane {\it must} have $\kappa=1$, that is `spherical'
spatial geometry. In $(R, Z)$ space, the brane trajectory is given by
\bml\label{critdr}\bea
Z &=& \frac{4}{\mu_nk_n} - |\zeta| \\
R &=& \mp \frac{1}{2k_n} \ln \left [ \frac{\mu_n^2+(4-k_n\mu_n|\zeta|)^2 }{ \mu_n^2 
+ 16 } \right]
\eea\eml
For the full spacetime we keep the region $Z<Z(\zeta)$ of the bulk:
\be
ds_n^2=Z^2\left[ -dt^2 +\cosh^2\! t\ d\Omega_{n-3}^2 \right]
+\frac{dZ^2}{k_n^2Z^2+1} +(k_n^2Z^2+1)dR^2
\ee
At first sight neither the trajectory nor bulk looks like the original
RS scenario, however, the coordinate transformation
\bml\bea
k_n u &=& e^{k_nR} / \sqrt{1+k_n^2 Z^2} \\
({\tilde t},{\tilde {\bf x}}) &=& k_n u Z (\sinh t,\cosh t\ {\bf n}_{n-2})
\eea\eml
(where ${\bf n}_{n-2}$ is the unit vector in $(n-2)$ dimensions) gives
\be
ds_n^2 = \frac{1}{ k_n^2u^2} \left [ -{\tilde t}^2 + d{\tilde {\bf x}}^2 + du^2 
\right ]
\ee
This is the familiar planar AdS metric in conformal coordinates. The
trajectory (\ref{critdr}) now becomes
\bml\begin{align}
\zeta<0 &: \quad ~u = u_0  \\
\zeta>0 &: \quad \left( u - \frac{1}{2k_n^2u_0} \right)^2+{\tilde{\bf x}}^2 - {\tilde t}^2
= \frac{1}{4 k_n^4 u_0^2}
\end{align}\eml
where 
\be
u_0 = \frac{\mu_n}{k_n \sqrt{16+\mu_n^2}}
\ee
The change of coordinates means that the trajectory is no longer
manifestly $\mathbb{Z}_2$ symmetric. However, the $\zeta<0$ branch now
becomes a subset of the RS planar domain wall, specifically, the
interior of the hyperboloid
\be\label{hyprad}
\frac{{\tilde{\bf x}}^2 - {\tilde t}^2}{k_n^2 u_0^2}
= \frac{16}{k_n^2 \mu^2_n} = \left [ 2 \pi G_{n-1} \mu \right ] ^{-2}
\ee
where we have used equation (\ref{eqn:NC}) with $\sigma_n=k_n$. Recall that the global spacetime structure
of a vacuum domain wall is that of two identified copies of the interior of a 
hyperboloid in Minkowski spacetime,  with proper radius $1/2 \pi G_{n-1} \mu$,
\cite{Gibbons:sugradw,Cvetic:dw, Ipser:repulsivedw}. We conclude that (\ref{hyprad}) corresponds identically with what we
would expect from $(n-1)$-dimensional Einstein gravity.
 
Meanwhile, the $\zeta>0$ branch is a hyperboloid in the bulk centered
on $u = 1/2k_n^2 u_0$ with comoving radius $1/2k_n^2 u_0$. 
As $\mu$ increases, more and more of the hyperboloid
is removed, with the spacetime `disappearing' only as
$\mu\to\infty$. This is the same behaviour as we found in
section~\ref{sec:vacuumbulk} for the domain ribbon in a vacuum
bulk. As before, this is normal behaviour for a domain wall, but very
different to what one would expect from a vortex.

In order to examine the geometry on the brane more carefully, we note
that the induced metric on the brane is given by
\be\label{indvac}
ds_{n-1}^2 = \left ( 1 - \frac{\mu_nk_n|\zeta|}{4}\right )^2 \left [ -d{\hat t}^2 +
\left(\frac{4}{\mu_nk_n}\right)^2 \cosh^2 \left(\frac{\mu_nk_n{\hat t}}{4}\right) 
d\Omega_{n-3}^2 \right] +d\zeta^2
\ee
where ${\hat t} = 4t/\mu_nk_n$. This is {\it precisely} the metric of 
a self-gravitating domain wall of tension $\mu$ in $(n-1)$-dimensional
Einstein gravity~\cite{Vilenkin:vacuumdw, Ipser:repulsivedw}. 
Again, this is best seen by examining the Israel equations (at $\zeta=0$) in $(n-1)$ dimensions. The jump in extrinsic curvature across
a wall of tension $T$ is given by
equation (\ref{eqn:tensionT}). However, from the metric (\ref{indvac}), the jump in extrinsic curvature at
$\zeta=0$ is given by
\be
\Delta K_{ab} = -\frac{k_n\mu_n}{2} \gamma_{ab}
\ee
If we once again use  equation (\ref{eqn:NC}) with $\sigma_n=k_n$,
we can conclude $T \equiv \mu$. This proves that the geometry on the
brane is indeed behaving in an $(n-1)$-dimensional way, just as it did
for the vacuum bulk in section~\ref{sec:vacuumbulk}. We have shown
that even in this non-perturbative case, the RS model exhibits exact
$(n-1)$-dimensional Einstein gravity on the brane, even though the
model is manifestly $n$-dimensional. 
\subsection{Nested braneworlds}
We now have the tools to construct nested Randall-Sundrum type
configurations, that is, a flat ($\kappa =0$) ribbon on an AdS brane with an
AdS bulk. Fortunately, we see from  (\ref{zsolns}) that a
subcritical (AdS) brane can sustain a flat ribbon. From equations
(\ref{Rtraject}), (\ref{zsolns}) and (\ref{muvalues}), the  brane
trajectory is given by
\be
Z=Z_0e^{-k_{n-1}|\zeta|}, \qquad k_nR=\pm
\frac{4}{k_n\mu_n}\left( Z^{-1}-Z_0^{-1} \right)
\ee
with $\mu=4 k_{n-1}/\sigma_n$.
Transforming to conformal coordinates  $(u, v)=(1/k_nZ, k_nR)$, the brane trajectories become 
\be
v = \pm \frac{4}{\mu_n}(u-u_0)
\ee
Each branch of this trajectory is a subcritical brane, which, if it were
not for the vortex at $(u_0,0)$ would reach the AdS boundary at
$v = \mp \sigma_n u_0/k_{n-1}$.

Notice that the induced metric on the braneworld
\be
ds_{n-1} = Z_0^2 e^{-2k_{n-1}|\zeta|} [ -dt^2 +dx_i^2 ] +d\zeta^2
\ee
is indeed that of a RS universe in $(n-1)$ dimensions.  We would
expect there to be an analogue of the criticality condition for flat
branes. Again this arises from the Israel equations at $\zeta=0$. As
expected, we find that
\be
k_{n-1} = \frac{4\pi G_{n-1} \mu}{ n-3}
\ee
where we have used the condition (\ref{eqn:NC}). This is precisely the RS criticality condition $\sigma_n = 
4\pi G_n \sigma /(n-2) = k_n$ adjusted for one dimension less.

We conclude this section by emphasizing its main result. In each of
the examples we have looked at, the geometry on the brane has been in
exact agreement with the geometry predicted by $(n-1)$-dimensional
Einstein gravity, without any knowledge of the bulk. This is a
remarkable result because it means that, at least in this highly
symmetric set up, RS braneworld models exhibit localisation of gravity
on the brane, even in the non-perturbative regime. 

We have had the added bonus that we have seen how to construct nested
braneworld configurations. In the next section we will use the same
tools to construct braneworld instantons.
\section{Instantons and tunneling on the brane}\label{sect:inst}
Traditionally, instantons correspond to classical Euclidean solutions to the
equations of motion. In many cases, they represent a quantum tunneling from a
metastable false vacuum to a true vacuum. In~\cite{Coleman:vacuumdecay}, Coleman and
de Luccia discussed the effect of gravity on these decays. Such processes,
of course have direct relevance for cosmology, as they correspond to
a first order phase transition, and hence a
dramatic change in the structure of our universe.

In~\cite{Coleman:vacuumdecay}, the authors evaluated the probability of  
nucleation of a true vacuum
bubble in a false vacuum background. They focussed on two particular
configurations: a flat bubble spacetime in a de Sitter false vacuum;
and an AdS bubble spacetime in a flat false vacuum. This was before
the idea of large extra dimensions was fashionable, so the analysis
was done in just the usual four dimensions.

We now have the tools to develop these ideas in a braneworld set
up. To replicate the configurations of~\cite{Coleman:vacuumdecay}, we just have to
patch together our brane trajectories in the right way. Recall that
these trajectories are given by equation (\ref{hyper}), along with the
critical brane solution,  $u=\text{const}$. In Euclidean signature, 
the former are shaped like spheres and were illustrated in figures 
\ref{fig:walltraj:subfig:a} to \ref{fig:walltraj:subfig:d}. However, when patching these solutions together, 
we should be aware of a slight subtlety. In equation (\ref{trajectb}), 
the $\mu_n \sigma_n \delta(\zeta)$ term does not make sense if we have 
branes of different type either side of the vortex. Suppose we have a 
brane of tension $\sigma_n^+$ in $\zeta>0$, and $\sigma_n^-$ in $\zeta<0$, 
we must then modify equation (\ref{trajectb}) by replacing $\sigma_n$ 
with $\bar{\sigma}_n$, where 
\be
\bar{\sigma}_n=\frac{\sigma_n^+  + \sigma_n^- }{ 2}
\ee
It is easy to see that this is the right thing to do. Regard the
vortex as a thin wall limit of some even energy
distribution. Mathematically, this corresponds to $\mu_n \delta(\zeta)$
being the limit of some even function $\mu_nf(\zeta)$. The weight of the
distribution is the same on either side of $\zeta=0$, so we pick up
the average of the brane tensions.

We are now in a position to reproduce the work of~\cite{Coleman:vacuumdecay} in our
higher dimensional environment. Let us consider first the decay of a
de Sitter false vacuum, and the nucleation of a flat bubble spacetime.
\subsection{Nucleation of a flat bubble spacetime in a de Sitter false
vacuum}
We now describe the braneworld analogue of the nucleation of 
a flat bubble spacetime in a
de Sitter false vacuum. The de Sitter false vacuum is given
by a supercritical brane of tension $\sigma_n^{dS}>k_n$ 
with no vortex (see figure \ref{fig:walltraj:subfig:d}). This
metastable state decays into a ``bounce'' configuration given by a
critical brane (tension $\sigma_n^{flat}=k_n$) 
patched on to a supercritical brane
(tension $\sigma_n^{dS}>k_n$). If we are to avoid generating an
unphysical negative tension vortex we must patch together trajectories
in the following way:
\be
Z=\begin{cases} 
\frac{1}{k_{n-1}^{dS}}\cos(k_{n-1}^{dS}\zeta-\zeta_0)
 &  \zeta>0 \\
\zeta+\frac{1 }{ k_{n-1}^{dS}}\cos\zeta_0 &  \zeta<0  
\end{cases}
\ee
where $(k_{n-1}^{dS})^2=(\sigma_n^{dS})^2-k_n^2$. The vortex tension
$\mu$, 
is related to the constant $\zeta_0$ in the following way:
\be \label{matchingcond}
\frac{\mu_n \bar{\sigma}_n }{ 2  k_{n-1}^{dS}}=\sec\zeta_0-\tan\zeta_0
\ee
It is useful to have a geometrical picture of this bounce solution. We
just patch together the $u=$ const.\ critical brane trajectory and the
supercritical brane trajectory given by figure~\ref{fig:walltraj:subfig:d} to get
figure \ref{fig:bounce1}. Note that we have two copies of the bulk
spacetime because we imposed $\mathbb{Z}_2$-symmetry across the branes.
\begin{figure}[ht]
\begin{center}
\includegraphics[width=8cm]{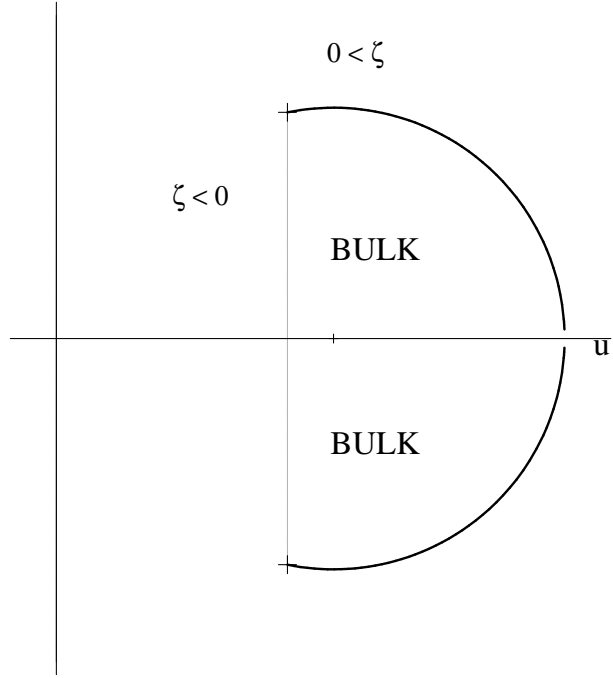}
\vskip 5mm
\caption{An example of a critical-supercritical brane ``bounce'' solution. 
This looks
like a flat bubble spacetime has nucleated on a de Sitter brane.} 
\label{fig:bounce1}
\end{center}
\end{figure}

It is natural to calculate the probability, $\mathcal{P}$, that this 
flat bubble spacetime does indeed nucleate on the de Sitter brane. 
\be
{\mathcal{P}}  \propto e^{-B}
\ee
where $B$ is the difference between the Euclidean actions of the bounce
solution and the false vacuum solution, that is:
\be
B=S_{bounce}-S_{false}
\ee
Given our geometrical picture it is straightforward to write down an
expression for the bounce action:
\begin{equation}
S_{bounce}=S_{bulk}+S_{flat}+S_{dS}+S_{vortex}
\end{equation}
where the contribution from the bulk, critical brane (flat),
supercritical brane (de Sitter), and vortex are as follows
\bml \label{bounceaction} \begin{eqnarray} 
S_{bulk} &=& -\frac{1}{16\pi G_{n}} \int_{bulk}
d^{n}x\sqrt{g}(R-2\Lambda_n) \label{bounceactiona} \\
S_{flat}&=& -\frac{1}{16\pi G_{n}}\int_{flat} d^{n-1}x
\sqrt{h}(-2\Delta K-4(n-2)\sigma_n^{flat})   \label{bounceactionb} \\
S_{dS} &=&  -\frac{1}{16\pi G_{n}}\int_{dS} d^{n-1}x
\sqrt{h}(-2\Delta K-4(n-2)\sigma_n^{dS})  \label{bounceactiond} \\
S_{vortex} &=& \mu \int_{vortex} d^{n-2}x \sqrt{\gamma}=
-\frac{1}{16\pi G_{n}}\int_{vortex} d^{n-2}x \sqrt{\gamma}(-2\mu_{n})  
\label{bounceactione}
\end{eqnarray}
\eml
Note that $\Delta K$, the jump in the trace of the extrinsic curvature 
across the brane, contains the Gibbons-Hawking boundary term~\cite{Gibbons:GHterm} for 
each side of the brane, and  $h_{ab}$, $\gamma_{ab}$ are the induced 
metrics on the brane and vortex respectively. We should point out 
that due to the presence of the vortex, there is a delta function in 
the extrinsic curvature that exactly cancels off the contribution 
of $S_{vortex}$.
The expression for $S_{false}$ is similar except that there is no flat
brane or vortex contribution, and no delta function in the extrinsic 
curvature. After some calculation (see appendix~\ref{app:instanton}), we find that 
our probability term, $B$ is given by:
\be\label{probterm}
B=\frac{4k_n^{2-n}\Omega_{n-2} }{ 16\pi G_n} \left( I_n-\left(\frac{1 }{
n-1} \right) \left(\frac{k_n\cos\zeta_0 }{
k_{n-1}^{dS}}\right)^{n-1} \right)
\ee
where $\Omega_{n-2}$ is the volume of an $(n-2)$ sphere and the integral
$I_n$ is given by:
\be \label{In}
I_n=\int_{u_0-u_1}^{u_c}  du \quad  \left ( u_0\frac{[\rho(u)]^{n-3} 
}{ u^{n-1}}-\frac{[\rho(u)]^{n-1} }{ u^n} \right )
\ee
where
\bml\label{dswallrelations}\bea
u_0 &=& \frac{\sigma_n^{dS} }{ (k_{n-1}^{dS})^2}(\sigma_n^{dS} +
k_n\sin\zeta_0)u_c \label{dswallrela} \\
u_1 &=& \frac{k_n }{ \sigma_n^{dS}}u_0  \label{dswallrelb}\\
\rho(u) &=& \sqrt{u_1^2-(u-u_0)^2} \label{dswallrelc} 
\eea\eml
Note that $u_c$ is an arbitrary constant so we are free to choose it as
we please (think of the flat brane as being at $u=u_c$). This integral is
non-trivial and although we can in principle solve it for any integer
$n$ it would not be instructive to do so. Instead, we will restrict
our attention to the case where $n=5$. This means that our braneworld
is four dimensional, so comparisons with~\cite{Coleman:vacuumdecay} are more
natural. Given that $\Omega_3=2\pi^2$, we find that:
\be
B = \frac{8\pi^2k_5^{-3} }{ 16\pi G_5} \Biggl[ 
\log \left[ \frac{\sigma_5^{dS}+k_5\sin\zeta_0 }{ \sigma_5^{dS}+k_5} \right] 
-\frac{1 }{ 2}  \left(\frac{k_5\cos\zeta_0 }{ k_4^{dS}}\right)^2 +\frac{k_5 \sigma_5^{dS} }{ (k_4^{dS})^2}
\left( 1-\sin\zeta_0 \right) \Biggr]
\ee
Equation (\ref{matchingcond}) in five dimensions enables us to replace
the trigonometric functions  using:
\bml
\bea
\cos\zeta_0 &=& \frac{2\lambda }{  1+\lambda^2} \\
\sin\zeta_0 &=& \frac{1- \lambda^2  }{ 1+\lambda^2}
\eea
\eml
where 
\be
\lambda=\frac{\mu_5\bar{\sigma}_5 }{ 2k_4^{dS}}
\ee
This leads to a complicated expression. It is perhaps more
instructive to examine the behaviour for small $\mu$ \ie\ in the regime
where we have a vortex with a low energy density. In this regime we
find that:
\be \label{Bsmallmu}
B=\frac{256\pi^5}{ (k^{dS}_4)^6}(G_5 \bar{\sigma}_5)^3\mu^4
+\mathcal{O} \rm (\mu^5)=\frac{256\pi^5}{ (k^{dS}_4)^6}(\bar{G}_4)^3\mu^4
+\mathcal{O} \rm (\mu^5)
\ee
where $\bar{G}_4$ is the average of the four dimensional Newton's 
constants on the flat brane and the de Sitter brane. The presence of 
this average as opposed to a single four dimensional Newton's constant 
is due to the difference in brane tension on either side of the vortex. 
From equation (\ref{eqn:NC}) we see that this induces a difference in 
the Newton's constants on each brane. 

We now compare this to the result we would have got had we assumed no
extra dimensions. The analogous probability term, $B^{\prime}$, is
calculated in~\cite{Coleman:vacuumdecay}. When the energy
density of the bubble wall, $\mu$,  is small, we now find
that\footnote{In order to reproduce equation (\ref{CDresult}) using
separating a flat bubble spacetime and a de Sitter spacetime whose
radius of curvature is $\frac{1 }{ k_4}$. Then substitute into the
relevant equations and take $\mu$ to be small.}:
\be \label{CDresult}
B^{\prime}=\frac{256\pi^5}{ (k_4)^6}(G_4)^3\mu^4 +\mathcal{O} \rm (\mu^5)
\ee
If we associate $G_4$ in equation (\ref{CDresult}) with $\bar{G}_4$ in equation (\ref{Bsmallmu}) we see that the approach of~\cite{Coleman:vacuumdecay}, where no extra dimensions are
present,  yields exactly the same result to the braneworld setup, at least for small $\mu$. 

Before we move on to discuss alternative instanton solutions we should
note that in the above analysis we have assumed $\frac{\pi }{ 2}>\zeta_0>0$. The
bounce solution presented is therefore really only valid if we have
$\lambda<1$. However, the extension to regions where $\lambda>1$
corresponds to allowing $\zeta_0$ to take negative values and
everything holds.
\subsection{Nucleation of an AdS bubble spacetime in a flat false vacuum} 
We now turn our attention to the decay of a flat false vacuum, and
the nucleation of an AdS bubble spacetime. The braneworld analogue of 
 the flat false vacuum is given by a critical brane of tension
$\sigma_n^{flat}=k_n$ with no vortex ($u=$ const). This 
decays into a new ``bounce'' configuration
given by a subcritical brane (tension $\sigma_n^{AdS}<k_n$) patched onto a
critical brane (tension $\sigma_n^{flat}=k_n$). Again, in order to avoid
generating an unphysical vortex, we must patch together trajectories
in the following way:
\be
Z=\begin{cases}
\zeta + \frac{1 }{ k_{n-1}^{AdS}}\sinh\zeta_0 & \zeta>0 \\
\frac{1 }{ k_{n-1}^{AdS}}\sinh(k_{n-1}^{AdS}\zeta +\zeta_0) &
\zeta<0
\end{cases}
\ee
where $(k_{n-1}^{AdS})^2=k_n^2-(\sigma_n^{AdS})^2$. The vortex tension
$\mu$ is related to the constant $\zeta_0$ in the following way:
\be \label{matchingcond2}
\frac{\mu_n\bar{\sigma}_n }{ 2k_{n-1}^{AdS}}=\textrm{coth}\zeta_0-
\textrm{cosech}\zeta_0
\ee
By patching together $u=$ const.\ and figure  \ref{fig:walltraj:subfig:a} we
again obtain a geometrical picture of our bounce solution (see figure
\ref{fig:bounce2}).
\begin{figure}[ht]
\begin{center}
\includegraphics[width=8cm]{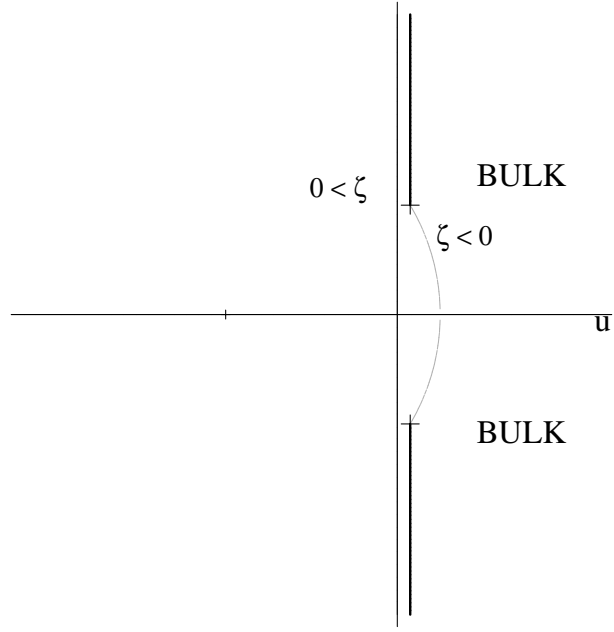}
\vskip 5mm
\caption{An example of a subcritical-critical brane ``bounce'' solution. 
This looks like an AdS  bubble spacetime has nucleated on a flat brane.} 
\label{fig:bounce2}
\end{center}
\end{figure}

As before we now consider the probability term, $B$, given by the
difference between the Euclidean actions of the bounce and the false
vacuum.  We shall not go into great detail here as the calculation is
very similar to that in the previous section. We should emphasize, 
however, that the bounce action will include as before 
an Einstein-Hilbert action  with negative cosmological constant in the
bulk, a Gibbons Hawking surface term on each brane, and tension
contributions from  each  brane and the vortex. Again we find that the delta function in the extrinsic curvature of the brane exactly cancels off the contribution from the vortex tension. The false vacuum action
omits the AdS brane and vortex contributions, and contains no delta functions from extrinsic curvature. Recall that 
in each case we have two
copies of the bulk spacetime because of $\mathbb{Z}_2$-symmetry across the
brane.

This time, we find that our probability term, $B$, is given by:
\be
B=\frac{4k_n^{2-n}\Omega_{n-2} }{ 16\pi G_n} \left( I_n+\left(\frac{1 }{
n-1} \right) \left(\frac{k_n\sinh\zeta_0 }{
k_{n-1}^{AdS}}\right)^{n-1} \right)
\ee
where the integral $I_n$ is now given by:
\be \label{newJn}
I_n=\int_{-u_0+u_1}^{u_c}  du \quad  \left ( u_0\frac{[\rho(u)]^{n-3} 
}{ u^{n-1}}+\frac{[\rho(u)]^{n-1} }{ u^n} \right )
\ee
where $u_c$ is an arbitrary constant corresponding to the position of
the flat brane, and
\bea
u_0 &=& \frac{\sigma_n^{AdS} }{ (k_{n-1}^{AdS})^2}(\sigma_n^{dS} +
k_n\cosh\zeta_0)u_c \\
u_1 &=& \frac{k_n }{ \sigma_n^{AdS}}u_0 \\
\rho(u) &=& \sqrt{u_1^2-(u+u_0)^2} 
\eea
Again, although we could in principle solve this integral for any
positive integer $n$, we shall restrict our attention to $n=5$. In
this case we now find that:
\be
B  =  \frac{8\pi^2k_5^{-3} }{ 16\pi G_5}
\Biggl[ -\log\left[\frac{\sigma_5^{AdS}+k_5\cosh\zeta_0 }{ 
\sigma_5^{AdS}+k_5} \right] +\frac{1 }{
2}  \left(\frac{k_5\sinh\zeta_0 }{
k_4^{AdS}}\right)^2  + \frac{k_5 \sigma_5^{AdS} }{
(k_4^{AdS})^2}\left(1-\cosh\zeta_0 \right) \Biggr]
\ee
We can now replace the hyperbolic functions using equation
(\ref{matchingcond2}):
\bml \bea
\sinh\zeta_0 &=& \frac{2\lambda }{  1-\lambda^2} \\
\cosh\zeta_0 &=& \frac{1+\lambda^2  }{ 1-\lambda^2}
\eea \eml
where 
\be
\lambda=\frac{\mu_5\bar{\sigma}_5 }{ 2k_4^{AdS}}
\ee
This is again an ugly expression. It is more interesting to examine 
the behaviour at small $\mu$:
\be \label{newBsmallmu}
B=\frac{256\pi^5}{ (k^{AdS}_4)^6}(G_5 \bar{\sigma}_5)^3\mu^4
+\mathcal{O} \rm (\mu^5)=\frac{256\pi^5}{ (k^{AdS}_4)^6}(\bar{G}_4)^3\mu^4
+\mathcal{O} \rm (\mu^5)
\ee
This is  very similar to what we had for the nucleation of a flat
bubble spacetime in a de Sitter false vacuum with $\bar{G}_4$  now 
representing the average of the Newton's constants on the AdS brane 
and the flat brane. Again we compare this to
the result from~\cite{Coleman:vacuumdecay} where we have no extra dimensions. When the
energy density of the bubble wall is small, the expression for the
probability term is again given by equation (\ref{CDresult}), where
$\frac{1 }{ k_4}$ now corresponds to the radius of curvature of the 
AdS spacetime. Once again we see that the braneworld result agrees exactly with~\cite{Coleman:vacuumdecay} in the small $\mu$ limit, provided we associate $G_4$ with $\bar{G}_4$.

Note that again we have assumed $\zeta_0>0$ and therefore, the
bounce solution given here is only valid for $\lambda<1$. The
extension to $\lambda>1$ is more complicated than for the nucleation
of the flat bubble in the previous section.  We now have to patch
together figure~\ref{fig:walltraj:subfig:b} and figure~\ref{fig:walltraj:subfig:c}. 
However, in~\cite{Coleman:vacuumdecay}, the
analogue of $\lambda>1$ violates conservation of energy as one tunnels
from the false vacuum  to the new configuration. In the braneworld set up we
should examine what happens as $\lambda$ approaches unity from
below. In this limit, $\zeta_0$ becomes infinite, and the AdS bubble
encompasses the entire brane. The probability,  $\mathcal{P}$, of
this happening is zero and so there is no vacuum decay. Beyond this, in
analogy with~\cite{Coleman:vacuumdecay}, we would suspect that the energy of the 
false vacuum is
insufficient to allow the nucleation of a bubble with a large wall
energy density.  This is indeed the case. When we calculate the
probability term, $B$,  for the AdS bubble in a flat, spherical 
false vacuum, we find that it is divergent and the probability of bubble 
nucleation vanishes. This divergence comes from the fact that the 
false vacuum touches the AdS boundary whereas the bounce solution does not.
 
Finally, we could also have created an AdS bubble in a flat spacetime using a
$\kappa=0$ vortex.  However, it is of no interest since the probability of
bubble nucleation is exponentially
suppressed by the vortex volume.
\subsection{Ekpyrotic Instantons}
The notion of the Ekpyrotic universe~\cite{Khoury:ekpyrotic} proposes that 
the Hot Big Bang came about as the result of a collision between 
two braneworlds. The model claims to solve many of the problems facing 
cosmology without the need for inflation. Although the authors work 
mainly in the context of heterotic M theory, they acknowledge that an 
intuitive understanding can be gained by considering Randall-Sundrum type 
braneworlds. In this context, we regard the pre-Big Bang era in the 
following way. We start off with two branes of equal and opposite tension: 
the hidden brane of positive tension, $\sigma$, and the visible brane of 
negative tension, $-\sigma$.  A bulk brane with a small positive 
tension, $\frac{\epsilon }{ 2}$, then ``peels off'' the hidden brane 
causing its tension to fall to $\sigma-\epsilon$. The bulk brane is then 
drawn towards our universe, the visible brane, until it collides with us, 
giving rise to the Hot Big Bang. 

The process of ``peeling off'' is not really considered in great detail 
in~\cite{Khoury:ekpyrotic}. They suggest that the hidden brane undergoes a 
small instanton transition with the nucleation of a bubble containing 
a new hidden brane with decreased tension, and the bulk brane. 
The walls of this bubble then expand at the speed of light until it 
envelopes all of the old hidden brane. Given that all the branes in 
this model are critical we can illustrate the instanton solution in the
simplified RS set-up by using 
a suitable combination of critical brane solutions. In conformal 
coordinates, critical branes look like planes ($u=\text{const}$) or 
spheres (see figure~\ref{fig:walltraj:subfig:c}). In describing the Ekpyrotic 
instanton we present the visible and hidden branes (old and new) as planes. 
The bulk brane is given by a sphere that intersects the hidden brane, 
separating the old and new branches (see figure~\ref{fig:Ekpyrotic}).
\begin{figure}[ht]
\begin{center}
\includegraphics[width=8cm]{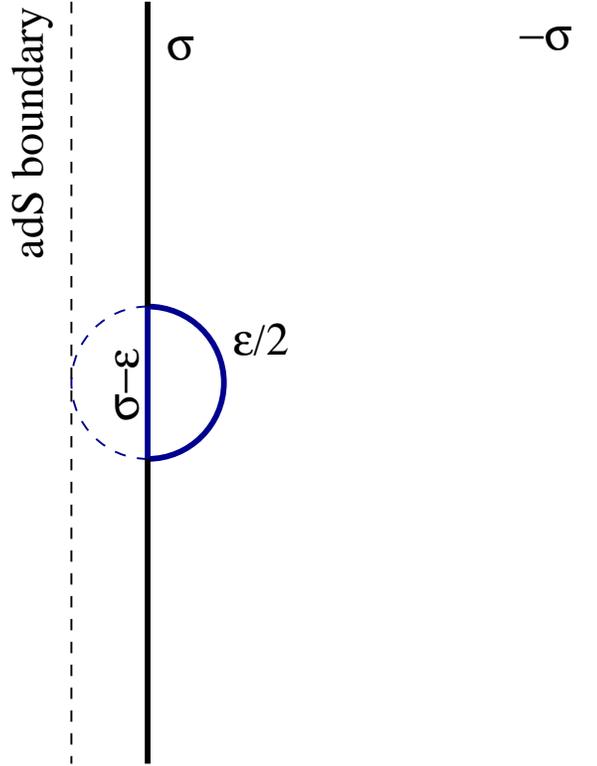}
\vskip 5mm
\caption{ The Ekpyrotic Instanton}
\label{fig:Ekpyrotic}
\end{center}
\end{figure}

Given this geometrical picture we can calculate the probability 
of tunneling to this configuration from the initial two brane state. 
We proceed much as we did in the previous section, and obtain the 
following expression for the probability term:
\be
B=\frac{\pi^2 \epsilon }{4}\left(\frac{3 }{ k^3}\ln(1+k^2Z_0^2)
-\frac{2k^{-2}Z_0^2 }{ 1+k^2Z_0^2}-\frac{Z_0^2 }{ k^2}-\frac{Z_0^4 }{4} \right) 
+\mathcal{O} \rm (\epsilon^2)
\ee
where $k$ is related to the cosmological constant in the bulk of the 
initial state ($\Lambda = -6k^2$), and $Z_0$ is a free parameter 
related to the ``size'' of the bubble: the larger the value of $Z_0$, 
the larger the bubble.  We should not be worried by this freedom 
in $Z_0$, as we are working with Randall-Sundrum  braneworlds which are 
much simpler than the M5 branes of heterotic M theory. When we return 
to the  M theory context, we lose a number of degrees of freedom and 
one might expect the value of $Z_0$ to be fixed. However, since we 
are dealing with a ``small'' instanton, we might expect $Z_0$ itself 
to be small, and the probability term approximates to the following:
\be
B=\frac{\pi^2 }{16}\epsilon Z_0^4 + \mathcal{O} \rm (\epsilon^2, Z_0^6)
\ee
We should once again stress however, that this is an extremely simple
and naive calculation that ignores any dynamics of the additional scalars,
or other fields, that result from a five-dimensional heterotic M-theory 
compactification~\cite{Lukas:Mtheory5}.
Another point to note is that while we can have a small brane peel off from the
positive tension braneworld, we cannot have one peel off from the negative
tension braneworld, as a quick glance at figure~\ref{fig:Ekpyrotic} shows.
Such a brane, being critical, must have the form of a sphere grazing 
the AdS boundary, which therefore necessarily would intersect the positive
tension brane as well. 
\section{The AdS soliton}
Recall that at the end of section~\ref{section:dr} we set the
integration constant, $c$, to zero. Now consider what happens when we allow for
non-zero values. We will assume that we have a negative cosmological
constant in the bulk given by equation (\ref{lamkn}). The bulk
spacetime is now described by the metric (\ref{canbulk}) with
\be
h(Z)=k_n^2Z^2+\kappa-\frac{c}{Z^{n-3}}
\ee
For $c<0$, the metric becomes singular at the AdS horizon, $Z=0$. Of
greater interest is the case $c>0$, when the metric takes the form of
the AdS soliton~\cite{Horowitz:AdSsoliton}. This is the double
analytic continuation of the AdS-Schwarzschild metric
(\ref{eqn:schmetric}). For this reason the $(R, Z)$ plane behaves like
a Euclidean AdS black hole, with a horizon at $Z_H$ where
$h(Z_H)=0$. In order to avoid a conical singularity at $Z_H$, we cut
the spacetime off there, and identify $R$ as a angular coordinate with
periodicity $\Delta R=4 \pi/h'(Z_H)$. The geometry (up to an AdS warp factor)
is therefore the familiar cigar shape with a smooth tip at $Z=Z_H$.


We can clearly try to  play the same game as before and investigate
branes and vortices in the AdS soliton background. The
equations of motion (\ref{srctraject}) for a $\mathbb{Z}_2$-symmetric brane become:
\bml\label{adssoleom}\bea
Z^{\prime 2} &=& -aZ^2 + \kappa - \frac{c}{ Z^{n-3}}\label{adssola}\\
Z'' &=& -a Z + \left(\frac{n-3 }{ 2} \right)\frac{c}{ Z^{n-2}} - 4\pi G_n \sum_i
\mu_i\sigma_n Z\delta(\zeta - \zeta_i)\label{adssolb}\\
R' &=& \frac{\sigma_n Z}{k_n^2Z^2+\kappa-\frac{c}{Z^{n-3}}}
\label{adssolc}
\eea\eml
where $a = \sigma_n^2-k_n^2$ as before, and (\ref{adssolb}) now allows for
a multitude of vortices of tension $\mu_i$ located at $\zeta_i$.

Although we will explicitly solve these equations for $n=5$, we will
simply describe the qualitative behaviour of solutions for arbitrary
values of $n$. The generic trajectory (which must be periodic in $R$) will consist of 
two segments of $Z(\zeta)$ of opposite gradient. These patch together at
a positive tension vortex at $R=0$, say, and a 
negative tension vortex at $R=\Delta R/2$. This is exactly analogous
to the usual situation with a domain wall spacetime when we need both
positive and negative tension walls to form a compact extra dimension. 
However, we see that with the bulk ``mass'' term, $c$, there are now also other
possibilities. This is because (\ref{adssola}) now has at least one root for
$Z>Z_H$, and for supercritical branes $(a>0)$ there are two roots. These roots correspond
to zeros of $Z'$, and enable a smooth transition from
the positive branch of $Z'$ to the negative branch. We can therefore form 
a trajectory which loops symmetrically around the cigar, and has only
one kink -- which we can fix to be a positive tension vortex. Of course,
the tension of this vortex will be determined by the other parameters
of the set-up: the bulk mass, cosmological constant and the braneworld
tension, but this is no worse a fine tuning than is already
present in conventional RS models. Note that this is now distinct from
a domain wall on a compact extra dimension, as we can construct a
domain ribbon spacetime with only a single positive tension vortex on
asymptotically de Sitter, flat and anti-de Sitter branes.
In addition, for a supercritical brane (asymptotically de Sitter) we have the possibility of dispensing
with the ribbon altogether. In this case we have a smooth trajectory with two
roots of $Z'$, where the brane smoothly wraps the cigar, although a
fine tuned mass term is required

In all cases the induced geometry on the brane has the form
\be
ds_{n-1}^2 = Z^2(\zeta) d{\bf x}^2_\kappa + d\zeta^2
\ee
(where of course $\zeta$ has a finite range). The energy momentum
tensor of this spacetime is
\bml\bea
8\pi G_{n-1} T^\mu_\nu &=& \left[ (n-3)\frac{Z''}{ Z} +\frac{(n-3)(n-4)}{2}
\frac{(Z^{\prime2} - \kappa)}{ Z^2}  \right]\delta^{\mu}_{\nu}
\nonumber\\
&=& -\frac{(n-3)}{2} \left [ (n-2)a -\frac{c}{ Z^{n-1}} +\mu_n\sigma_n\delta(\zeta) \right ]\delta^{\mu}_{\nu}\\
8\pi G_{n-1} T^\zeta_\zeta &=& \frac{(n-2)(n-3)}{2}
\frac{(Z^{\prime2} - \kappa)}{ Z^2} \nonumber \\
&=& -\frac{(n-2)(n-3)}{2}
\left ( a +\frac{c}{ Z^{n-1}} \right)
\eea\eml
which has three distinct components. There is a cosmological constant (the $a$-term)
which reflects the lack of criticality of the braneworld when it is
non-vanishing. The domain ribbon terms ($\mu_i$) when present indicate
the presence of a nested $(n-3)$-brane -- note the normalisation is
precisely correct for the induced $(n-1)$-dimensional Newton's constant.
Finally, the $c$-term corresponds to a negative stress-energy tensor and can be directly associated to the Casimir 
energy of field theory living in the braneworld. We will
discuss holographic interpretations like this in more detail in chapters~\ref{chap:holography} and~\ref{chap:exact}.

\subsection{The AdS soliton in five dimensions}

We will now present explicit solutions to
equations~(\ref{adssola}) and (\ref{adssolc}) when we specialise to
$n=5$. We also restrict attention to the case where $\kappa=1$, which
in any case is the only possibility for supercritical and critical branes. So as not
to be littered with confusing suffices let us adopt the notation of
chapter~\ref{chap:RSbw} and replace  $k_5$ and $\sigma_5$ with $k$ and
$\tilde \sigma$ respectively.  The set of equations we wish to solve are therefore just:

\bea
Z^{\prime 2}&=&-aZ^2+1-\frac{c}{Z^2}  \label{brane1}\\
R^{\prime} &=& \frac{{\tilde \sigma} Z}{k^2Z^2+1-\frac{c}{Z^2}} \label{brane2}
\eea
where $a={\tilde \sigma}^2-k^2$. We can easily solve (\ref{brane1}) to obtain $Z(\zeta)$. The solutions are:
\be \label{solitonsol}
Z(\zeta)^2= \begin{cases} \frac{1}{2|a|}\left[-1+\sqrt{1-4ac}\cosh(2\sqrt{|a|}(\zeta-\zeta_0))\right]  & a<0 \\
c+(\zeta-\zeta_0)^2 & a=0 \\
  \frac{1}{2a}\left[1+\sqrt{1-4ac}\cos(2\sqrt{a}(\zeta-\zeta_0))\right] & a>0
\end{cases}
\ee
where $\zeta_0$ is just a constant of integration. Notice that the
solution for the supercritical wall is only valid when $c \leq
\frac{1}{4a}$. As a consistency check we observe that
(\ref{solitonsol}) gives (\ref{genzsolns}) when $c=0$. In order to
construct branes containing domain ribbons we patch together solutions
with the opposite sign in $Z'$. This corresponds to taking the
opposite sign in the square root of (\ref{solitonsol}).

We now tackle the more interesting problem of expressing $R$ in terms of $Z$. The governing equation is given by:
\be \label{R and Z ode}
\frac{dR}{dZ}=\pm \frac{{\tilde \sigma} Z}{k^2Z^2+1-\frac{c}{Z^2}} \frac{1}{\sqrt{-aZ^2+1-\frac{c}{Z^2}}}
\ee
Consider first critical branes with $a=0$. Define:
\bea
x_{\pm}&=&\frac{-1 \pm \sqrt{1+4k^2c}}{2k^2} \\
\mu_{\pm} &=& 1-\frac{2x_{\pm}}{c} \\
\nu_{\pm} &=& \frac{2}{c}\sqrt{\pm x_{\pm}(c-x_{\pm})}
\eea
Given that for critical branes, ${\tilde \sigma}=k$, the solution is:
\bea
R(Z) &=& R_0 \pm\frac{1}{k}\Bigg[ \cosh^{-1}\left(\frac{Z}{\sqrt{c}}\right) \nonumber \\
& &+\frac{c-x_+}{ \sqrt{1+4k^2c}}\left( \frac{2}{\nu_+ c} \right)\tan^{-1}\left(\frac{u(Z)+\mu_+}{\nu_+}\right) \nonumber \\
& &-\frac{c-x_-}{ \sqrt{1+4k^2c}}\left( \frac{1}{\nu_- c}\right)\log\left| \frac{u(Z)+\mu_- - \nu_-}{u(Z)+\mu_- + \nu_-}\right| \Bigg]
\eea
where 
\be
u(Z)=\exp\left[2\cosh^{-1}\left(\frac{Z}{\sqrt{c}}\right)\right]
\ee
and $R_0$ is an integration constant. When we consider the non critical branes we find that equation (\ref{R and Z ode}) gives an elliptic integral. The best we can do is express the solution in terms of canonical forms for elliptic integrals. We will require the incomplete elliptic integrals of the first and the third kind. They are defined below for $0 \leq x \leq 1$~\cite{Abramowitz:elliptic}:

\bea
F(x|t) &=& \int_0^x \frac{dz}{\sqrt{(1-z^2)(1-tz^2)}} \\
\Pi(n;x|t) &=& \int_0^x \frac{dz}{\sqrt{(1-z^2)(1-tz^2)}}\left( \frac{1}{1-nz^2} \right)
\eea
where $0<t<1$. We will also need to define the following:
\bea
\lambda_{\pm} &=& \frac{1 \pm \sqrt{1-4ac}}{2a} \\
n_{\pm} &=& \frac{\lambda_+-\lambda_-}{\lambda_+-x_{\pm}} \\
m_{\pm} &=& \frac{x_{\pm}}{x_{\pm}-\lambda_-} \\
q &=& \frac{\lambda_+-\lambda_-}{\lambda_+}
\eea
Consider now the supercritical branes with $a>0$. The solution is:
\bea
R(Z) &=& R_0 \mp \frac{{\tilde \sigma}}{k^2 \sqrt{a\lambda_+}}\Bigg[ F(v(Z)|q) \nonumber \\
& & +\frac{1}{\sqrt{1+4k^2c}}\left(\frac{c-x_+}{\lambda_+-x_+}\right) \Pi(n_+;v(Z)|q) \nonumber \\
& & -\frac{1}{\sqrt{1+4k^2c}}\left(\frac{c-x_-}{\lambda_+-x_-}\right) \Pi(n_-;v(Z)|q)\Bigg]
\eea
where 
\be
v(Z)=\sqrt{\frac{\lambda_+-Z^2}{\lambda_+-\lambda_-}}
\ee
For subcritical branes, with $a<0$, the solution is: 

\bea
R(Z) &=& R_0 \pm \frac{{\tilde \sigma} \lambda_-}{k^2\sqrt{|a|(\lambda_--\lambda_+)(1+4k^2c)}} \Bigg[ \frac{1}{x_+}\left(\frac{c-x_+}{\lambda_--x_+}\right)\Pi(m_+;w(Z)|q^{-1}) \nonumber \\
& & \qquad  -\frac{1}{x_-}\left(\frac{c-x_-}{\lambda_--x_-}\right)\Pi(m_-;w(Z)|q^{-1}) \Bigg]
\eea
where 
\be
w(Z)=\sqrt{\frac{Z^2-\lambda_-}{Z^2}}.
\ee

As a mathematical exercise, the derivation of these solutions has been
of some use. However, do we gain any further understanding of
braneworld physics?  We suggested earlier that for the supercritical
brane, we might be able to place a brane on  this compact soliton background,
without any need for a vortex. This is because the supercritical
brane solution is periodic, so we can have a smooth brane trajectory
wrapping the soliton cigar. We are now in a position to investigate
this more closely.

If such a configuration does exist, then the following would be true:
\be
2N[R(Z_{max})-R(Z_{min})]=\Delta R
\ee
where $N=1,2,3...$, and  $Z_{max}=\sqrt{\lambda_+}$, $Z_{min}=\sqrt{\lambda_-}$ are the maximum and minimum values of $Z$
respectively. This amounts to the fine tuning conditon on
the mass term. In particular, the value of $c$ that satisfies
\be \label{finetunec}
\mathcal{F}(c)=2[R(Z_{max})-R(Z_{min})]-\Delta R=0
\ee
gives us the fine tuned valued for $N=1$. It is not obvious that
(\ref{finetunec}) has a solution in the allowed range, $0 \leq c \leq
1/4a$. However, we can choose $\tilde \sigma$ and $k$ and then hope to solve
$\mathcal{F}(c)=0$ numerically.
\begin{figure}[t]
\begin{center}
\includegraphics[width=7cm, height=7cm]{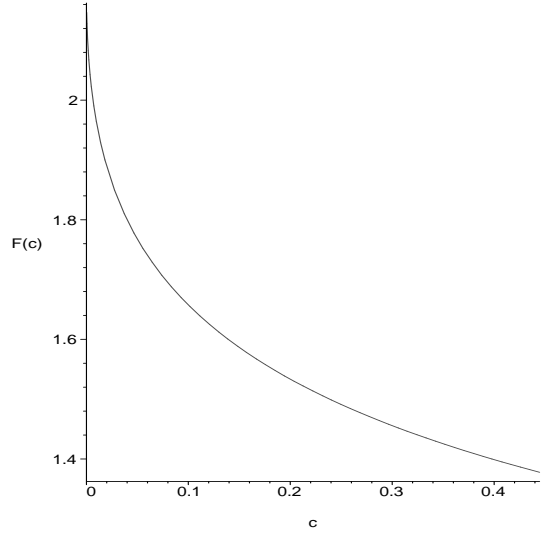}
\vskip 5mm
\caption{A plot of $\mathcal{F}(c)$ for $0 \leq c \leq
1/4a$ when $\tilde \sigma=1.25$ and $k=1$. Note that there is no
solution to $\mathcal{F}(c)=0$.}
\label{fig:test}
\end{center}
\end{figure}
For example, in figure~\ref{fig:test}, we see that
there is no solution for $\tilde \sigma=1.25$ and $k=1$. Whether or not this behaviour is true for all choices of  $\tilde
\sigma$ and $k$ is unclear and would require a detailed analytic investigation.
\chapter{Braneworld holography} \label{chap:holography}
\setcounter{equation}{0}
We will now change direction in our study of braneworlds, and focus on
how they fit into the realms of {\it holography}. We will discover that
the brane cosmology described in chapter~\ref{chap:bwcosmo} can be
understood from a ``holographic'' point of view, by adjusting the
properties of the bulk geometry. Before describing this in detail, it
is important we review some of the fundamental ideas behind {\it the
holographic principle}~\cite{'tHooft:holog,Susskind:holog} as well as its most celebrated example, the
AdS/CFT correspondence~\cite{Maldacena:adscft,
Gubser:adscft,Witten:adscft}.
\section{The holographic principle}
The holographic
principle is a radical idea that rose from attempts to
understand gravity and quantum field theory simultaneously. The
natural tool with which to do this is the black hole. As we move close
enough to the black
hole singularity, the curvature of spacetime becomes of order the
Planck scale. At this point the gravitational interactions become as
strong as the weak interactions, and the classical description of
gravity is inadequate. The time has come to apply quantum physics.

There are two very important results that arise from a quantum
description of black holes. The first concerns the black hole entropy,
which rather surprisingly turns out to be~\cite{Bekenstein:2ndlaw1,Bekenstein:entropy,Bekenstein:2ndlaw2,Hawking:explosions}
\be \label{eqn:BHentropy}
S=\frac{A}{4G_n}
\ee
where $A$ is the area of the black hole horizon. The second result is
due to Hawking~\cite{Hawking:radiation} who noticed that black holes are not as black
as they seem. They emit thermal radiation (Hawking radiation!) and can
eventually evaporate.

Now consider a spherical region, $\Gamma$, of volume $V$,  in an asymptotically flat spacetime. We
will place no restrictions on the matter contained within, and will
only state that the boundary, $\delta \Gamma$ has area $A$. We begin by using local quantum field theory to calculate
the maximum entropy, $S_{max}$, of the quantum mechanical
system contained in $\Gamma$. By definition,
\be
S_{max}=\ln[N_{states}]
\ee
where $N_{states}$ is the  total number of possible states of
$\Gamma$. We can think of the maximum entropy as counting the total
number of degrees of freedom. Locality tells us that
there is at least one degree of freedom at each spatial point, so we
conclude that $S_{max}= \infty$. Even if we say that $\Gamma$ is not
continuous but discrete we still find that $S_{max} \propto V$, as we will
now explain. Suppose that $\Gamma$ is really a lattice with lattice spacing
$\alpha$. The number of cells is approximately
$V/\alpha^{n-1}$, where $n$ is the spacetime dimension. We assume that
each cell has $m$ possible states, and
deduce that
\be \label{eqn:QFTstates}
N_{states}=m^{V/\alpha^{n-1}}
\ee
The maximum entropy is then
\be
S_{max}=\ln[N_{states}]=\frac{V\ln m}{\alpha^{n-1}} \propto V
\ee

We now use our knowledge of gravity and black holes to calculate the maximal
entropy.   First consider how much mass can be contained in
$\Gamma$. We cannot continue to add mass to  $\Gamma$ indefinitely
because eventually we will start to form a black hole. As we wish to
avoid gravitational collapse  we have an upper bound on the mass. It
corresponds to the mass, $M$ of the black hole that just fits inside
$\Gamma$, with its horizon
coinciding exactly with the boundary $\delta \Gamma$. Such a black
hole has entropy given by equation (\ref{eqn:BHentropy}). If the
mass is smaller than $M$ we can avoid gravitational collapse. If it is
$M$ or larger, gravitational collapse is inevitable.

Now suppose that $\Gamma$ starts of with mass, $E$ and 
entropy $S$. We must have $E < M$ to ensure that gravitational collapse
has not already taken place. Now consider a spherically symmetric shell of matter with
entropy $S'$. The combined system has initial entropy
\be
S_{initial}=S+S'
\ee
We now assume that the shell is collapsing to form a black hole inside $\Gamma$. If the shell
has mass $M-E$, then we might expect that the final state will be the black hole described in
the last paragraph. The final entropy of the combined system is
therefore given by
\be
S_{final}= \frac{A}{4G_n}
\ee
However, the second law of thermodynamics tells
us that the entropy of a thermodynamical system  cannot decrease. This
means that $S_{initial} \leq S_{final}$, and since $S' \geq 0$ we
conclude that
\be
S \leq \frac{A}{4G_n} \qquad \Rightarrow \qquad S_{max}= \frac{A}{4G_n}.
\ee
The gravitational approach and the QFT approach are clearly at odds
with one another. Gravity predicts $S_{max} \propto A$ whereas QFT
predicts $S_{max} \propto V$. It turns out that its the QFT approach
that is wrong because it badly over-estimates the number of degrees of
freedom. Each cell of the lattice described earlier has volume
$V_{cell}=\alpha^{n-1}$. How much mass, $E_{cell}$, can be contained in a
particular cell without the threat of gravitational collapse? Again, we
can have no more than the mass of the largest black hole that can fit
into  the cell. The mass of a black hole is given by its radius, so we
see that $E_{cell} \lesssim \alpha$. This implies that the total mass contained
in $\Gamma$, is no greater than
\be
E_{max}= \alpha \frac{V}{\alpha^{n-1}} =\frac{V}{\alpha^{n-2}}
\ee
However, the mass, $M$ of the largest black hole that can fit inside
$\Gamma$ is given by the radius of $\Gamma$, so that
\be
M \sim V^{1/n-1}
\ee
We require $E_{max} \leq M$ which gives $V \lesssim \alpha^{n-1}$. So the
upper mass limit $E_{max}$ is only valid if $\Gamma$ is the size of a
single cell. The lattice approach permits total energies
that exceed the mass, $M$ of the largest black hole. This means that
although black holes will not form in each individual cell, they will
form on larger scales.
 
We could of course reject the gravitational approach if we accept the
possibility of gravitational collapse. Let us suppose that the number
of possible states of $\Gamma$ is indeed given by equation
(\ref{eqn:QFTstates}). If $\Gamma$ contains total mass $M$, the matter
within will collapse to form a black hole. After collapse, the number
of possible states is given by $e^{A/4G_n}$. This violates unitarity
because the number of initial states is greater than the number of
final states. Hawking argued that unitarity broke down in black
holes~\cite{Hawking:predict}. If we accept that the maximum entropy of
a spatial
region is proportional to the area of its boundary, rather than its volume, then we can retain
unitarity in black holes. This is how the holographic principle was
first formulated.

Note that we have made various assumptions so far, such as spherical
volumes and asymptotic flatness. We might think that the maximum
entropy of a spatial region is still given by $A/4G_n$, even when we
drop these assumptions. This is known as the spacelike entropy bound,
but it is clearly not valid. Suppose we have a contracting universe. Entropy does not
decrease but the boundary area of a given region does. As we shrink
smaller and smaller the entropy will eventually exceed the boundary
area. 

We can however form the covariant entropy bound~\cite{Susskind:holog,Bousso:holog}
using light sheets and the focusing theorem of General
Relativity (for a nice review see~\cite{Bigatti:TASI} or~\cite{Bousso:review}). Briefly this states that given a (codimension 2) boundary
surface, $\delta \Gamma$, of area $A$, the entropy on any light sheet of $\delta
\Gamma$ cannot exceed $A/4G_n$. A light sheet is made up of the light rays passing through
$\delta \Gamma$, as long as they are not expanding. Note that a light
sheet is a null surface whereas $\Gamma$ is a spacelike surface.

Depending on the structure of the spacetime, we can use the covariant
entropy bound to place bounds on spacelike surfaces. Consider anti-de
Sitter space (see figure~\ref{fig:holog}). 
\begin{figure}[ht]
\begin{center}
\includegraphics[width=13cm, angle=-90]{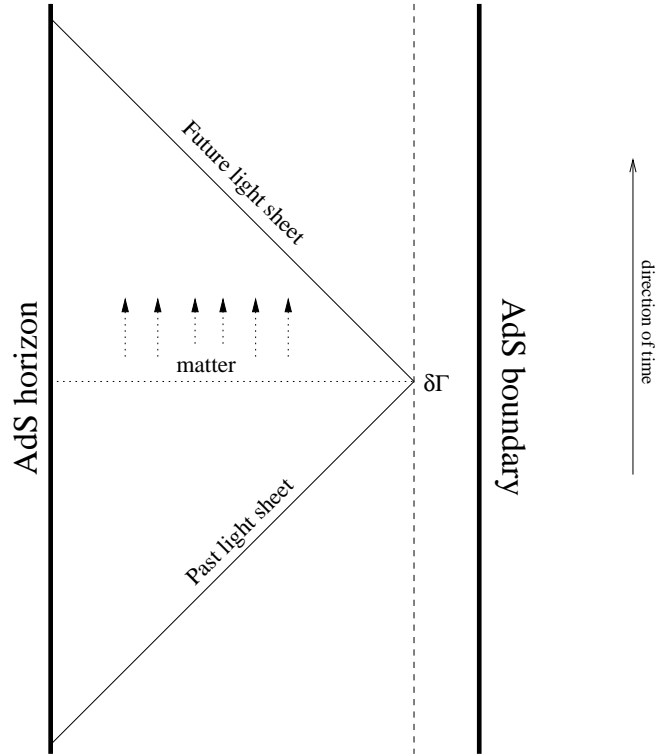}
\vskip 5mm
\caption{In AdS space, matter contained within $\Gamma$ crosses the
future light sheet of the boundary $\delta \Gamma$.} \label{fig:holog}
\end{center}
\end{figure}
Since light sheets are not allowed to expand, the warped
geometry of AdS means that light sheets point away from the AdS boundary
towards the AdS horizon. Now take a  static (codimension 2) surface, $\delta
\Gamma$, near
the  AdS boundary and consider the region $\Gamma$ bounded by $\delta
\Gamma$, including the AdS horizon. Since the future light sheet of
$\delta \Gamma$ points towards the horizon, matter contained in
$\Gamma$ will eventually pass through it.  Suppose the entropy
contained in $\Gamma$ is $S$. When the matter in  $\Gamma$ passes through the light
sheet its entropy is $S' \geq S$, in accordance with the second law of
thermodynamics. By  the covariant entropy bound we have $S' \leq
A/4G_n$, and therefore
\be
S \leq \frac{A}{4G_n}
\ee
This is an important property of AdS space. There is a timelike Killing
vector (so we can define static surfaces like $\delta \Gamma$) and the entire geometry can be
foliated by spacelike surfaces satisfying the holographic bound. This
means  that we have the counterintuitive result: the total number of
physical degrees of freedom in a region, $\Gamma$ of AdS  is proportional to the
area of the boundary, $\delta \Gamma$,  rather than the volume. Note that we can take $\delta \Gamma$  to be as close to the AdS
boundary as we wish, so in this sense the holographic principle
applies to the whole of
AdS. With all this in mind, it is
no surprise that the first concrete example of holography
involves anti-de Sitter space. This is the AdS/CFT
correspondence~\cite{Maldacena:adscft, Witten:adscft, Gubser:adscft}, which we will now
describe, albeit very briefly.
\section{The AdS/CFT correspondence}
Consider type IIB string theory on 10 dimensional Minkowski
space. The fundamental dimensionful parameter of the theory is the
string tension, $T \propto l_s^{-2}$, where  $l_s$ is a stringy length scale. We define  $g_s$ to be the string coupling, which we
will hold fixed.  IIB string theory
contains objects known as D$p$-branes~\cite{Johnson:primer}. A D$p$-brane is a timelike brane of
$p$ dimensions where the ends of open strings can terminate. Its world
volume is therefore $(p+1)$-dimensional. The low
energy physics
of a single brane is described by a $U(1)$ gauge theory. For $N$
distinct branes we naturally have a $U(1)^N$, although $N$ coincident
branes are described by an $SU(N)$, where we neglect an overall centre
of mass degree of freedom. Furthermore, we can think of
a D-brane as a source of energy momentum in the bulk spacetime as
well as a source of other supergravity fields. It couples to the bulk
by absorbing and emitting closed strings.

Now consider $N$ coincident D3-branes and 
take the following low energy supergravity limit,
\be
l_s \to 0, \qquad u \equiv r/l_s^2 =\textrm{fixed}.
\ee
Here $u$ represents a typical energy scale corresponding to an open
string stretched by an amount $r$. In this limit, the closed string
physics in the bulk can be shown to decouple from the open string
physics on the brane~\cite{Maldacena:adscft}. The
open string physics is described by $\mathcal{N}=4$ $SU(N)$ super Yang Mills on Minkowski
space in 3+1 dimensions. 

As we stated earlier we can think of D-branes as sources of
10-dimensional supergravity fields, in this case the metric, dilaton
and 4-form potential, $C_{(4)}$, with field strength $F_{(5)}=dC_{(4)}$. The {\it extremal} black D3-brane
solution is given by~\cite{Horowitz:pbranes}
\be \label{eqn:Dbranes}
ds^2_{10} = H^{-1/2}(r)\eta_{\mu\nu}dx^{\mu}dx^{\nu}
+H^{1/2}(r)\left[dr^2+r^2d\Omega_5^2 \right]
\ee
where the dilaton is constant and
\be
H(r)=1+4\pi g_sN\left(\frac{l_s}{r} \right)^4.
\ee
$d\Omega_5^2$ is the metric on $S^5$ and  $x^{\mu}$, for
$\mu=0,1,2,3$, are the D-brane coordinates.  The D-brane stack is located at $r=0$, which also corresponds to the
horizon in the extremal case. Finally, we note that the 5-form flux
through the 5-sphere surrounding the D-brane source is integer valued,
\be
\int_{S^5} F_{(5)}=N.
\ee

As before we define $u=r/l^2_s$. In the low energy supergravity limit
we have $l_s \to 0$, so holding $u$ fixed corresponds to taking the
{\it near horizon limit}. The limiting form of the metric is just
$AdS_5 \times S^5$,
\be
ds_{10}^2=l_s^2\left[ \sqrt{4\pi g_sN}\frac{du^2}{u^2}+\frac{u^2}{\sqrt{4\pi g_sN}}\eta_{\mu\nu}dx^{\mu}dx^{\nu}+\sqrt{4\pi g_sN}d\Omega_5^2 \right]
\ee
Note that the $AdS_5$ and the $S^5$ both have the same radius given by
\be
l^2=l_s^2\sqrt{4\pi g_sN}.
\ee
and that we still have integer 5-form flux across the $S^5$. Can we really trust this supergravity description? We can if the
curvature is small compared to the string scale. This means
\be \label{largeN}
l \gg l_s \qquad \Rightarrow \qquad g_sN \gg 1
\ee
For classical supergravity we also wish to suppress stringy loop corrections
so we assume $g_s <1$. This means we really need $N \gg 1$.

The preliminary conjecture is that IIB supergravity on $AdS_5 \times
S^5$ describes the same physics as a large $N$ Yang Mills
theory. However, when $N$ is not large, we can no longer trust the
supergravity description and need to go to the full string theory.
We now formally state the Maldacena conjecture:

\begin{quote}
{\it The following theories are equivalent 
\begin{itemize}
\item Type IIB string theory on $AdS_5 \times S^5$ where both the
$AdS_5$ and the $S_5$ have the same radius, and the 5-form has integer
flux, $N$, across the $S^5$.
\item $\mathcal{N}=4$ super Yang Mills on 3+1 dimensional Minkowski
space, with gauge group $SU(N)$.
\end{itemize}}
\end{quote}

Naturally, if we are to make sense of the correspondence we ought to
provide a dictionary that translates the gauge theory language into
the gravity language, and vice-versa. Two important entries are
\bea
g_{YM}^2 &=& g_s \\
\left(\frac{l}{l_s}\right)^4 &=& 4 \pi g_{YM}^2N
\eea
where $g_{YM}$ is the 
Yang Mills coupling constant. The quantity $g_{YM}^2N$
is known as the 't Hooft coupling. This is the natural loop counting
parameter and we note from (\ref{largeN}) that it should be large. It was 't Hooft who initiated the study of large $N$ gauge
theories in an attempt to understand their behaviour at strong coupling~\cite{'tHooft:planar}.

The boundary of $AdS_5 \times S^5$ is given by Minkowski space in 3+1
dimensions, and is invariant under conformal transformations of the
metric.   $\mathcal{N}=4$ super Yang Mills  is also
conformally invariant and we think of it
as living on this boundary. It is decoupled from gravity in
the bulk. This means that the
correspondence is indeed holographic, as all the degrees of freedom
of the bulk gravity theory are projected on to the boundary.

As we will see later on, from the point of view of braneworlds, the
most important feature of the AdS/CFT correspondence is {\it the UV/IR
connection}~\cite{Susskind:uv/ir}. This states that the ultra-violet
degrees of freedom in the CFT correspond to the infra-red in the bulk
theory. How can we see this? Consider a string stretched from a D-brane
probe in the AdS bulk all the way to the boundary. From the CFT perspective the
string looks like a point charge. The mass of the string is
proportional to its proper length, which is divergent near the
boundary. On the CFT side this corresponds to the divergent
self-energy of the point charge. In order to regularize the divergence
in the bulk the string is only allowed to approach to within some
finite distance of the boundary. This is a long distance, or infra-red
cut-off in the length of the string. In the
CFT, this turns out to be equivalent to cutting out a shell of small radius around the point
charge. This time we have a short distance, or ultra-violet cut-off. 
\subsection{AdS-Schwarzschild/Finite temperature CFT} \label{sec:SAdS/CFT}
We will now change the picture slightly by relaxing the condition that
the D branes should be extremal. Instead we will consider {\it near
extremal} D branes. The supergravity solution for a non-extremal
black brane is given by
\be \label{eqn:nonextDbranes}
ds^2_{10}=H^{-1/2}(r)\left[ -f(r)dt^2+\delta_{ij}dx^idx^j\right] +H^{1/2}(r)\left[ f^{-1}(r)dr^2+r^2d\Omega_5^2 \right]
\ee
where 
\be
H(r)=1+\left(\frac{l}{r} \right)^4, \qquad f(r)=1-\left(\frac{r_0}{r}
\right)^4
\ee
and the constants $r_0$ and $l$ are related to the overall brane tension
 and Ramond-Ramond charge. Again the brane is located at $r=0$,
 although this time it is hidden behind a horizon at $r=r_0$. For a
 near extremal brane we take $r_0 \ll l$ and the near
horizon limit corresponds to taking $r_0<r \ll l$. This gives
\be
ds_{10}^2=-h(r)dt^2+\frac{dr^2}{h(r)}+\left(\frac{r}{l}\right)^2\delta_{ij}dx^idx^j+l^2d\Omega_5^2
\ee
where 
\be
h(r)=\left(\frac{r}{l}\right)^2\left[1-\left(\frac{r_0}{r}\right)^4 \right]
\ee 
This is Schwarzschild-$AdS_5 \times S^5$. When we rotate to Euclidean
signature we get a conical singularity at the horizon, $r_0$. In order to avoid this we cut the spacetime off at the
horizon, and identify time $t$ with time $t+\beta$, where 
\be
\beta=\frac{4\pi}{h'(r_0)}=\frac{\pi l^2}{r_0}.
\ee
This black hole is at temperature $T=1/\beta$, and its
entropy is given by the area of the horizon
\be \label{eqn:AdSBHentropy}
S_{BH} =\frac{A}{4G}
\ee
where
\be
A=\left(\frac{r_0}{l}\right)^3V_3~.~\Omega_5 l^5 \sim V_3T^3
\ee
and $V_3=\int_{\mathbb{R}^3} dx^1dx^2dx^3$.

Staying in Euclidean signature, the boundary of this black hole
spacetime has topology $\mathbb{R}^3 \times S^1$, where the $S^1$ has
radius $\beta/2\pi$. A gauge theory living on
this boundary would be heated to a finite temperature, $T$, due to
Hawking radiation from the bulk black hole. A large $T$ we would
expect the entropy of the gauge theory to scale like the spatial
volume, ie $S \sim V_3$.  In the case of $\mathcal{N}=4$ super Yang
Mills we started out with a {\it conformal} field theory, and although
conformal invariance is broken at finite temperature,  the only scale
we have introduced in $T$. On dimensional grounds we conclude then
that
\be
S_{YM} \sim V_3T^3
\ee
which is in agreement with the black hole entropy (\ref{eqn:AdSBHentropy}).

Here we have only given an intuitive argument but more precise
calculations of the CFT entropy have been carried
out~\cite{Gubser:entropy1,Gubser:entropy2}. To sum up, we find that when we switch on a finite temperature, we can
associate the temperature, entropy and indeed the energy of the CFT with
the corresponding black hole quantities in the
bulk~\cite{Witten:thermal, Witten:adscft}. 

So vast is the subject, we have not been able to present the AdS/CFT
correspondence in all its
glory\footnote{See~\cite{D'Hoker:adscft} or ~\cite{Aharony:adscft} for a more extensive review.}. The hope is that we
now have a feeling for AdS/CFT and can embark on a
study of holography in the context of braneworlds.
\section{Braneworld holography}
We can think of the RS2 braneworld model as two identical copies of AdS space
patched together in such a way as to form a brane of given
tension. Consider one of these copies of AdS. Notice that we have cut
the spacetime off before reaching the AdS boundary. From the point of
view of AdS/CFT this corresponds to a long distance, or infra-red
cut-off in the bulk. We learnt from the UV/IR connection that an
infra-red cut-off in the bulk corresponds to an ultra-violet cut-off
in the CFT. Therefore when studying braneworlds we might expect some
version of AdS/CFT whereby the gravity theory in the bulk is dual to a
CFT with a UV cut-off~\cite{Gubser:gravity}.

At this point we note that our language is rather misleading. The
notion of a {\it conformal} field theory with a momentum cut-off is
paradoxical. What we really have is a {\it broken} conformal field
theory. By chopping off part of AdS near the boundary, we broke
translational invariance in the ``radial'' direction. Since we have
introduced a scale, this corresponds
to breaking conformal invariance in the dual field theory.

There is, however, a twist in the tale. Recall that in the traditional
picture of AdS/CFT, the CFT on the boundary is decoupled from gravity
in the bulk. We can understand this in the following way. Consider a
bulk graviton propagating towards the boundary. It cannot reach the
boundary because the background AdS metric blows up there. Gravity is therefore decoupled from the boundary theory.

The situation for braneworlds, meanwhile, is slightly
different. The metric at the brane does not blow up. This time, the bulk
graviton {\it can} reach the brane, and gravity is coupled to the
field theory there.

Braneworld holography can be summed up in the following statement:
\begin{quote}
{\it Randall-Sundrum braneworld gravity is dual to a CFT with a UV
cut-off, coupled to gravity on the brane}. 
\end{quote}
This is nothing more than a conjecture, and is far from proven. One of the difficulties in studying
this type of holography is our lack of knowledge regarding the dual
field theory. It is some abstract field theory which we know very
little about. 

However, consider what happens when we switch on a finite
temperature. We have seen how this corresponds to creating a black
hole in the bulk, where we now have a non-zero Weyl tensor. Casting our
mind back to the Einstein equations on the brane
(\ref{eqn:Etensor3}), we see that the presence of the bulk Weyl tensor
affects the geometry on the brane. The hope is that we can understand
this effect from a holographic perspective. Intuitively we might think
that Hawking radiation from the bulk black hole heats up the brane,
giving energy to the dual field theory. We can then examine how this
energy enters (say) the cosmological equations on the brane, and
compare this to what happens when there is no bulk black hole and we
put mass on the brane by hand. If we find the same behaviour we have
evidence for braneworld holography. The remainder of this thesis will
be devoted to this problem.
\section{CFTs on critical branes}\label{sec:critical}
Consider two $n$-dimensional spacetimes with negative cosmological
constant
\be
\Lambda_n=-\frac{1}{2}(n-1)(n-2)k_n^2
\ee
and glue them together across an $(n-1)$-dimensional brane of tension
$\sigma$.  We saw in section~\ref{section:bulkbranecos} that a
generalised Birkhoff's theorem admits the following solution for the
bulk metric
\be \label{bulk metric}
ds^2_n=-h(Z)dt^2+\frac{dZ^2}{h(Z)}+Z^2d\Omega_{n-2}^2  
\ee
where 
\be
h(Z)=k_n^2Z^2+1-\frac{c}{Z^{n-3}}.
\ee
Here we have taken the $\kappa=1$ slicing, with $d \Omega_{n-2}^2$ giving the
metric on a unit $(n-2)$-sphere. Recall that $c=0$ corresponds to pure AdS in the bulk, whereas $c>0$
corresponds to AdS-Schwarzschild. We wish to see the effect when there
is a non-vanishing Weyl tensor, so we will consider the latter. 

As in section~\ref{section:bulkbranecos}, we  parametrise the brane using
the affine parameter $\tau$.  The brane is then given by the section
$({\bf x}^{\mu}, t(\tau), Z(\tau))$ of the bulk metric. Since
$\tau$ corresponds to the proper time of an observer comoving with the
brane, we have the condition
\be
-h\dot t^2+\frac{\dot Z^2}{h}=-1
\ee
where dot denotes $\partial /\partial \tau$. This
condition ensures that the induced metric on the brane takes the standard
FRW form (\ref{eqn:FRWbranemetric}). Again, we treat $Z(\tau)$ as the
scale factor of our brane universe, and construct the Hubble parameter
$H=\dot Z/Z$.

Now suppose that we have a {\it critical} brane, ie 
\be
\sigma_n=\frac{4 \pi G_n \sigma}{n-2}=k_n
\ee
so that the induced cosmological constant, $\Lambda_{n-1}=0$. We
further assume that there is no additional matter on the brane so that
the brane energy-momentum consists only of brane tension. We can read
off the cosmological evolution equations from equations
(\ref{eqn:tdot}) to (\ref{eqn:evolSADS2}) by
setting $\kappa=1$, $a=\sigma_n^2-k_n^2=0$ and $\rho=p=0$. The brane
evolution is therefore given by
\bea 
\dot t &=& \frac{k_n Z}{h} \label{eqn:tdotcrtical} \\
H^2 &=&-\frac{1}{Z^2} +\frac{c}{Z^{n-1}} \label{eqn:evolSADS1critical}
\\
\dot H &=&
\frac{1}{Z^2}-\left(\frac{n-1}{2}\right)\frac{c}{Z^{n-1}} \label{eqn:evolSADS2critical}
\eea
This cosmology is very similar to the standard $\kappa=1$ cosmology
of closed FRW universes. We start off with a Big Bang at $Z=0$ and
experience a period of cosmological expansion, crossing the black hole
horizon\footnote{Recall that the horizon of the bulk black holes is
given by $Z=Z_H$, where $h(Z_H)=0$.}. Eventually, the rate of
expansion slows down and we reach a maximum value of $Z$. After this,
the brane starts to contract until Armageddon, when we disappear
with a Big Crunch. The shape of the brane trajectory is shown in
figure~\ref{fig:critcosmo}.
\begin{figure}[ht]
\centering
\includegraphics[width=10cm, angle=-90]{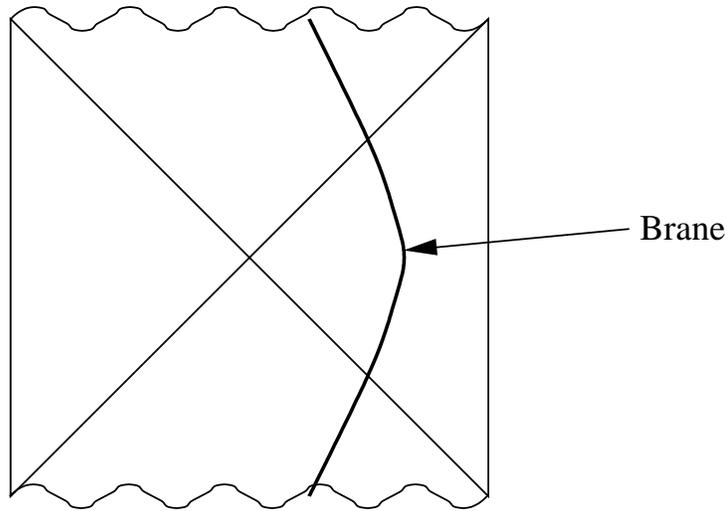}
\vskip 5mm
\caption{The Penrose diagram showing the trajectory of a critical
brane in an AdS-Schwarzschild bulk. We have two copies of the bulk
glued together at the brane. We have only shown one of those copies here.} \label{fig:critcosmo}
\end{figure}

It is clear from equations (\ref{eqn:evolSADS1critical}) and
(\ref{eqn:evolSADS2critical}) that the brane cosmology is driven by
the terms like $c/Z^{n-1}$. These come from the mass of the bulk black
holes. How should we understand them from a braneworld perspective?
Motivated by braneworld holography, we might expect them to correspond to
the energy density and pressure of a dual field theory. Given that they
go like $1/Z^{n-1}$, this field theory will probably look like
radiation. However, the conformal nature of radiation suggests that
this will only be the case when there is only a small UV cut-off, and the brane is near the AdS boundary.
\subsection{Energy density and pressure of the dual CFT} 
We will now attempt to calculate the energy density/pressure of the
dual field theory, at least when the brane is near the boundary
of AdS~\cite{Savonije:braneCFT}. We can think of these as being the energy density/pressure
measured by a braneworld observer. 

If we use the bulk time, $t$ as our time coordinate, we would measure the
total energy to be given by the sum of the black hole masses, that is
\be
E=2M
\ee
where the mass of an AdS black hole is given by the standard formula~\cite{Hawking:AdSblackholes}:
\be \label{BH mass}
M= \frac{(n-2)\Omega_{n-2}c}{16 \pi G_n} 
\ee
and $\Omega_{n-2}$ is the volume of the unit $(n-2)$-sphere. However, an
observer on the brane uses the CFT time, $\tau$ as his time coordinate, and will
therefore measure the energy differently. To arrive at the CFT energy,
$E_{CFT}$, we need to scale $E$ by
$\dot t$. We have assumed we are near the AdS boundary. This means
that $Z$ is large and we can say that $\dot t \approx 1/k_n
Z$. The CFT energy is therefore given by
\be
E_{CFT}=E\dot{t} \approx \frac{(n-2)\Omega_{n-2}c}{8\pi G_n}\left(\frac{1}{k_nZ}\right)
\ee
In order to calculate the energy {\it density} we must first evaluate
the spatial volume of the CFT, which is just the spatial volume of the brane,
\be
V_{CFT}=\Omega_{n-2}Z^{n-2}
\ee
The CFT energy density is given by the ratio of energy to
volume\footnote{Since we are concerned with this ratio, the extension
of these ideas to $\kappa \neq 1$ is presumably trivial.}, 
\be \label{eqn:rhocft}
\rho_{CFT}=\frac{E_{CFT}}{V_{CFT}} \approx \frac{(n-2)}{8 \pi G_n k_n}\left( \frac{c}{Z^{n-1}} \right)
\ee
To calculate the pressure of the CFT, we use the standard formula
from thermodynamics\footnote{This expression is easily derived from $p=-\partial
E/\partial V$, using $E=\rho V$ and $V \sim Z^{n-2}$.}
\be \label{eqn:TD}
p_{CFT}=-\left(\frac{Z}{n-2}\right)\frac{\partial \rho_{CFT}}{\partial
Z}-\rho_{CFT}
\ee
Using the expression (\ref{eqn:rhocft}) in (\ref{eqn:TD}) we see that the equation of
state for the CFT is indeed that of radiation.
\be \label{eqn:pcft}
p_{CFT} \approx \frac{1}{8 \pi G_n k_n}\left(\frac{c}{Z^{n-1}}\right) \approx \frac{\rho_{CFT}}{n-2}
\ee
\subsection{The cosmological evolution equations}  \label{sec:critcosmo}
Now that we know the CFT energy density and pressure in terms of $c$,
we can substitute back into (\ref{eqn:evolSADS1critical}) and
(\ref{eqn:evolSADS2critical}) and examine the brane
cosmology in terms of CFT quantities. Before we do this, we
recall that for a critical brane, the induced Newton's constant
is given by
\be
G_{n-1}=\frac{G_nk_n(n-3)}{2}
\ee
We now obtain a more useful expression for $\rho_{CFT}$,
\be
\rho_{CFT} \approx \frac{(n-2)(n-3)}{16 \pi G_{n-1}}\left(\frac{c}{Z^{n-1}}\right).
\ee
Substituting this and equation (\ref{eqn:pcft}) into
(\ref{eqn:evolSADS1critical}) and (\ref{eqn:evolSADS2critical}) gives
the cosmological evolution equations for the brane.
\bea 
H^2 &=& -\frac{1}{Z^2}+\frac{16 \pi G_{n-1}}{(n-2)(n-3)}\rho_{CFT} \label{evolution1crit} \\ 
\dot{H} &=& \frac{1}{Z^2}- \frac{8 \pi G_{n-1}}{(n-3)}(\rho_{CFT}+p_{CFT})\label{evolution2crit}
\eea
These are the standard FRW equations in $(n-1)$ dimensions.  They
correspond to a spatially spherical universe, with no cosmological
constant. The braneworld observer sees the normal cosmological
expansion driven by the dual CFT. The CFT behaves like radiation in
this instance. 

We should emphasise that we now have two very different ways of
interpreting this cosmology. On the ``gravity'' side, we think of the
cosmological expansion/contraction as being driven by the bulk black
holes. On the ``field theory'' side we think of it being driven by the
dual CFT, in the standard way. 
\section{CFTs on non-critical branes} \label{sec:CFTonnoncrit}
We will now attempt to generalise the above analysis to de Sitter and
anti-de Sitter branes. This corresponds to relaxing the criticality
condition so that
\be
\sigma_n \neq k_n.
\ee
We proceed exactly as before, except this time we allow for
$a=\sigma_n^2-k_n^2 \neq 0$.  Equations (\ref{eqn:tdotcrtical}) to (\ref{eqn:evolSADS2critical})
generalise to
\bea 
\dot t &=& \frac{\sigma_n Z}{h} \label{eqn:tdotnoncrit} \\
H^2 &=& a-\frac{1}{Z^2} +\frac{c}{Z^{n-1}} \label{eqn:evolSADS1noncrit}
\\
\dot H &=&
\frac{1}{Z^2}-\left(\frac{n-1}{2}\right)\frac{c}{Z^{n-1}}.
\label{eqn:evolSADS2noncrit}
\eea
For $a<0$, we have subcritical branes, which are asymptotically anti-de
Sitter.  In this case the brane evolves in much the same way as for
critical branes. We start off with a Big Bang and expand to some maximum
value of $Z$, and then contract back to the Big Crunch. As before, the brane
crosses black hole horizon. The Penrose diagram for this trajectory is
more or less the same as the critical brane trajectory given in figure~\ref{fig:critcosmo}.

For $a>0$, we have supercritical branes, which are asymptotically de
Sitter. This time there are four different possible trajectories for
the brane depending on the various parameters. This is summarised in
the following table,

\begin{table}[ht]
\begin{tabular}{p{1cm}|p{7cm}|p{4cm}}
Case & Trajectory & Conditions \\
\hline
a & $Z$ runs from $0$ to $\infty$. & $a \geq a_{crit}$, $Z$ starts out
small. \\ \hline
b & $Z$ runs from $\infty$ to $0$. & $a \geq a_{crit}$, $Z$ starts out
large. \\ \hline
c & $Z$ runs from $0$ up to a maximum, and then down to $0$. & $a
\leq a_{crit}$, $Z$ starts out small. \\ \hline
d & $Z$ runs from $\infty$ down to a minimum, and then up to
$\infty$. & $a
\leq a_{crit}$, $Z$ starts out large.
\end{tabular}
\end{table}
where
\be
a_{crit}=\left(\frac{n-3}{n-1}\right)\left(\frac{2}{(n-1)c}
\right)^{\frac{2}{n-3}}
\ee

For cases (a) to (c) the brane crosses the black hole horizon. Case
(d) is sometimes known as the ``bounce'' solution, and in this case
the brane does not cross the horizon. Each of these possible trajectories are shown in
figures~\ref{fig:dscosmo:subfig:a} to~\ref{fig:dscosmo:subfig:d}.
\begin{figure}
\centering
\subfigure[$Z$ starts out
small, ~~$a \geq a_{crit}$.]{
    \label{fig:dscosmo:subfig:a}
\begin{minipage}[b]{0.3\textwidth}
   \centering
   \includegraphics[width=6cm, angle=-90]{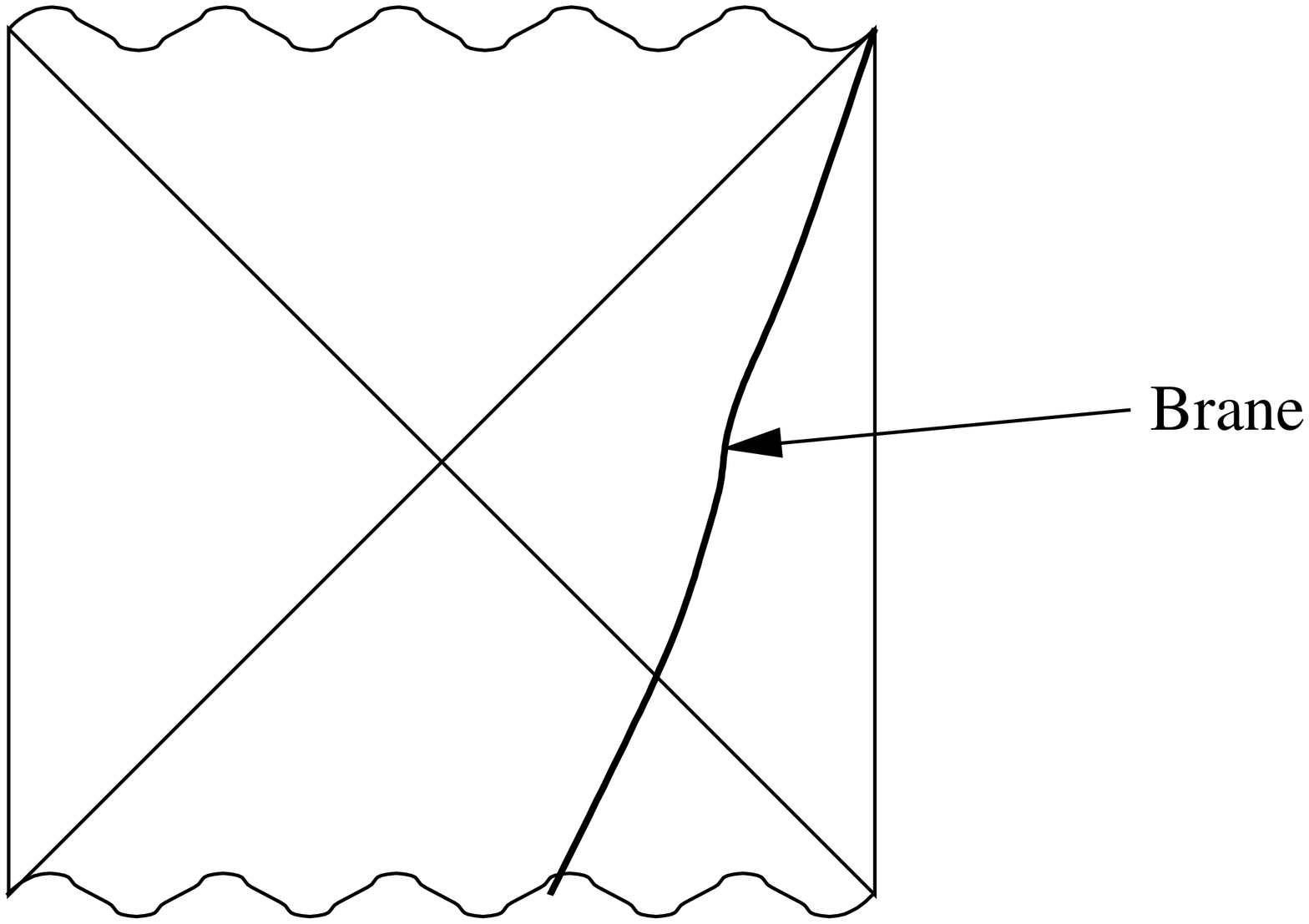}
\end{minipage}}%
\hspace{0.2\textwidth}%
\subfigure[$Z$ starts out large, ~~$a \geq a_{crit}$.]{
   \label{fig:dscosmo:subfig:b}
\begin{minipage}[b]{0.3\textwidth}
  \centering
    \includegraphics[width=6cm, angle=-90]{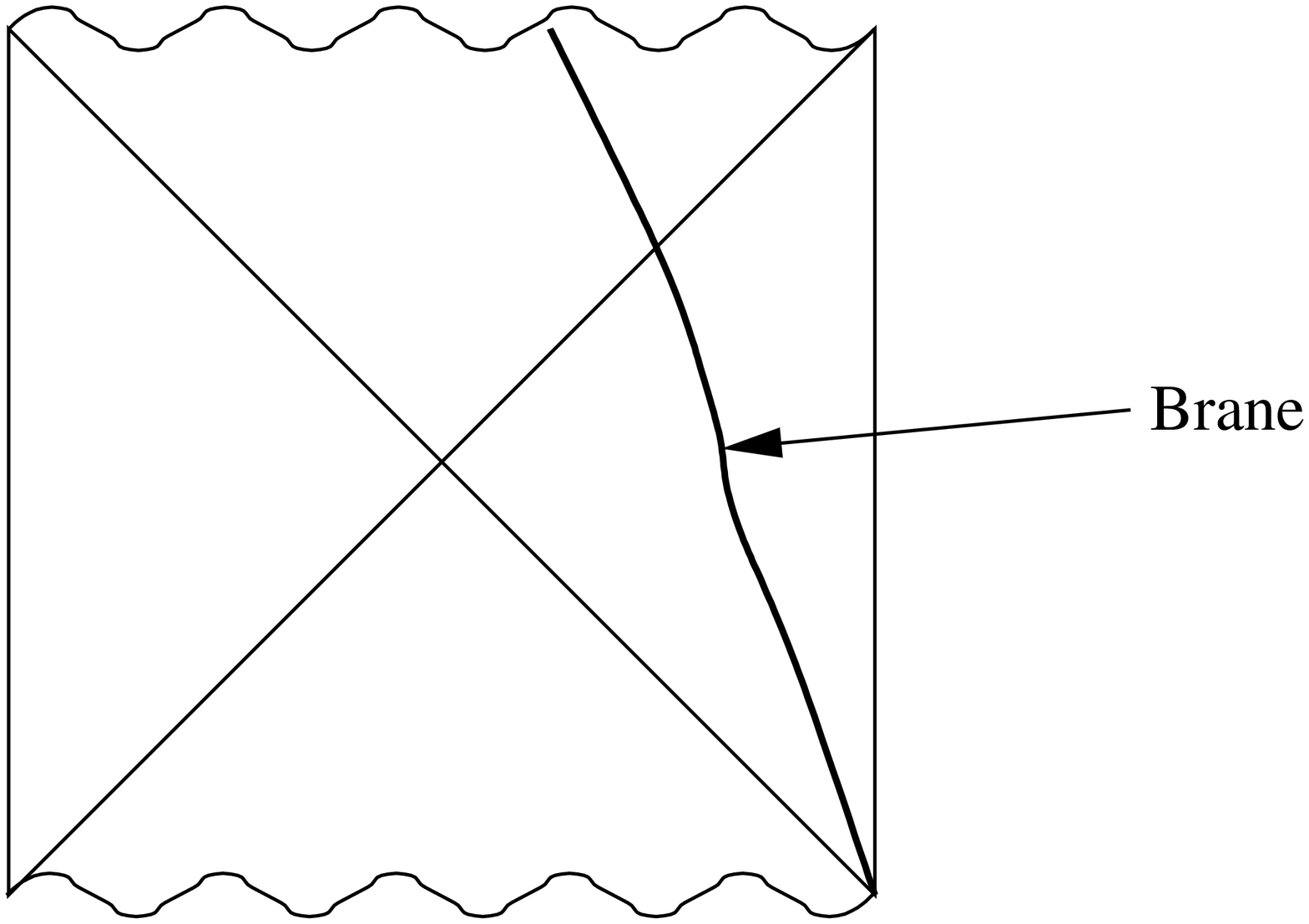} 
\end{minipage}} \\[20pt]
\subfigure[$Z$ starts out
small, ~~$a \leq a_{crit}$.]{
    \label{fig:dscosmo:subfig:c}
\begin{minipage}[b]{0.3\textwidth}
   \centering
     \includegraphics[width=6cm, angle=-90]{critcosmo.ps}
\end{minipage}}%
\hspace{0.2\textwidth}%
\subfigure[$Z$ starts out large, ~~$a \leq a_{crit}$.]{
 \label{fig:dscosmo:subfig:d}
\begin{minipage}[b]{0.3\textwidth}
   \centering
     \includegraphics[width=6cm, angle=-90]{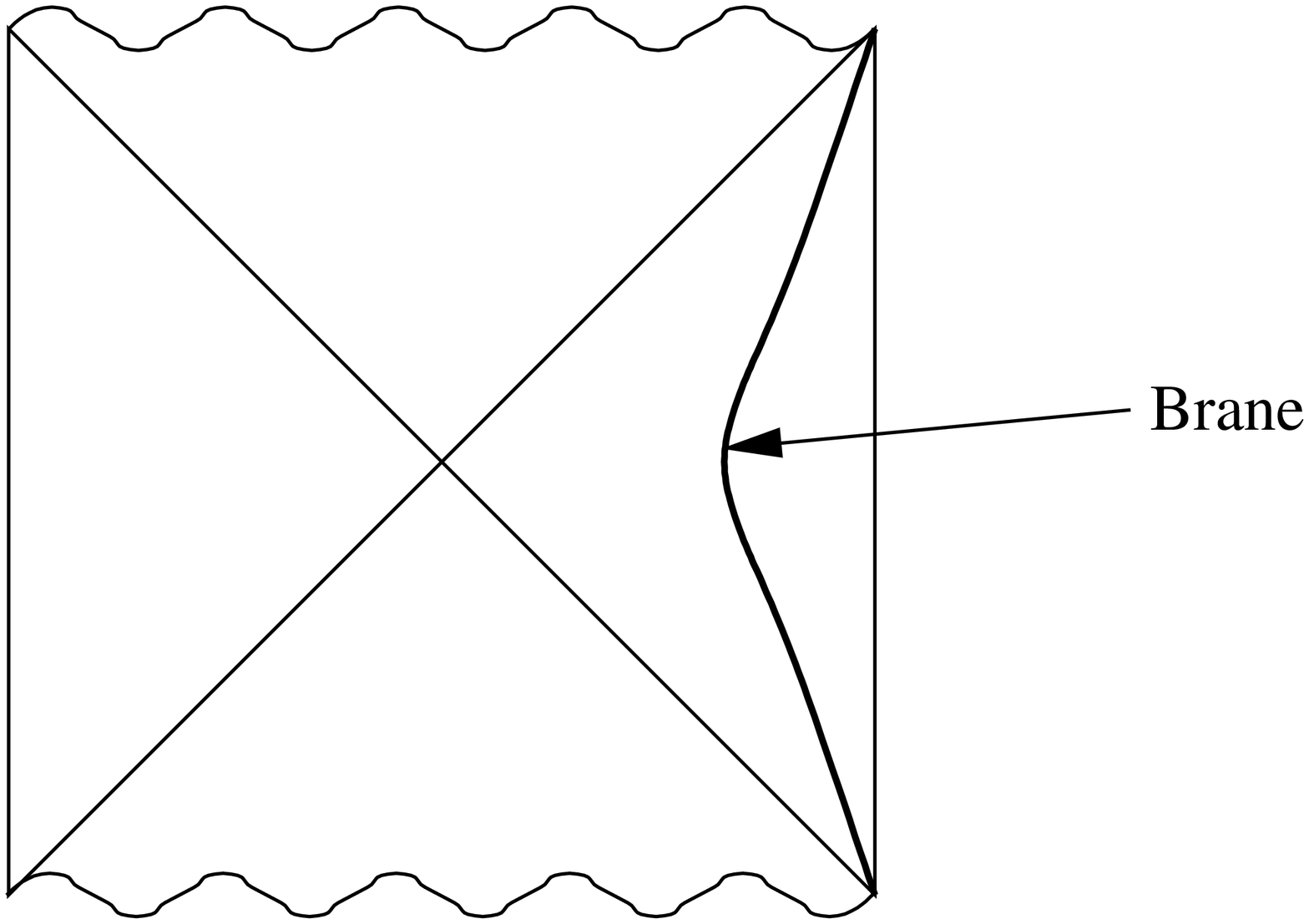}
\end{minipage}}
\caption{Penrose diagrams showing possible trajectories for
     supercritical (de Sitter) branes in an AdS-Schwarzschild bulk.} \label{fig:dscosmo:subfig}
\end{figure}

Notice that if $a=a_{crit}$, we can have
either cases (a) and (c), or (b) and (d), depending  on how $Z$
starts out. We can also have $Z=\textrm{const}$, although this solution is
presumably very unstable.

Once again, our goal is to understand the terms like $c/Z^{n-1}$ in
the evolution equations (\ref{eqn:evolSADS1noncrit}) and (\ref{eqn:evolSADS2noncrit}), from the point of view
of  AdS/CFT. Can we think of this cosmology as being driven by a dual
field theory? We will start by blindly adopting the approach
of~\cite{Savonije:braneCFT}, as described in the last section. We will run
into problems, but it is nevertheless illustrative to see how things
go wrong. We will then give a correct approach which agrees
with~\cite{Savonije:braneCFT}   for critical branes, but not for
non-critical branes.
\subsection{CFT energy density/pressure: naive approach} \label{sec:naive}
As in section~\ref{sec:critical}, we assume that the energy of bulk spacetime is given by
\be
E=2M
\ee
In order to calculate the energy of the CFT, we should once again
scale $E$ by $\dot t$ so that it is measured with respect to the CFT
time, $\tau$, rather than the bulk time, $t$.  However, for large $Z$, we have from equation (\ref{eqn:tdotnoncrit}),
\be
\dot t=\frac{\sigma_nZ}{k_n^2Z^2+1-\frac{c}{Z^{n-3}}} \approx  \frac{\sigma_n}{k_n^2Z}
\ee
The energy of the CFT is then
\be
E_{CFT} = E  \dot t \approx 2M \left(  \frac{\sigma_n}{k_n^2Z} \right)
\ee
Since the spatial volume of the CFT is just $V_{CFT}=\Omega_{n-2}Z^{n-2}$, we have the following expression for the energy density of the CFT.
\be
\rho_{CFT} \approx  \frac{2M}{\Omega_{n-2}Z^{n-2}} \left
(  \frac{\sigma_n}{k_n^2Z} \right) = \frac{(n-2)}{8 \pi G_n \sigma_n}
\left(\frac{c}{Z^{n-1}} \right) \left( \frac{\sigma_n^2}{k_n^2}
\right) \label{rhocftnaive}
\ee
where we have used equation (\ref{BH mass}). 

At this stage we note an important feature of non-critical branes: the induced
Newton's constant on the brane is proportional to the brane
tension. More precisely, from (\ref{eqn:NC}),
\be
G_{n-1}=\frac{G_n\sigma_n(n-3)}{2}
\ee
We can insert this back into (\ref{rhocftnaive})  to give
\be
\rho_{CFT} \approx \frac{(n-2)(n-3)}{16 \pi G_{n-1}}\left(\frac{c}{Z^{n-1}}\right)\left( \frac{\sigma_n^2}{k_n^2} \right)
\ee
Now consider what happens when we express the evolution equations in
terms of $\rho_{CFT}$. In particular, equation (\ref{eqn:evolSADS1noncrit}) now reads 
\be
H^2 = a-\frac{1}{Z^2}+\frac{16 \pi
G_{n-1}}{(n-2)(n-3)}\rho_{CFT}\left( \frac{k_n^2}{\sigma_n^2} \right) \label{naivecosmo}
\ee
Note that for critical branes the factor of $k_n^2/\sigma_n^2$ disappears and we recover the standard
Friedmann equation.  However, for non-critical branes,
$k_n^2/\sigma_n^2 \neq 1$. This means that equation (\ref{naivecosmo})
does not resemble the standard FRW cosmology. Either holography has
failed, or we have tackled the problem in the wrong way. We shall now
see that it is the latter.
\subsection{CFT energy density/pressure: better approach} \label{sec:better}
Unlike in flat space, when one derives the mass of an AdS black hole (\ref{BH mass}), the leading order contribution comes from the
bulk~\cite{Hawking:AdSblackholes}. Furthermore, this derivation
includes contributions from the AdS-Schwarzschild spacetime all the way
up to the AdS boundary. In our case, we have a brane that has cut off
our bulk spacetime before it was able to reach the boundary. We should
not therefore include contributions from ``beyond'' the brane and must
go back to first principles in order to calculate the energy of the
bulk~\cite{Padilla:CFT}.

We will begin by Wick rotating to Euclidean signature.
\be
t \to t_E=it, \qquad
\tau \to\tau_E=i\tau \nonumber
\ee
This analytic continuation is well defined for the subcritical brane,
critical brane
and for the supercritical brane, cases (c) and (d). For cases (a) and
(b) we find that $Z(\tau_E)$ is not a real function, so they are
excluded from this analysis. 

Our bulk metric is now given by
\be \label{Euclidean bulk metric}
ds^2_n=h(Z)dt_E^2+\frac{dZ^2}{h(Z)}+Z^2d\Omega^2_{n-2}  
\ee
As discussed in section~\ref{sec:SAdS/CFT}, we wish to avoid a conical
singularity at the horizon, $Z=Z_H$. In order to do this we cut the
spacetime off at the horizon and associate $t_E$ with $t_E+\beta$
where $\beta=4\pi/h^{\prime}(Z_H)$. The brane is now given by the
section $(\bf x \rm^{\mu}, t_E(\tau_E), Z(\tau_E))$ of the Euclidean
bulk. The new equations of motion of the brane are the following:
\bea \label{Euclidean brane EOM}
\frac{dt_E}{d\tau_E} &=& \frac{\sigma_n Z}{h}  \label{Euclidean brane
EOM3} \\
\left(\frac{dZ}{d\tau_E}\right)^2 &=& -aZ^2+1-\frac{c}{Z^{n-3}}   \label{Euclidean brane EOM1} \\
\frac{d^2Z}{d\tau_E^2} &=&-aZ +\left(\frac{n-3}{2}\right) \frac{c}{Z^{n-2}}  \label{Euclidean brane EOM2} 
\eea
It is not difficult to see that for both critical and non-critical
branes, $Z(\tau_E)$ has a minimum value. In contrast to Lorentzian
signature, in Euclidean signature none of these branes cross the black hole horizon. The supercritical branes have a maximum value of $Z$, whilst the critical and subcritical branes may stretch to the AdS boundary. This will not be a problem because the integrand in our overall action will remain finite, as we shall see. 

In calculating the energy we could go ahead and evaluate the Euclidean action of this solution and then differentiate with respect to $\beta$. We must however, remember to take off the contribution from a reference spacetime~\cite{Hawking:hamiltonian}. In this context, the most natural choice of the reference spacetime would be pure AdS cut off at a surface, $\Sigma$ whose geometry is the same as our braneworld.

The bulk metric of pure AdS is given by the following:
\be \label{Euclidean pure ADS metric}
ds^2_n= h_0(Z)dT^2+\frac{dZ^2}{h_0(Z)}+Z^2d\Omega^2_{n-2}
\ee
where
\be
h_0(Z)=k_n^2Z^2+1
\ee
As we said earlier, the cut-off surface, $\Sigma$,  should have the same geometry as our braneworld. The induced metric on this surface is therefore
\be \label{surface metric}
ds^2_{n-1}=d\tau_E^2+Z(\tau_E)^2d\Omega^2_{n-2}
\ee
To achieve this, we must regard our cut-off surface as a section  $(\bf x \rm^{\mu}, T(\tau_E), Z(\tau_E))$, where
\be \label{T condition}
h_0\left(\frac{dT}{d\tau_E}\right)^2+\frac{1}{h_0}\left(\frac{dZ}{d\tau_E}\right)^2=1
\ee
Let us now evaluate the difference $\Delta I$  between the Euclidean action of our AdS-Schwarzschild bulk, $I_{BH}$ and that of our reference background, $I_{AdS}$.
\bea
I_{BH} &=& -\frac{1}{16 \pi G_n}\int_{bulk} d^nx \sqrt{g}(R-2\Lambda_n) - \frac{1}{8 \pi G_n} \int_{brane} d^{n-1}x \sqrt{h} \ 2K \label{BH action} \\
I_{AdS} &=& -\frac{1}{16 \pi G_n}\int_{ref. \ bulk} d^nx \sqrt{g}(R-2\Lambda_n) - \frac{1}{8 \pi G_n} \int_{\Sigma} d^{n-1}x \sqrt{h} \ 2K_0 \label{reference action}
\eea
where $K$ and $K_0$ are the trace of the extrinsic curvature of the
brane and $\Sigma$ respectively. Now recall that we have the Einstein equations in the
bulk
\be 
R_{ab} - \half R g_{ab}=-\Lambda_ng_{ab}
\ee
and the ($\mathbb{Z}_2$-symmetric) Israel equations across the brane
\be
K_{ab}= \sigma_n h_{ab}.
\ee
Given that $\Lambda_n=-\frac{1}{2}(n-1)(n-2)k_n^2$, we can immediately read off the following:
\bea
R-2\Lambda_n &=& -2(n-1)k_n^2  \label{ricci}\\
2K &=& 2(n-1) \sigma_n \label{extrinsic}
\eea
The unit normal to the cut-off surface, $\Sigma$ is given by $n_a=(\bf
0 \rm, -\frac{dZ}{d\tau_E},  \frac{dT}{d\tau_E})$. We use this to find
\be
2K_0=(n-1)\frac{2\sigma_n^2Z(\tau_E)+cZ(\tau_E)^{2-n}}{h_0 \frac{dT}{d\tau_E}}.
\ee
We will also need the correct form of the measures and the limits in each case. If we say that $-\frac{\beta}{2} \leq t_E \leq \frac{\beta}{2}$, then we obtain the following (see appendix~\ref{app:measures} for a detailed derivation):
\begin{align}
\int_{bulk} d^nx \sqrt{g} \ (R-2\Lambda_n) &= 2\Omega_{n-2}\int_{-\frac{\beta}{2}}^{\frac{\beta}{2}} dt_E \frac{Z(\tau_E)^{n-1}-Z_H^{n-1}}{n-1}  \ (R-2\Lambda_n) \label{action1} \\
\int_{ref. \ bulk} d^nx \sqrt{g}  \ (R-2\Lambda_n) &= 2\Omega_{n-2}\int_{-\frac{\beta}{2}}^{\frac{\beta}{2}} dt_E \left(\frac{ \frac{dT}{d\tau_E}}{ \frac{dt_E}{d\tau_E}}\right)\frac{Z(\tau_E)^{n-1}}{n-1}  \ (R-2\Lambda_n)  \label{action2} \\
\int_{brane} d^{n-1}\sqrt{h} \ 2K &= \Omega_{n-2} \int_{-\frac{\beta}{2}}^{\frac{\beta}{2}} dt_E  \left(\frac{ 1}{ \frac{dt_E}{d\tau_E}}\right)Z(\tau_E)^{n-2} \ 2K \label{action3} \\
\int_{\Sigma} d^{n-1}\sqrt{h} \ 2K_0 &= \Omega_{n-2} \int_{-\frac{\beta}{2}}^{\frac{\beta}{2}} dt_E  \left(\frac{ 1}{ \frac{dt_E}{d\tau_E}}\right)Z(\tau_E)^{n-2} \ 2K_0 \label{action4}
\end{align}
The factor of two in equations (\ref{action1}) and (\ref{action2}) just comes from the fact that we have two copies of the bulk spacetime in each case. Notice that the expressions for the integrals over the brane and the cut-off surface $\Sigma$ are the same. This is a consequence of the two surfaces having the same geometry. 
Also using equations (\ref{Euclidean brane EOM3}) and (\ref{T condition}), we put everything together and arrive at the following expression for the difference in the Euclidean action:

\bea
\Delta I &=& \frac{\Omega_{n-2}k_n^2}{4 \pi G_n}  \int_{-\frac{\beta}{2}}^{\frac{\beta}{2}} dt_E Z^{n-1}\left[1-\left(1+\frac{cZ^{1-n}}{\sigma_n^2}\right)^{\half}\left(1-\frac{cZ^{1-n}}{k_n^2}\left(1+\frac{1}{k_n^2Z^2}\right)^{-1}\right) \right] \nonumber \\
& & - \frac{\Omega_{n-2}}{4 \pi G_n} \int_{-\frac{\beta}{2}}^{\frac{\beta}{2}} dt_E \ (n-1)h(Z)Z^{n-3} \left[1-\half \left(1+\frac{cZ^{1-n}}{\sigma_n^2}\right)^{\half}-\half \left(1+\frac{cZ^{1-n}}{\sigma_n^2}\right)^{-\half} \right] \nonumber \\
& & - \frac{\Omega_{n-2}k_n^2}{4 \pi G_n}\beta Z_H^{n-1}
\eea

To proceed further, we are going to have to make things a little bit
simpler. In the spirit of AdS/CFT, we want the brane to be close to
the AdS boundary. This corresponds to taking $c$ to be large, so our bulk is at a very high temperature. By considering this regime we guarantee that we focus on the ``holographic'' energy density, and can ignore contributions from matter on the brane. We have not included any such contributions in our analysis so it is appropriate for us to assume that we are indeed working at large $c$.
To leading order:
\bea
Z_H &\approx& \left(\frac{c}{k_n^2}\right)^{\frac{1}{n-1}} \\
\beta &\approx& \frac{4\pi}{(n-1)k_n^2} \left(\frac{k_n^2}{c}\right)^{\frac{1}{n-1}}
\eea
For supercritical and critical branes we can assume $Z(\tau_E) \gg c^{\frac{1}{n-1}}$. For subcritical branes this is true provided $|a| \ll 1$ (see appendix~\ref{app:largec}). Given this scenario, we now evaluate $\Delta I$ to leading order in $c$.
\be
\Delta I = -\frac{\Omega_{n-2}c\beta}{4 \pi G_n}+\frac{\Omega_{n-2}k_n^2c}{4 \pi G_n}  \int_{-\frac{\beta}{2}}^{\frac{\beta}{2}} dt_E \left(\frac{1}{k_n^2}-\frac{1}{2\sigma_n^2}\right) + \ldots = -\frac{\Omega_{n-2}c\beta}{8 \pi G_n}\left(\frac{k_n^2}{\sigma_n^2}\right)+\ldots
\ee
The entire leading order contribution comes from the bulk rather than the brane, which is consistent with~\cite{Hawking:AdSblackholes}. We can now determine the energy of our bulk spacetime.
\be
E= \frac{d \Delta I} {d \beta} \approx \frac{(n-2)\Omega_{n-2}c}{8\pi G_n}\left(\frac{k_n^2}{\sigma_n^2}\right)
\ee
Notice that in this large $c$ limit, $E \approx 2M\left(\frac{k_n^2}{\sigma_n^2}\right)$, so for critical branes the choice $E=2M$ would indeed have worked.
Our aim was to calculate the energy of the dual CFT, rather than the
bulk AdS-Schwarzschild. We must therefore scale $E$, by $\dot{t}$, so
that it is measured with respect to the CFT time $\tau$. Recall that when $Z$ is large,  $\dot{t} \approx \sigma_n/k_n^2Z$ and the energy of the CFT is given by:
\be
E_{CFT}=E\dot{t} \approx \frac{(n-2)\Omega_{n-2}c}{8\pi G_n}\left(\frac{k_n^2}{\sigma_n^2}\right)\left(\frac{\sigma_n}{k_n^2Z}\right)=\frac{(n-2)\Omega_{n-2}c}{8\pi G_n}\left(\frac{1}{\sigma_nZ}\right)
\ee
We divide this by the spatial volume of the CFT,
$V_{CFT}=\Omega_{n-2}Z^{n-2}$ to find the CFT energy density.
\be
\rho_{CFT}=\frac{E_{CFT}}{V_{CFT}} \approx \frac{(n-2)}{8 \pi G_n \sigma_n}\left( \frac{c}{Z^{n-1}} \right)
\ee
To calculate the pressure of the CFT, we just use equation
(\ref{eqn:TD}). This yields an expression that is consistent with the CFT corresponding to radiation:
\be \label{pressure}
p_{CFT} \approx \frac{1}{8 \pi G_n \sigma_n}\left(\frac{c}{Z^{n-1}}\right) \approx \frac{\rho_{CFT}}{n-2}
\ee
\subsection{The cosmological evolution equations}
As before, we want to understand the cosmological  equations (\ref{eqn:evolSADS1noncrit}) and
(\ref{eqn:evolSADS2noncrit}) for the brane in terms of braneworld quantities only. This
means making use of the correct expression for the induced Newton's constant
(\ref{eqn:NC}). The CFT energy density is now given by
\be
\rho_{CFT} \approx \frac{(n-2)(n-3)}{16 \pi
G_{n-1}}\left(\frac{c}{Z^{n-1}}\right).
\ee
We are now ready to insert this and equation (\ref{pressure}) into equations (\ref{eqn:evolSADS1noncrit}) and (\ref{eqn:evolSADS2noncrit}) to derive the cosmological evolution equations for  our braneworld.
\bea 
H^2 &=& a-\frac{1}{Z^2}+\frac{16 \pi G_{n-1}}{(n-2)(n-3)}\rho_{CFT} \label{evolution1} \\ 
\dot{H} &=& \frac{1}{Z^2}- \frac{8 \pi G_{n-1}}{(n-3)}(\rho_{CFT}+p_{CFT})\label{evolution2}
\eea
As in section~\ref{sec:critcosmo}, these are the standard FRW
equations in $(n-1)$ dimensions, although this time we have a
cosmological constant term $a$. As was the case for flat branes, we can think
of the cosmology as being driven by a dual CFT corresponding to
radiation. Alternatively, from a ``gravity'' perspective, the  brane
cosmology is driven by the bulk black holes.

The important thing about this analysis was that went beyond the work
of~\cite{Savonije:braneCFT}, which concentrated only on
flat braneworlds. Recent observations that we may live in a universe
with a small positive cosmological constant~\cite{Perlmutter:astro,
Riess:astro} suggest that it is important that we extend the
discussion at least to de Sitter braneworlds. We have considered de
Sitter branes satisfying $a \leq a_{crit}$. In the large $c$
limit, $a_{crit} \ll 1$, so we actually have $a \ll 1$. Our analysis
also applies to anti-de Sitter branes satisfying $|a| \ll 1$.

Given the mounting evidence for holography in the literature, we are not really surprised by our result. What is interesting is the way in which we were forced to prove it. The proof offered by~\cite{Wang:FRW} is unacceptable because it relies on the assumption that:
\be
G_{n-1}=\frac{G_nk_n(n-3)}{2}
\ee
This is true for critical branes, but one should replace $k_n$ in the
above expression with $\sigma_n$ when one considers non-critical
branes. We also see in section~\ref{sec:naive} that if we had applied the  approach of~\cite{Savonije:braneCFT} to non-critical branes, a factor of $k_n^2/\sigma_n^2$ would have appeared in front of the CFT terms in equations (\ref{evolution1}) and (\ref{evolution2}). This comes from assuming that the bulk energy is just given by the sum of the black hole masses. As we stated in section 3, this involves an over-counting because it includes energy contributions from ``beyond'' the brane. The correct calculation of the bulk energy given in this paper ensures that the undesirable factor of  $k_n^2/\sigma_n^2$ does not appear.

Finally, we end with a note of caution. In the spirit of AdS/CFT we
have consistently assumed large $Z$, and for various reasons, large
$c$. This means that our results are only
approximate. We suspect that one could find corrections to higher
orders in $1/Z$ and $1/c$. Clearly we should be more
careful, and seek an alternative approach that gives us {\it exact}
results, even at finite values of $Z$ and $c$. Furthermore, because of
the limitations imposed by Wick rotation, we were not able to say
anything about cases (a) and (b) for supercritical branes. In the next
chapter we will adopt a new approach to braneworld holography that
does not suffer from any of these limitations or approximations.

\chapter{Exact braneworld holography} \label{chap:exact}
\setcounter{equation}{0}
\section{Introduction}
In the last section, we tried to interpret the Weyl tensor
contribution to the Einstein equation induced on a brane. Specifically,
we embedded the brane in a AdS-Schwarzschild spacetime so that the
non-trivial Weyl tensor manifested itself as a ``radiation'' term in
the FRW equations for the brane universe. Using the ideas of AdS/CFT,
we could interpret this term in two ways: (i) it came from the mass of
the bulk black holes or (ii) it came from the energy-momentum tensor of
some dual conformal field theory.

However, our analysis was based on the assumption that the brane
probed deep into AdS, near to the boundary. This allowed us to assume that the
energy density of the braneworld  was small, and the true
holographic description of an $(n-1)$ dimensional braneworld in an $n$
dimensional bulk was understood.  Unfortunately, these results were all
approximations in the sense that for a general brane evolution it is
not necessary for the brane to remain close to the boundary.  In this
chapter, we will undertake a new study
in which we calculate the energy of the  field theory on the brane
exactly, regardless of the brane's position in the
bulk~\cite{Padilla:exact}. 

In order to emphasize the full generality of these results, we will
allow the bulk black holes to couple to an electromagnetic field. We
are therefore generalising from AdS-Schwarzschild in the bulk, to
Reissner-Nordstr\"om AdS. We will also allow
the brane tension to be arbitrary, thereby including both critical and
non-critical branes. 
\section{Branes in a charge black hole background}
Consider an $(n-1)$ dimensional brane of tension $\sigma$
sandwiched in between two $n$ dimensional black holes.  Although our
brane will be uncharged, we will allow the black holes to be
charged. Since this means that
lines of flux must not converge to or diverge from the brane, we must
have black holes of equal but opposite charge. In this case, the flux
lines will pass through the brane since one black hole will act as a
source for the charge whilst the other acts as a sink. It should be
noted that although we do not have $\mathbb{Z}_2$ symmetry across the
brane for the electromagnetic field, the geometry \it{is} \rm
$\mathbb{Z}_2$ symmetric.

We denote our two spacetimes by $\mathcal{M}^+$ and $\mathcal{M}^-$
for the positively and negatively charged black holes respectively.
Their boundaries, $\partial\mathcal{M}^+$ and $\partial\mathcal{M}^-$,
both coincide with the brane.  This scenario is 
described by the following action:
\begin{eqnarray}
\label{eq:action}
S & = & \frac{1}{16\pi G_n}\int_{\mathcal{M}^+ + \mathcal{M}^-} d^n x
\sqrt{g}\left(R - 2\Lambda_n - F^2\right) + \frac{1}{8\pi
G_n}\int_{\partial\mathcal{M}^+ +\partial\mathcal{M}^-}
d^{n-1}x\sqrt{h} K \nonumber \\ & & + \frac{1}{4\pi
G_n}\int_{\partial\mathcal{M}^+ +\partial\mathcal{M}^-} d^{n-1}
x\sqrt{h} F^{a b} n_a A_b + \sigma\int_{brane} d^{n-1}x \sqrt{h} ,
\end{eqnarray}
where $g_{a b}$ is the bulk metric and $h_{ab}$ is the induced metric
on the brane.  $K$ is the trace of the extrinsic curvature  of the
brane, and $n_a$ is the unit normal to the brane pointing from
$\mathcal{M}^+$ to $\mathcal{M}^-$. Notice the presence of the
Hawking-Ross term in the action (\ref{eq:action}) which is necessary
for black holes with a fixed charge \cite{Hawking:ross}.

The bulk equations of motion which result from this action are given by
\begin{eqnarray}
R_{ab} - \frac{1}{2}R g_{ab} & = & -\Lambda_n g_{ab} + 2F_{ac}{F_b}^c
- \frac{1}{2}g_{ab} F^2 \\  \partial_a\left(\sqrt{g} F^{ab}\right) & =
& 0
\end{eqnarray}
These admit the following 2 parameter family of electrically charged
black hole solutions for the bulk metric
\begin{equation} \label{eqn:metric}
{ds_n}^2 = -h(Z)dt^2 + \frac{dZ^2}{h(Z)} + Z^2 d\Omega^2_{n-2},
\end{equation}
in which
\begin{equation} \label{eqn:h(Z)}
h(Z) = {k_n}^2 Z^2 + 1 - \frac{c}{Z^{n-3}} + \frac{q^2}{Z^{2n-6}},
\end{equation}
and the electromagnetic field strength
\begin{equation} \label{eqn:fieldstrength}
F = dA \quad \textrm{where} \quad A =
\left(-\frac{1}{\kappa}\frac{q}{Z^{n-3}} + \Phi\right) dt \quad
\textrm{and} \quad \kappa = \sqrt{\frac{2(n-3)}{n-2}}.
\end{equation}
Recall that $d\Omega^2_{n-2}$ is the metric on a unit $(n-2)$
sphere. $k_n$ is related to the bulk cosmological constant by
$\Lambda_n=-\frac{1}{2}(n-1)(n-2)k_n^2$, whereas $c$ and $q$ are
constants of integration. If $q$ is set to zero in this solution, we
regain the AdS-Schwarzschild solution discussed in the last chapter, where $c$ introduces a black
hole mass.  The presence of $q$ introduces black hole charge for which
$\Phi$ is an electrostatic potential difference.  In this general
metric, $h(Z)$ has two zeros, the larger of which, $Z_+$, represents
the event horizon of the black hole.

Here, charge is a localised quantity. It can be evaluated from a
surface integral on any closed shell wrapping the black hole (Gauss'
Law). In $\mathcal{M}^{\pm}$ the total charge is
\begin{equation}
Q=\pm \frac{(n-2)\kappa \Omega_{n-2}}{8\pi G_n} q,
\end{equation}  
The mass of each black hole, meanwhile, is same as for the uncharged case~\cite{Hawking:AdSblackholes}.
\begin{equation}
M=\frac{(n-2)\Omega_{n-2}c}{16 \pi G_n}.
\end{equation}

Let us now consider the dynamics of our brane embedded in this
background of charged black holes. Once again, we use the affine
parameter, $\tau$ to parametrise the
brane so that it is given by the section $({\bf{x} \rm}^{\mu}, t(\tau),
Z(\tau))$ of the bulk metric.  The Israel equations for the jump in
extrinsic curvature across the brane give the brane's equations of
motion. One might suspect that the presence  of the Hawking-Ross term
in the action will affect the form of these equations. However, since
the charge on the black holes is fixed, the flux across the brane does
not vary and the Israel equations take their usual form
\begin{equation}
K_{ab}=\sigma_nh_{ab},
\end{equation}
where
\begin{equation}
K_{ab} = h_{a}^c h_{b}^d \nabla_{\left(c\right.}n_{\left.d\right)}
\quad \textrm{and} \quad n_a=({\bf 0}, -\dot{Z},\dot{t}).
\end{equation}
As in the uncharged case, we also have the condition
\begin{equation} \label{eqn:condition}
-h(Z)\dot{t}^2+\frac{\dot{Z}^2}{h(Z)}=-1
\end{equation}
This ensures that the induced metric on the brane once again takes the
standard FRW form (\ref{eqn:FRWbranemetric}). Again we think of
$Z(\tau)$ as the scale factor on the brane, and $H=\dot Z/Z$, is the
Hubble parameter. We find that the cosmological evolution equations
are given by
\begin{subequations}
\begin{eqnarray} \label{eqn:FRW}
\dot t & = & \frac{\sigma_n Z}{h(Z)} 
\label{eqn:FRW:c}\\ 
H^2 & = & a - \frac{1}{Z^2} + \frac{c}{Z^{n-1}} - \frac{q^2}{Z^{2n-4}}
\label{eqn:FRW:a} \\
\dot{H} & = & \frac{1}{Z^2} -
\left(\frac{n-1}{2}\right)\frac{c}{Z^{n-1}}
+(n-2)\frac{q^2}{Z^{2n-4}}. \label{eqn:FRW:b}
\end{eqnarray}
\end{subequations}
Let us examine these equations in more detail. Equation
(\ref{eqn:FRW:a}) contains the cosmological constant term $a=\sigma_n^2-k_n^2$. For subcritical and critical branes, $Z$ has a maximum and minimum
value. For supercritical branes, we have two possibilities: either $Z$
is bounded above and below or it is only bounded below and may stretch
out to infinity. All trajectories cross the horizon, except the
unbounded supercritical one.

Our real interest in equations (\ref{eqn:FRW:a}) and
(\ref{eqn:FRW:b}), lies in understanding the $c$ and $q^2$ terms. If
we take the brane to be close to the AdS boundary, we have already
seen how the $c$ term behaves like radiation from a dual CFT. If we
make the same approximations, we find that the $q^2$ term behaves like
stiff matter\footnote{Stiff matter has the equation of state
$\rho_{CFT} \approx p_{CFT}$.}~\cite{Biswas:stiffmatter}. In the next
section we will not make any of these approximations. We will modify
the Hamiltonian approach of~\cite{Hawking:hamiltonian} to calculate
the energy density and pressure of the field theory on the brane
exactly.
\section{Energy density on the brane}
Consider an observer living on the brane. He measures time using the
braneworld coordinate, $\tau$, rather than the bulk time coordinate,
$t$. We saw in the last chapter how this can affect his measurement
of the energy density. Since we are trying to  understand
physics on the brane, we will calculate the energy with respect to $\tau$.

We begin by focusing on the contribution from the positively charged
black hole spacetime, $\mathcal{M}^+$ and its boundary,
$\partial\mathcal{M}^+$.  This boundary of course coincides with the
brane. Consider the timelike vector field defined on $\partial \mathcal{M}^+$

\begin{equation} \label{eqn:tau}
\tau^a=(\mathbf{0}, \dot{t}, \dot{Z}).
\end{equation}
This maps the boundary/brane onto itself, and satisfies $\tau^a
\nabla_a \tau=1$.  In principle we can extend the definition
of $\tau^a$ into the bulk, stating only that it approaches the form
given by equation (\ref{eqn:tau}) as it nears the brane. We now introduce a family of spacelike surfaces,
$\Sigma_\tau$, labelled by $\tau$ that are always normal to
$\tau^a$. This family provide a slicing of the spacetime,
$\mathcal{M}^+$ and each slice meets the brane orthogonally. As usual
we decompose $\tau^a$ into the lapse function and shift vector,
$\tau^a=N r^a +N^a$, where $r^a$ is the unit normal to
$\Sigma_\tau$. However, when we lie on the brane, $\tau^a$ \it{is} \rm
the unit normal to $\Sigma_\tau$, because there we have the condition
(\ref{eqn:condition}). Therefore, on $\partial \mathcal{M}^+$, the lapse function, $N=1$ and
the shift vector, $N^a=0$. Before we consider whether or not we need
to subtract off a background energy, let us first state that the
relevant part of the action at this stage of our analysis is the
following:

\begin{equation}
I^+ =  \frac{1}{16\pi G_n}\int_{\mathcal{M}^+} R - 2\Lambda_n - F^2 +
\frac{1}{8\pi G_n}\int_{\partial \mathcal{M}^+}  K  + \frac{1}{4\pi
G_n}\int_{\partial\mathcal{M}^+} F_{a b} n^a A^b.
\end{equation}

As stated earlier, we do not include any contribution from
$\mathcal{M}^-$ or $\partial \mathcal{M}^-$, nor do we include the
term involving the brane tension. This is because we want to calculate
the gravitational Hamiltonian, without the extra contribution of a
source. The brane tension has already been included in the analysis as
a cosmological constant term, and it would be wrong to double count.

Given the slicing $\Sigma_\tau$, the Hamiltonian that we derive from
$I^+$ is given by

\begin{eqnarray} \label{eqn:hamiltonian}
H^+ &=& \frac{1}{8 \pi G_n} \int_{\Sigma_{\tau}} N\mathcal{H} + N^a
\mathcal{H}_a - 2NA_{\tau} \nabla_a E^a \nonumber \\ & & -\frac{1}{8
\pi G_n} \int_{S_{\tau}} N \Theta + N^a p_{ab}n^b - 2NA_{\tau}n_a E^a
+ 2NF^{ab}n_aA_b
\end{eqnarray}
where $\mathcal{H}$ and $\mathcal{H}_a$ are the Hamiltonian and momentum constraints respectively~\cite{Hawking:hamiltonian}. $p^{ab}$ is the canonical momentum
conjugate to the induced metric on $\Sigma_{\tau}$ and $E^a$ is the
momentum conjugate to $A_a$.  The surface $S_\tau$ is the intersection
of $\Sigma_{\tau}$ and the brane, while $\Theta$ is the trace of the
extrinsic curvature of $S_{\tau}$ in $\Sigma_\tau$ (see
figure~\ref{fig:hamiltonian}).
\begin{figure}[ht]
\hskip -2cm
\includegraphics[width=13cm, angle=-90]{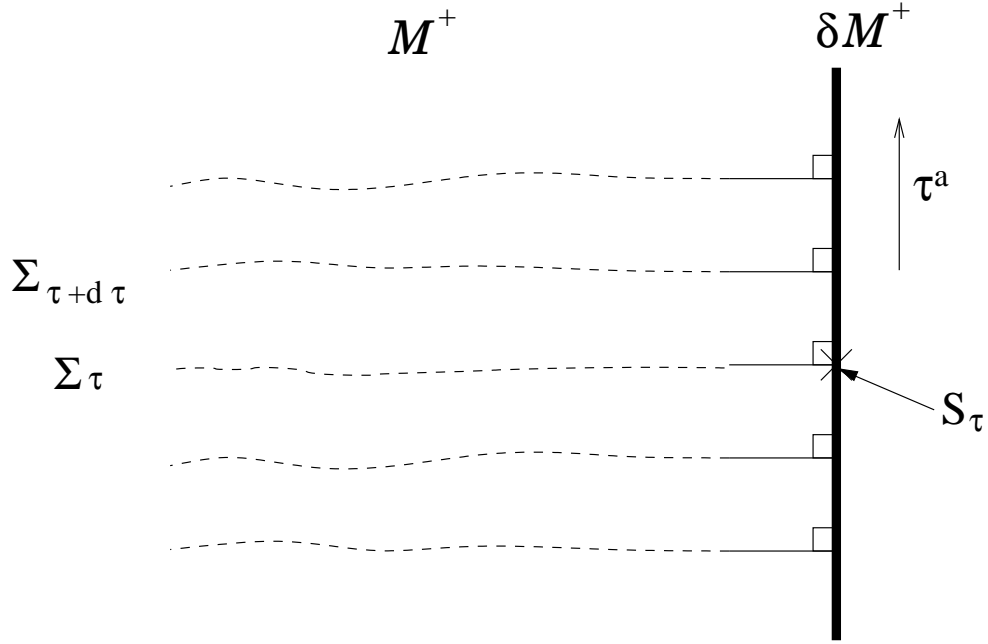}
\caption{Foliation of $\mathcal{M}^+$ into spacelike surfaces
$\Sigma_\tau$. These surfaces meet the brane orthogonally as shown. } \label{fig:hamiltonian}
\end{figure}

Note that the momentum $E^a=F^{a \tau}$. In particular, $E^{\tau}=0$
and we regard $A_{\tau}$ as an ignorable coordinate. We will now
evaluate this Hamiltonian for the RNAdS spacetime described by
equations (\ref{eqn:metric}), (\ref{eqn:h(Z)}) and
(\ref{eqn:fieldstrength}). Each of the constraints vanish because this
is a solution to the equations of motion.

\begin{equation}
\mathcal{H}=\mathcal{H}_a=\nabla_a E^a=0.
\end{equation}
The last constraint is of course Gauss' Law. When evaluated on the
surface $S_{\tau}$,  the potential,
$A=\left(-\frac{1}{\kappa}\frac{q}{Z(\tau)^{n-3}} + \Phi\right)
\dot{t} \ d\tau$. The important thing here is that it  only has
components in the $\tau$ direction. This ensures that the last two
terms in the Hamiltonian cancel one another. Since $N=1$ and $N^a=0$
on $S_{\tau}$, it only remains to evaluate the extrinsic curvature
$\Theta$. If $\gamma_{ab}$ is the induced metric on $S_{\tau}$, it is
easy to show that

\begin{equation}
\Theta=\Theta_{ab} \gamma^{ab}=K_{ab} \gamma^{ab} = (n-2)\frac{h(Z)
\dot{t}}{Z}.
\end{equation}
The energy is then evaluated as
\begin{equation}
\mathcal{E}= -\frac{1}{8 \pi G_n} \int_{S_{\tau}} (n-2)\frac{h(Z)
\dot{t}}{Z}.
\end{equation}
We will now address the issue of background energy. This is usually
necessary to cancel divergences in the Hamiltonian.  In our case, the
brane cuts off the spacetime. If the brane does not stretch to the AdS
boundary there will not be any divergences that need to be
cancelled. However it is important to define a zero energy
solution. In this work we will choose pure AdS space. This is because
the FRW equations for a brane embedded in pure AdS space would include
all but the holographic terms that appear in equations
(\ref{eqn:FRW:a}) and (\ref{eqn:FRW:b}). These are the terms we are
trying to interpret with this analysis.

We will denote the background spacetime by $\mathcal{M}_0$.  We have
chosen this to be pure AdS space cut off at a surface $\partial
\mathcal{M}_0$ whose geometry is the same as our brane. As is
described in section~\ref{sec:better}, this means we have the bulk metric
given by

\begin{equation} \label{eqn:AdSmetric}
{ds_n}^2 = -h_{AdS}(Z)dT^2 + \frac{dZ^2}{h_{AdS}(Z)} + Z^2
d\Omega_{n-2},
\end{equation}
in which
\begin{equation} \label{eqn:h_AdS(Z)}
h_{AdS}(Z) = {k_n}^2 Z^2 + 1.
\end{equation}
There is of course no electromagnetic field. The surface  $\partial
\mathcal{M}_0$ is described by the section $(\mathbf{x}^{\mu},
T(\tau), Z(\tau))$ of the bulk spacetime. In order that this surface
has the same geometry as our brane we impose the condition

\begin{equation} \label{eqn:AdScondition}
-h_{AdS}(Z)\dot{T}^2+\frac{\dot{Z}^2}{h_{AdS}(Z)}=-1
\end{equation} 
which is analogous to the condition given in equation
(\ref{eqn:condition}).

We now repeat the above evaluation of the Hamiltonian for the
background spacetime. This gives the following value for the
background energy

\begin{equation}
\mathcal{E}_0= -\frac{1}{8 \pi G_n} \int_{S_{\tau}}
(n-2)\frac{h_{AdS}(Z) \dot{T}}{Z}.
\end{equation}
Making use of equations (\ref{eqn:FRW:c}),
(\ref{eqn:FRW:a}) and (\ref{eqn:AdScondition}) we find that the
energy of $\mathcal{M}^+$ above the background $\mathcal{M}_0$ is
given by

\begin{equation}
E_+=\mathcal{E}-\mathcal{E}_0=\frac{(n-2)}{8 \pi G_n} \int_{S_{\tau}}
\sqrt{\sigma_n^2-\frac{\Delta h}{Z^2}} -\sigma_n
\end{equation}
where
\begin{equation}
\Delta h=h(Z)-h_{AdS}(Z)=-\frac{c}{Z^{n-3}}+\frac{q^2}{Z^{2n-6}}.
\end{equation}
In this relation $\Delta h$ is negative everywhere outside of the black hole
horizon and so it is clear that $E_+$ is positive.
We now turn our attention to the contribution to the energy from
$\mathcal{M}^-$. Since the derivation of $E_+$ saw the cancellation of
the last two terms in the Hamiltonian (\ref{eqn:hamiltonian}) we note
that the result is purely geometrical. Even though $\mathcal{M}^+$ and
$\mathcal{M}^-$ have opposite charge, they have the same geometry and
so $E_+=E_-$. We deduce then that the total energy

\begin{equation} \label{eqn:energy}
E=E_++E_-= \frac{(n-2)}{4 \pi G_n} \int_{S_{\tau}}
\sqrt{\sigma_n^2-\frac{\Delta h}{Z^2}} -\sigma_n
\end{equation}
Since the spatial volume of the braneworld
$V=\int_{S_{\tau}}=\Omega_{n-2}Z^{n-2}$, we arrive at the exact
expression for the energy density measured by an observer living on
the brane

\begin{equation} \label{eqn:rho}
\rho= \frac{(n-2)\sigma_n}{4 \pi G_n} \left( \sqrt{1-\frac{\Delta
h}{\sigma_n^2Z^2}} -1 \right)
\end{equation}
where we have pulled out a factor of $\sigma_n$.
\section{Pressure on the brane and equation of state}
Using equation (\ref{eqn:TD}), we can derive the pressure, $p$,
measured on the brane:
\be \label{eqn:pexact}
p =
-\rho +\frac{1}{8\pi G_n\sigma_n}\left(1-\frac{\Delta
h}{\sigma_n^2Z^2}\right)^{-\frac{1}{2}}
\left[\frac{(n-1)c}{Z^{n-1}} - \frac{2(n-2)q^2}{Z^{2n-4}}\right]
\ee
This is not very illuminating as it stands. If we take $\Delta
h/\sigma_n^2Z^2 \ll 1$, we recover the approximate results for when
the brane is near the AdS boundary:
\begin{subequations}
\bea
\rho &\approx& \frac{(n-2)}{8 \pi
G_n\sigma_n}\left(\frac{c}{Z^{n-1}}-\frac{q^2}{Z^{2n-4}} \right) \\
p &\approx& \frac{(n-2)}{8 \pi
G_n\sigma_n}\left(\frac{c}{(n-2)Z^{n-1}}-\frac{q^2}{Z^{2n-4}}\right)
\eea
\end{subequations}
Here we can clearly see how the pressure is made up of a ``radiation''
and a ``stiff matter'' contribution:
\be \label{eqn:pcombo}
p\approx\frac{\rho_{\textrm{rad}}}{n-2}+\rho_{\textrm{stiff}}
\ee
However, the equation of state in our exact analysis is far more
complicated. In the simpler case when $q^2=0$, we can express the
equation of state in the following way: 
\be \label{eqn:radstate}
p=-\rho+\frac{(n-1)\sigma_n}{8 \pi G_n}\left[\left(1+\frac{4 \pi
G_n}{(n-2)\sigma_n}\rho\right)-\left(1+\frac{4 \pi
G_n}{(n-2)\sigma_n}\rho\right)^{-1}\right]
\ee
This simplifies to the radiation state $p=\frac{\rho}{n-2}$ when $\rho
\ll 1$.  The $c/Z^{n-1}$ term that appears in the FRW equations is
often referred to as the radiation term. We have shown that this is only
true when $\rho$ is small, and the brane is near to the AdS
boundary. More generally, the equation of state is not as simple as
that of radiation, and nor should we expect it to be. By introducing a
significant UV cut-off in our field theory on the brane, we
have completely lost the conformal properties of the theory, and
therefore its resemblance to radiation.

When we consider non-zero values of $q^2$ it is even harder to write
down an expression like (\ref{eqn:radstate}). In the limit of
small $\rho$, we have shown that the equation of state simplifies to (\ref{eqn:pcombo}),
but we cannot say much more. 
\section{The cosmological evolution equations}
We shall now insert our expressions for the braneworld energy density
(\ref{eqn:rho}) and pressure (\ref{eqn:pexact}) into the cosmological
evolution equations (\ref{eqn:FRW:a}) and (\ref{eqn:FRW:b}). We find
\bml
\bea
H^2 &=& a - \frac{1}{Z^2} + \frac{8 \pi G_n \sigma_n}{n-2} \rho + \left (
\frac{4 \pi G_n}{n-2} \right)^2 \rho^2 \label{eqn:FRWft:a} \\
\dot{H} &=& \frac{1}{Z^2}-4 \pi G_n \sigma_n (\rho+p) -(n-2)\left (
\frac{4 \pi G_n}{n-2} \right)^2\rho  (\rho+p) \label{eqn:FRWft:b}
\eea
\eml
These are clearly \underline{not} the standard Friedmann equations for an $(n-1)$
dimensional universe with energy density $\rho$ and pressure
$p$. This should come as no surprise. We have not made any
approximations in arriving at these results so it is possible that we
would see non-linear terms. What is exciting is that the quadratic
terms we see here have exactly the same form as the unconventional
terms that we discussed in section~\ref{sec:unconventional}. In that case, one places extra matter on the brane to
discover this unconventional cosmology. We have no extra matter on the
brane but by including a bulk black hole, we get exactly the same type
of cosmology. Clearly there is an alternative description.

We also note that in section~\ref{sec:unconventional}, the
energy momentum tensor on the brane is split between tension and
additional matter in an arbitrary way.  In the analysis we have just
carried out, the tension is the \it
{only} \rm explicit source of energy momentum on the brane so there is
no split required.  With this in mind we are able to interpret each
term in the FRW equations more confidently, in particular, the
cosmological constant term.  Furthermore, we have not yet  made any
assumptions on the form of the braneworld Newton's constant. 

Finally, we see that for small $\rho$ and $p$, we can neglect the
$\rho^2$ and $\rho p$  terms and recover the standard Friedmann
equations for an $(n-1)$ dimensional universe:
\begin{eqnarray}
H^2 &=& a - \frac{1}{Z^2} + \frac{16 \pi G_{n-1}}{(n-2)(n-3)} \rho
\label{eqn:linearFRW:a} \\
\dot{H} &=& \frac{1}{Z^2}- \frac{8 \pi G_{n-1}}{(n-3)} (\rho+p)
\label{eqn:linearFRW:b}
\end{eqnarray}
where we have taken the induced  Newton's constant on the brane to be
given by (\ref{eqn:NC}). We see, then, how the relationships noticed in
sections~\ref{sec:critical} and~\ref{sec:CFTonnoncrit} are just an approximation of the relationship
described here.

\chapter{Discussion} \label{chap:discuss}
\setcounter{equation}{0}
Having been on a long, and sometimes difficult journey through the  braneworld, we might wonder whether or not such objects really exist in Nature. Moreover, do we actually live on a brane? It is highly unlikely that the Randall-Sundrum models~\cite{Randall:hierarchy,Randall:compactification} accurately describe the structure of our universe. As we emphasized in chapter~\ref{chap:RSbw}, these are  merely toy models. Nature, meanwhile, is far more complicated than this. In particular, neither RS1 nor RS2 includes any supersymmetry, which, although yet to be discovered, is commonly thought to exist. Furthermore, if we believe that something like M-theory represents a ``theory of everything'' we have to accept that we might have more than just five dimensions. However, despite their simplicity, the RS models have contributed in at least two very important ways:
\begin{itemize}
\item they provide a viable ``alternative to compactification''.

\item they give us new tools with which to study holography and its applications.
\end{itemize}
We will now discuss each point in turn, with emphasis on the relevance to this thesis.

\section{An alternative to compactification}
In chapter~\ref{chap:intro}, we noted that to be consistent at a quantum level, superstring theory and M-theory need
to live in 10 and 11 dimensions respectively.
We have generally believed that the reason we do not see more than four dimensions
is that the extra dimensions are very small, and we require very high energies
to probe them. In RS2, we have seen that generically this need not be the case.
In RS2, the extra dimension is infinite, and yet preliminary results
suggest that an
observer on the brane would see four-dimensional physics up to at least a few TeV.
This is achieved in the following way: standard model fields are bound to a domain
wall, or brane, although gravity can propagate into the bulk. The bulk geometry is 
warped, and this warp factor ensures that gravitational perturbations are
damped as they move away from the brane. This is known as localisation of gravity.

In this thesis, we began a study of gravity localisation at a non-perturbative level.
In chapter~\ref{chap:bwcosmo} we discussed cosmology on the brane. The most 
interesting feature of this was the quadratic energy-momentum terms that 
appeared in the FRW equations~\cite{Shiromizu:einstein, Binetruy:branecos1,Binetruy:branecos2}.
We can neglect the effect of these terms at low density. However, if the universe
was very small at some time, these terms become important. This does not disagree with 
the idea that extra dimensions might show up in the very early universe.

We should mention at this stage that some braneworld cosmologies do not possess a Big Bang singularity. In chapters~\ref{chap:holography} and~\ref{chap:exact}
we saw that there exist brane trajectories that do not pass through $Z=0$, where $Z$
is the scale factor of the brane universe. These ``bounce'' solutions are made
possible by modifying the structure of the bulk space-time. By introducing a non-trivial
Weyl tensor in the bulk we obtain ``dark matter'' terms in the FRW equations that
prevent the brane from shrinking to zero size. We will discuss ``dark matter'' terms
more in the section on holography.

In chapter~\ref{chap:bubbles} we attacked the issue of non-perturbative gravity
in a very different way. Our approach was to place a strongly gravitating object on
the brane and examine how that affected the geometry there. When we think of
a strongly gravitating object, we immediately think of a black hole. However, finding a
solution for a black hole  bound to the brane is an outstanding problem.
Instead, we chose to study a domain wall on the brane. This is a codimension
two object living entirely on the brane. For this reason, we refer to it as a vortex.
Because there are only two dimensions transverse to the vortex, the transverse
part of the bulk metric is conformally flat. This conformal flatness ensures that
our equations of motion are completely integrable and we can find an exact solution
for the bulk geometry. Remarkably, when we examine the geometry induced on the brane 
we find that it behaves as if there were no extra dimensions. This is one of 
the main results of this thesis: the geometry on a brane containing
a vortex of tension $T$ is the same as the geometry that arises from a
domain wall of the same tension in $(n-1)$-dimensional Einstein gravity. In this non-perturbative example,
gravity is localised on the brane {\it exactly}. This exactness is probably
due to the high degree of symmetry in the problem. Nevertheless, the result has added
to the claims that $(n-1)$-dimensional gravity can be reproduced even when a large $n$-th dimension is present.

Finally, the techniques used in this analysis opened up a number of possibilities. 
Firstly, we were able to construct {\it nested} Randall-Sundrum scenarios where the geometry on the 
{\it brane} is the traditional RS geometry (in $(n-1)$-dimensions) and we live
on the vortex. We could try and play the whole Randall-Sundrum game again and see
if ``an alternative to compactification'' can work with two large extra dimensions.
We might also consider the implications this has for holography although more on that later. 
We also have the tools to construct {\it braneworld instantons}. This means we can
discuss first order phase transitions whereby a true vacuum bubble nucleates in a false vacuum, and then grows.
In particular we have shown how one could start off with a de Sitter false vacuum which corresponds to an inflationary
era. We have calculated the probability that a flat bubble universe nucleates in this background. The result
seems to agree with~\cite{Coleman:vacuumdecay}, where we have no extra dimensions.
\section{A tool for holography}
We have seen how the presence of the AdS warp factor in the bulk ensures that gravity is localised on a braneworld. In chapter~\ref{chap:holography}, we came across another important property of AdS space: it can be foliated by a family of spacelike surfaces, each of which satisfy the holographic entropy bound. This makes AdS space a prime candidate for a holographic description. The first concrete example of this is the AdS/CFT correspondence, where we have a duality relating gravity in the bulk to a conformal field theory on the boundary. Specifically, type IIB superstring theory on $AdS_5 \times S^5$ is dual to $\mathcal{N}=4$ super Yang Mills on the boundary.

Braneworld holography is not so precise. We have Einstein gravity with
a negative cosmological constant in the bulk. This is thought to be
dual to a field theory on the brane that is cut-off in the
ultra-violet. We do not know what the field theory is. However,
whereas in the original Maldacena conjecture, gravity decouples from
the CFT, this is not the case for the braneworld theory. Although we
know very little about this field theory, we can use its coupling to
gravity to derive some of its properties. To study braneworld
holography we usually require two things: a FRW brane and a black hole in the bulk.

The intuition is as follows: the bulk black hole emits Hawking radiation that heats the brane to a finite temperature. If the braneworld theory exists, it should be hot, and have a non-zero energy density and pressure. In the original work of Verlinde and Savonije~\cite{Savonije:braneCFT}, they found that we could interpret the brane cosmology in two different ways. Either it is driven by the bulk black hole or it is driven by a dual field theory. In the latter case, the FRW equations are those of the standard cosmology. If the bulk black hole is uncharged, the field theory behaves like radiation. 

In chapter~\ref{chap:holography}, we saw that the extension of these ideas to de Sitter and anti-de Sitter branes was non-trivial. We need to be careful when using our AdS/CFT dictionary. The method of Verlinde and Savonije was to take the black hole mass and calculate the energy of the dual CFT by scaling with some appropriate red-shift. Although this method works for flat braneworlds, it does not quite work for dS and AdS branes. The AdS/CFT dictionary should really state that bulk energy, rather than black hole mass, translates into the energy of the field theory. Since we have a brane present, the bulk space-time is cut-off before it reaches the AdS boundary. The presence of the bulk cosmological constant ensures that this can affect the calculation of the bulk energy. In chapter~\ref{chap:holography}, we use Euclidean quantum gravity techniques to calculate the bulk energy properly. We find that the bulk energy differs from the black hole mass in just the right way. The dual description described at the end of the last paragraph for flat branes carries over to de Sitter and anti-de Sitter branes.

From a phenomenological point of view, a study of the de Sitter brane is important as recent observations suggest our universe has a small positive cosmological constant~\cite{Perlmutter:astro, Riess:astro}. However, from a  holographic point of view, we might be more interested in the anti-de Sitter brane. We have already discussed the nested Randall-Sundrum scenario described in chapter~\ref{chap:bubbles}. Perhaps in this case we could do holography twice and project all degrees of freedom on to the vortex.

We could criticise this kind of braneworld holography for being too imprecise. However, in chapter~\ref{chap:exact} we saw that we can actually  do much more exact calculations. In the approximate braneworld holography of chapter~\ref{chap:holography}, we assumed that the brane was close to the AdS boundary.  We can relax this assumption if we use a hamiltonian approach to calculate the energy on the brane. By allowing the brane trajectory to move far away from the boundary, we can see the effect of the UV cut-off in the dual field theory. Although the field theory is nowhere near being conformal, braneworld holography survives. This is another very important result of this thesis. It enables us to make the following exact statement:

The cosmological evolution equations on the brane have the same form whether we have

(i) a black hole in the bulk with no additional matter on the brane.
\\
or (ii) no bulk black hole with additional matter placed on the brane by hand. 

For case (ii), we saw in chapter~\ref{chap:bwcosmo} how the evolution
equations contain quadratic energy density/pressure terms. When we
calculate the energy density/pressure of the dual field theory in (i)
we find that they contribute to the evolution equations in exactly the
same way. A braneworld observer cannot tell whether the energy that
drives his cosmology comes from additional brane matter or a bulk
black hole. In this way, the bulk black hole behaves like ``dark matter'' on the brane: you cannot see it, but you can tell it is there.

\bibliographystyle{utphys}
\addcontentsline{toc}{chapter}{\numberline{}Bibliography}
\bibliography{padilla}

\appendix
\setcounter{chapter}{0}
\renewcommand{\chaptername}{Appendix}
\renewcommand{\theequation}{\Alph{chapter}.\arabic{section}.\arabic{equation}}
\addcontentsline{toc}{chapter}{\numberline{}Appendix}
\setcounter{equation}{0}
\chapter{Detailed Calculations}
\section{Green's function in RS2} \label{app:Greens}
In order to construct the full Green's function in the RS2 model, we
will use Sturm
Liouville theory techniques.  We begin by reintroducing the negative
tension brane at $z=z_c$ so that it acts as a regulator, and an
additional boundary condition is imposed
\begin{equation} \label{eqn:regulatorbc}
\left( \partial_z+2k \right)\Big\vert_{z=z_c^-}
h_{\mu\nu}=0
\end{equation}
This places new constraints on the (regulated) Green's function so we  modify equation (\ref{eqn:Greens}) appropriately
\begin{equation} \label{eqn:regGreens}
\left[e^{2k|z|}\Box^{(4)} + \partial_z^2
-4k^2+4k\delta(z)-4k\delta(z-z_c)\right]G_R(x,z; x^{\prime}, z^{\prime})=\delta^{(4)}(x-x^{\prime}) \delta(z-z^{\prime}).
\end{equation}
We now take Fourier transforms with respect to $x^{\mu}$,
\begin{equation} \label{eqn:FTGreens}
\left[-e^{2k|z|}p^2 + \partial_z^2
-4k^2+4k\delta(z)-4k\delta(z-z_c)\right]\tilde G_R(p;z,z')=\delta(z-z^{\prime})
\end{equation}
where 
\begin{equation} \label{eqn:FT}
\tilde G_R(p;z,z')=\int d^4x e^{-ip_{\mu}(x^{\mu}-x'^{\mu})} G_R(x,z;
x^{\prime}, z^{\prime}).
\end{equation}
For $z \neq z'$,  the Green's function satisfies the following Sturm
Liouville equation
\begin{equation} \label{eqn:SL}
\left(\partial_z^2
-4k^2\right)\tilde G_R=p^2e^{2k|z|} \tilde G_R 
\end{equation}
with boundary conditions
\begin{equation} \label{eqn:SLbc}
\left( \partial_z+2k \right) \Big\vert_{z=0^+}\tilde G_R=0,\qquad 
\left( \partial_z+2k \right) \Big\vert_{z=z_c^-}\tilde G_R=0
\end{equation}
We wish to find eigenstates, $u_m(z)$, for this problem, with
eigenvalues $p^2=-m^2$. The zero mode eigenstate is trivially given by
\begin{equation}
u_0(z)=N_0 e^{-2k|z|}
\end{equation}
where $N_0$ is some normalisation constant. Note that we have inserted
the $\mathbb{Z}_2$ symmetry about $z=0$ explicitly. In order to
determine the massive  eigenstates we will change variables to
$y=me^{k|z|}/k$, so that equation (\ref{eqn:SL}) is transformed into
Bessel's equation with $n=2$~\cite{Abramowitz:Bessel}
\begin{equation} \label{eqn:Bessel}
\left[y^2 \partial^2_y+y\partial_y +(y^2-4)\right]\tilde G_R=0
\end{equation}
with boundary conditions
\begin{equation} \label{eqn:Besselbc}
\left( y\partial_y+2 \right) \Big\vert_{y=m/k}\tilde G_R=0,\qquad 
\left( y\partial_y+2 \right) \Big\vert_{y=y_c}\tilde G_R=0
\end{equation}
where $y_c=me^{kz_c}/k$. Equation (\ref{eqn:Bessel}) has solutions $J_2(y)$ and
$Y_2(y)$ which satisfy the following recurrence
relations~\cite{Abramowitz:Bessel}
\begin{equation}
\left(y\partial_y+2\right)J_2(y)=yJ_1(y), \qquad
\left(y\partial_y+2\right)Y_2(y)=yY_1(y)
\end{equation}
We deduce then that the massive eigenstates are given by
\begin{equation}
u_m(z)=N_m\left[J_1(m/k)Y_2(y)-Y_1(m/k)J_2(y)\right]
\end{equation}
where $N_m$ is the normalisation constant. Note that the boundary
condition at $y=y_c ~(z=z_c)$ is only satisfied for quantised values
of $m$ satisfying the following condition
\begin{equation} \label{eqn:qu}
J_1(m/k)Y_1(me^{kz_c}/k)-Y_1(m/k)J_1(me^{kz_c}/k)=0
\end{equation}
For large $z$, the asymptotic behaviour of Bessel's functions is
given by
\begin{eqnarray}
J_n(me^{kz}/k) &\sim& ~\sqrt{\frac{2ke^{-kz}}{\pi
m}}\cos\left(\frac{me^{kz}}{k}-\frac{n \pi}{2}-\frac{\pi}{4} \right) \nonumber \\ 
Y_n(me^{kz}/k) &\sim& \sqrt{\frac{2ke^{-kz}}{\pi
m}}\sin\left(\frac{me^{kz}}{k}-\frac{n \pi}{2}-\frac{\pi}{4} \right). \label{eqn:lgz}
\end{eqnarray}
As we send the regulator brane towards infinity $(z_c \to \infty )$, equations (\ref{eqn:qu}) and (\ref{eqn:lgz}) imply that $m$ is
quantised in units of $\pi k e^{-kz_c}$. The normalisation constants,
meanwhile, are determined by the following normalisation condition
\begin{equation} \label{eqn:normalisation}
\int_{-z_c}^{z_c} dz ~e^{2k|z|} u_m(z)u_n(z)=\delta_{mn}.
\end{equation}
For the zero mode, it is easy to see that this gives
\begin{equation}
N_0^2=k \left( 1-e^{-2kz_c} \right)^{-1}
\end{equation}
The normalisation for the heavy modes is less obvious. However, we
note that for large $z_c$, the dominant contribution to the integral
(\ref{eqn:normalisation}) lies near $|z| =z_c$. Using the asymptotic
behaviour (\ref{eqn:lgz})   we find that
\begin{equation}
N_m^2 = \frac{\pi m}{2}e^{-kz_c}\left[J_1(m/k)^2+Y_1(m/k)^2
\right]^{-1}+\mathcal{O}(e^{-2kz_c})
\end{equation}
The (Fourier transformed) Green's function satisfies
\begin{equation}
\left(\partial_z^2-4k^2-p^2e^{2k|z|} \right)\tilde G_R =\delta(z-z')
\end{equation}
and can be expressed in
terms of the complete set of eigenstates, $\{ u_m(z) \}$.
\begin{equation} \label{eqn:sumofestates}
\tilde G_R(p; z, z')= -\frac{u_0(z)u_0(z')}{p^2}-\sum_{m} \frac{u_m(z)u_m(z')}{m^2+p^2}
\end{equation}
where we ensure $p^2 \neq -m^2$ by adding a small imaginary part in
the ``time'' direction, ie. $p^{\mu}=(\omega+i\epsilon, {\bf p})$. 
We now remove the regulator brane completely by sending $z_c \to
\infty$. The quantisation in $m$ disappears and we go to a continuum
limit, replacing the sum in equation
(\ref{eqn:sumofestates}) with the following integral
\begin{equation}
 \sum_{m} \frac{u_m(z)u_m(z')}{m^2+p^2} \longrightarrow
 \int_{0}^{\infty} dm~\lim_{z_c \to \infty} \frac{1}{\pi k e^{-kz_c}}
 \left(\frac{u_m(z)u_m(z')}{m^2+p^2}\right)
\end{equation}
The extra term appearing in the integral is just a ``density of
states'' factor that will cancel the  vanishing part of the
normalisation constant. Inverting the Fourier transform
(\ref{eqn:FT}), we find that the full Green's function is given by
\begin{equation} 
G_R(x,z;x',z')=-\int \frac{d^4 p}{(2\pi)^4}
e^{ip_{\mu}(x^{\mu}-x'{}^{\mu})}\Biggl[
\frac{e^{-2k(|z|+|z'|)} k}{ {\bf p}^2-(\omega+i\epsilon)^2}
+\int_0^{\infty} dm\,
 \frac{v_m(z) v_m(z')}{m^2+{\bf p}^2-(\omega+i\epsilon)^2}\Biggr],
\end{equation}
where
\begin{equation}
v_m(z)=\frac{\sqrt{m/2k} \left[J_1(m/k)Y_2(me^{k|z|}/k) - 
Y_1(m/k) J_2(me^{k|z|}/k)\right]}{\sqrt{J_1(m/k)^2+Y_1(m/k)^2}}. 
\end{equation}
Finally we
should note that we did not include eigenstates satisfying
$p^2=m^2>0$. These would be linear combinations of ``modified''
Bessel's functions, but would not be normalisable and are therefore omitted.

\section{Warp factor around non-critical branes} \label{app:noncrit}
Given the ansatz (\ref{eqn:noncritansatz}) we need to solve the bulk
equations of motion with cosmological constant, $\Lambda=-6k^2$. Our
solution must then satisfy the boundary conditions imposed at the
brane of (positive) tension $\sigma$, sitting at $z=0$.  

The bulk equations of motion are just given by the Einstein equations with the appropriate cosmological
constant.
\begin{equation}
R_{ab}-\frac{1}{2}Rg_{ab}=-\Lambda g_{ab} 
\end{equation}
If we define $\lambda$ to be the cosmological constant {\it on the
brane}, this gives
\begin{align}
\mu\nu ~\textrm{equation} &:& \frac{\lambda}{a^2}-3\left(\frac{
a'}{a}\right)^2-\frac{a''}{a} = -4k^2, \\
zz ~\textrm{equation}&:& -4\frac{a''}{a} =  -4k^2.
\end{align}
where `prime' denotes differentiation with respect to $z$.
These equations have three classes of solutions, depending on whether
$\lambda$ is positive, negative or zero.
\begin{align}
\lambda>0 &:& a(z)&=\frac{1}{k}\sqrt{\frac{\lambda}{3}}\sinh(\pm kz+c) \\
\lambda=0 &:& a(z)&=e^{\pm kz+c} \\
\lambda<0 &:& a(z)&=\frac{1}{k}\sqrt{-\frac{\lambda}{3}}\cosh(\pm kz+c)
\end{align}
where $c$ is a constant of integration.

The boundary conditions are given by the Israel junction
conditions~\cite{Israel:junction} at the brane.
\begin{equation}
\Delta K_{ab}=-\frac{8 \pi G_5}{3}\sigma g_{0ab}
\end{equation}
where $g_{0ab}$ is the induced metric on the brane. Given our ansatz
(\ref{eqn:noncritansatz}) and the fact that we have
$\mathbb{Z}_2$ symmetry across the brane, we find that
\begin{equation}
\frac{a'}{a}\Big\vert_{z=0^+}=-\frac{4 \pi G_5}{3}\sigma 
\end{equation}
Since we are assuming $\sigma>0$ we find that we are left with
\begin{align}
\lambda>0 &:& a(z)&=\frac{1}{k}\sqrt{\frac{\lambda}{3}}\sinh(-k|z|+c) \\
\lambda=0 &:& a(z)&=e^{-k|z|+c} \\
\lambda<0 &:& a(z)&=\frac{1}{k}\sqrt{-\frac{\lambda}{3}}\cosh(-k|z|+c)
\end{align}
with the following conditions
\begin{eqnarray} 
\lambda>0 &:& \qquad \tilde \sigma =k\coth c>k \label{eqn:cond1a}\\
\lambda=0 &:& \qquad \tilde \sigma =k \label{eqn:cond1b}\\
\lambda<0 &:& \qquad \tilde \sigma =k\tanh c<k \label{eqn:cond1c}
\end{eqnarray}
where $\tilde \sigma = 4 \pi G_5 \sigma/3$. We are also free to set
$a(0)=1$ in each case giving
\begin{eqnarray} 
\lambda>0 &:& \qquad k=\sqrt{\frac{\lambda}{3}}\sinh c \label{eqn:cond2a}\\
\lambda=0 &:& \qquad c=0 \label{eqn:cond2b}\\
\lambda<0 &:& \qquad k=\sqrt{-\frac{\lambda}{3}}\cosh c \label{eqn:cond2c}
\end{eqnarray}
Equations (\ref{eqn:cond1a}) to (\ref{eqn:cond2c}) fix the cosmological
constant on the brane to be
\begin{equation}
\lambda=3(\tilde \sigma^2-k^2)
\end{equation}
with the final solutions given by equations
(\ref{eqn:DSsoln}),(\ref{eqn:flatsoln}) and (\ref{eqn:ADSsoln}).
\section{Extrinsic curvature of a dynamic brane} \label{app:extrinsic}
Suppose we have a bulk spacetime whose metric is given by
\begin{equation}
ds_n^2=-h(Z)^2dt^2+\frac{dZ^2}{h(Z)}+Z^2d{\bf x}_{\kappa}^2
\end{equation}
cut off at a brane given by the section
\begin{equation}
X^a=(\bf x\rm^{\mu}, t(\tau), Z(\tau))
\end{equation}
where $\tau$ is the proper time for an observer comoving with the
brane. This gives the condition
\begin{equation} \label{eqn:app3condition}
-h\dot t ^2+\frac{\dot Z ^2}{h}=-1
\end{equation}
so that the induced metric on the brane is given by equation
(\ref{eqn:FRWbranemetric}). Now suppose the normal to the brane is
defined as
\begin{equation}
n_a==\epsilon({\bf 0}, -\dot Z(\tau), \dot t(\tau))
\end{equation}
and define the extrinsic curvature of the brane to be
$K_{ab}=h_a^ch_b^d\nabla_{(c}n_{d)}$. We first find that
\begin{equation}
K_{\mu\nu}=\nabla_{(\mu} n_{\nu)}=-\Gamma_{\mu\nu}^a n_a=\frac{\epsilon h \dot t}{Z}h_{\mu\nu}
\end{equation}
The components of $\partial /\partial \tau$ are given by 
\begin{equation}
\tau^a=({\bf 0}, \dot t(\tau), \dot Z(\tau))
\end{equation}
which is normal to $n_a$. The last non-zero component of the extrinsic
curvature is then
\begin{eqnarray}
K_{\tau \tau}&=&\tau^a\tau^b\nabla_a n_b=-\tau^a n_b \nabla_a \tau^b=-n_c (\dot \tau^c + \Gamma_{ab}^c \tau^a\tau^b) \nonumber \\
&=& \epsilon \dot Z \left[ \ddot t + \frac{h'}{h} \dot t \dot Z
\right] -\epsilon \dot t \left[ \ddot Z +\left(\frac{h'}{2}\right)h\dot t ^2-
\left(\frac{h'}{2}\right)\frac{\dot Z ^2}{h} \right] \nonumber \\
&=& \frac{\ddot Z+\frac{1}{2}h'}{\epsilon h \dot
t}
\end{eqnarray}
where we have used equation ({\ref{eqn:app3condition}).
\section{Probability of bubble nucleation on the brane} \label{app:instanton}
In section~\ref{sect:inst} we calculated the probability of bubble nucleation 
in a number of braneworld situations. The details of these calculations 
are remarkably similar for both the flat bubble and the AdS bubble. 
In this section we shall present the calculation for the flat bubble 
spacetime forming in a de Sitter false vacuum.

Consider now equations (\ref{bounceaction}a-d). Our solution 
satisfies the equations of motion both in the bulk and on the brane. 
The bulk equations of motion are just the Einstein equations (in 
Euclidean signature) with a negative cosmological constant:
\be \label{einstein}
R_{ab}-\frac{1}{ 2}Rg_{ab}=-\Lambda_n g_{ab}
\ee
from which we can quickly obtain 
\be\label{bulkint}
R-2\Lambda_n=\frac{4\Lambda_n}{n-2}=-2(n-1)k_n^2
\ee
where we have used the relation (\ref{lamkn}). 
The brane equations of motion are just the Israel equations given that 
we have a brane tension and a nested domain wall:
\be \label{Israel}
\Delta K_{ab}-\Delta K h_{ab}=8\pi G_n\sigma h_{ab} 
+8\pi G_n\mu \delta(\zeta) \gamma_{ab}
\ee
where $\sigma$ is $\sigma^{flat}$ and $\sigma^{dS}$ on the flat 
and de Sitter branes respectively. We can therefore read off 
the following expression:
\be\label{Kdiff}
\Delta K=-2(n-1)\sigma_n-\mu_n \delta(\zeta)
\ee
where we have also used (\ref{eqn:sigman}) and $\mu_n=8\pi G_n\mu$. 
We are now ready to calculate the action. Inserting (\ref{bulkint}) and 
(\ref{Kdiff}) in (\ref{bounceaction}), 
we immediately see that the contribution from the vortex is cancelled 
off by the delta function in the extrinsic curvature and we are left with
\be
S_{bounce}=\frac{2(n-1)k_n^2}{ 16\pi G_n}\int_{bulk} d^n x \sqrt{g}
-\frac{4\sigma_n^{flat}}{ 16\pi G_n}\int_{flat} d^{n-1} x \sqrt{h}
-\frac{4\sigma_n^{dS}}{ 16\pi G_n}\int_{dS} d^{n-1} x \sqrt{h}
\ee
The expression for $S_{false}$ is similar except that there is of course no 
flat brane contribution and the limits for the bulk and de Sitter brane 
integrals run over the whole of the de Sitter sphere interior and surface
respectively.

Working in Euclidean conformal coordinates (\ie\ the metric (\ref{confmet})
rotated to Euclidean signature) the bulk measure is simply
\be \label{bulkmeasure}
\sqrt{g} d^n x= \frac{\rho^{n-2} }{ (k_nu)^n}
du\, d\rho\, d \Omega_{n-2}
\ee
where $d\Omega_{n-2}$ is the measure on a unit $n-2$ sphere.
From (\ref{hyper}) and (\ref{dswallrelations}), the de Sitter brane 
is given by 
\be \label{dSwall}
(u-u_0)^2+\rho^2= u_1^2
\ee
so the induced metric on this brane is given by:
\be \label{dSinduced}
ds^2_{n-1}=\frac{1 }{ k_n^2u^2} \left [ \left ( 
\frac{k_n u_0 }{ \sigma^{dS}_n \rho(u)} \right )^2 du^2
+\rho(u)^2d\Omega^2_{n-2} \right]
\ee
where $\rho(u)$ is given in (\ref{dswallrelc}). 
As we did for the bulk, we can now read off the de Sitter brane measure:
\be \label{dSmeaure}
\sqrt{h} d^{n-1} x=  u_1\frac{ \rho(u)^{n-3}}{ (k_nu)^{n-1}}\, du\, 
d\Omega_{n-2}\,.
\ee
Now consider the flat brane. This is given by $u=u_c$ where $u_c$ is 
given by (\ref{dswallrela}), and the measure can be easily seen
to be
\be \label{flatmeasure}
\sqrt{h} d^{n-1}x = \frac{ \rho^{n-2}}{ (k_nu_c)^{n-1}}\, d\rho\,d\Omega_{n-2}.
\ee

Now we are ready to evaluate the probability term $B=S_{bounce}-S_{false}$. 
Given each of the measures we have just calculated and taking care 
to get the limits of integration right for both the bounce action 
and the false vacuum action, we arrive at the following expression:
\bea \label{bounceexpression}
B &=& -\frac{4(n-1)k_n^2}{ 16\pi G_n}\Omega_{n-2}\int_{u_0-u_1}^{u_c}du  
\int_0^{\rho(u)} d \rho  \frac{\rho^{n-2} }{ (k_nu)^n} \nonumber \\
& & -\frac{4\sigma_n^{flat}}{ 16\pi G_n}\Omega_{n-2}\int_0^{\rho(u_c)}
d \rho \frac{ \rho^{n-2}}{ (k_nu_c)^{n-1}}
+\frac{4\sigma_n^{dS}}{ 16\pi G_n}\Omega_{n-2}\int_{u_0-u_1}^{u_c} d u  
u_1\frac{ \rho(u)^{n-3}}{ (k_nu)^{n-1}} \nonumber \\
\eea
We should note that we have a factor of two in the bulk part of the 
above equation arising from the fact that we have two copies of the 
bulk spacetime. If we use the fact that:
\be
\rho(u_c)=\frac{k_n u_c }{ k_{n-1}^{dS}}\cos\zeta_0
\ee
along with $\sigma_n^{flat}=k_n$ and equation (\ref{dswallrelb}), we 
can simplify (\ref{bounceexpression}) to arrive at 
equation (\ref{probterm}).
\section{Limits and measures for action integrals} \label{app:measures}
Let us consider in more detail each contribution to the action integrals given in equations (\ref{BH action}) and (\ref{reference action}). We will start by looking at the bulk integral for the black hole action:

\be
\int_{bulk}=\int_{bulk} d^n x \sqrt{g}(R-2\Lambda_n)
\ee
From equation (\ref{ricci}), we see that $R-2\Lambda_n$ is constant and so does not cause us any problems. Given that the AdS-Schwarzschild bulk is cut off at the brane, $Z(\tau_E)$, and the horizon, $Z_H$,  we find that:

\be
\int_{bulk}=2\Omega_{n-2}\int_{-\frac{\beta}{2}}^{\frac{\beta}{2}} dt_E \int_{Z_H}^{Z(\tau_E)} dZ \ Z^{n-2} (R-2\Lambda_n)=2\Omega_{n-2}\int_{-\frac{\beta}{2}}^{\frac{\beta}{2}} \frac{Z(\tau_E)^{n-1}-Z_H^{n-1}}{n-1} (R-2\Lambda_n)
\ee
which is just equation (\ref{action1}). The factor of two comes in because we have two copies of AdS-Schwarzschild. The factor of $\Omega_{n-2}$ just comes from integrating out $\int d\Omega_{n-2}$. We now turn our attention to the bulk integral for the reference action:

\be
\int_{ref. \ bulk}=\int_{ref. \ bulk} d^n x \sqrt{g}(R-2\Lambda_n)
\ee
Again, $R-2\Lambda_n$ is constant and does not worry us. This time the AdS bulk is cut off at $\Sigma$ (given by $Z=Z(\tau_E)$), and at $Z=0$. The periodicity of the $T$ coordinate is $\beta^{\prime}$ rather than $\beta$. The bulk integral for the reference action is then:

\be
\int_{ref. \ bulk}=2\Omega_{n-2}\int_{-\frac{\beta^{\prime}}{2}}^{\frac{\beta^{\prime}}{2}} dT \int_{0}^{Z(\tau_E)} dZ \ Z^{n-2} (R-2\Lambda_n) =2\Omega_{n-2}\int_{-\frac{\beta^{\prime}}{2}}^{\frac{\beta^{\prime}}{2}} \frac{Z(\tau_E)^{n-1}}{n-1} (R-2\Lambda_n)
\ee
$\beta^{\prime}$ is fixed by the condition that the geometry of $\Sigma$ and the brane should be the same. This just amounts to saying that $T^{-1}(\pm \frac{\beta^{\prime}}{2})=\pm \tau_{max}=t_E^{-1}(\pm \frac{\beta}{2})$ where $-\tau_{max} \leq \tau_E \leq \tau_{max}$ on both $\Sigma$ and the brane. As illustrated below by changing coordinates to $\tau_E$ and then $t_E$, we arrive at equation (\ref{action2}):

\bea
\int_{ref. \ bulk} &=& 2\Omega_{n-2}\int_{-\tau_{max}}^{\tau_{max}} d\tau_E \ \frac{dT}{d\tau_E} \frac{Z(\tau_E)^{n-1}}{n-1} (R-2\Lambda_n) \nonumber \\
&=& 2\Omega_{n-2}\int_{-\frac{\beta}{2}}^{\frac{\beta}{2}} dt_E \frac{d\tau_E}{dt_E} \frac{dT}{d\tau_E} \frac{Z(\tau_E)^{n-1}}{n-1} (R-2\Lambda_n)
\eea
Consider now the brane integral:
\be
\int_{brane}= \int_{brane} d^{n-1}x \sqrt{h}\ 2K
\ee
We will use the coordinate $\tau_E$ to begin with and then change to $t_E$, thus arriving at equation (\ref{action3}):
\be 
\int_{brane}=\Omega_{n-2}\int_{-\tau_{max}}^{\tau_{max}}d\tau_E \ Z(\tau_E)^{n-2} \ 2K =\Omega_{n-2}\int_{-\frac{\beta}{2}}^{\frac{\beta}{2}} dt_E \frac{d\tau_E}{dt_E} Z(\tau_E)^{n-2} \ 2K
\ee
The procedure for arriving at equation (\ref{action4}) is exactly the same, owing to the fact that $\Sigma$ and the brane have the same geometry.

\section{Justifying $Z(\tau_E) \gg c^{\frac{1}{n-1}}$ in large $c$ limit} \label{app:largec}

Let us consider the claim made in section~\ref{sec:better} that for most brane solutions,  $Z(\tau_E) \gg c^{\frac{1}{n-1}}$ in the large $c$ limit. The governing equation for the branes in Euclidean AdS-Schwarzschild is given by equation (\ref{Euclidean brane EOM1}):

\be 
\left(\frac{dZ}{d\tau_E}\right)^2 = -aZ^2+1-\frac{c}{Z^{n-3}}  
\ee 
Now in each case, $Z \geq Z_{min}$ where $Z_{min}$ is the minimum value of $Z$ on the brane. It is sufficient to show that $Z_{min} \gg c^{\frac{1}{n-1}}$. At $Z=Z_{min}$, $\frac{dZ}{d\tau_E}=0$.
For $a=0$,  we have:

\be
Z_{min}=c^{\frac{1}{n-3}} \gg c^{\frac{1}{n-1}}
\ee
For $a>0$, we have:

\be 
Z_{min} \geq c^{\frac{1}{n-3}} \gg c^{\frac{1}{n-1}}
\ee
We see that our claim holds for supercritical and critical branes. For subcritical branes with $a<0$ we need to be more careful. $Z_{min}$ satisfies:

\be
Z_{min}^{n-3}(1+|a|Z_{min}^2)=c
\ee
If $Z_{min}^2 \ll |a|^{-1}$ then $Z_{min} \approx c^{\frac{1}{n-3}}$. If $Z_{min}^2 \sim |a|^{-1}$ then $(1+|a|Z_{min}^2) \sim c^0$ and therefore $Z_{min} \sim  c^{\frac{1}{n-3}}$. In each case we have $Z_{min} \gg  c^{\frac{1}{n-1}}$.
Finally, when $Z_{min}^2 \gg |a|^{-1}$:

\be
Z_{min} \approx \left(\frac{c}{|a|}\right)^{ \frac{1}{n-1}}
\ee 
Provided $|a| \ll 1$ we can say:
\be
Z_{min} \gg c^{ \frac{1}{n-1}}
\ee
We see, therefore that the claim made in section~\ref{sec:better} was indeed valid: $Z(\tau_E) \gg c^{\frac{1}{n-1}}$ for subcritical branes with $|a| \ll 1$ and  all supercritical and  critical branes.
\end{document}